\documentclass[preprint,floats,floatfix,amsmath,amssymb,aps,12pt,prb]{revtex4}
\usepackage{times} 
\usepackage{color} 

\usepackage{graphicx,afterpage}
\graphicspath{{.},{./Fig/}}

\usepackage{subfigure,epsfig,psfig,rotating,minitoc}
\usepackage[english,activeacute]{babel}
\usepackage{ae}

\renewcommand{\vec}[1]{\mathbf{#1}}

\newcommand{\tens}[1]{\mathbf{\underline{#1}}}

\newcommand{\bea}{\begin{eqnarray}}
\newcommand{\eea}{\end{eqnarray}}
\newcommand{\etal}{{\em et al.}}

\newcommand{\pt}{\partial}

\newcommand{\no}[1]{{\tt #1}\\}
\newcommand{\mylab}[1]{\label{#1}}
\newcounter{countuwe}

\newlength{\htw}
\begin{document}

\title{Decomposition driven interface evolution for layers of
binary mixtures: {II}. Influence of convective transport on linear stability}
\author{Santiago Madruga}
\email{santiago.madruga@upm.es,smadruga@gmail.com}
\affiliation{Universidad Polit\'ecnica de Madrid, ETSI Aeron\'auticos,
  Plaza Cardenal Cisneros 3, 28040 Madrid, Spain}
\author{Uwe Thiele}
\email{u.thiele@lboro.ac.uk}
\homepage{http://www.uwethiele.de}
\affiliation{School of Mathematics, Loughborough University,
Loughborough, Leicestershire, LE11 3TU, UK}
\begin{abstract}
  We study the linear stability with respect to lateral perturbations
  of free surface films of polymer mixtures on solid
  substrates. The study focuses on the stability properties of the
  stratified and homogeneous steady film states studied in Part I
  [U. Thiele, S. Madruga and L. Frastia, Phys. Fluids 19, 122106 (2007)]. To this aim, the
  linearized bulk equations and boundary equations are solved using
  continuation techniques for several different cases of energetic
  bias at the surfaces, corresponding to linear and quadratic solutal
  Marangoni effects.
  For purely diffusive transport, an increase in film thickness either
  exponentially decreases the lateral instability or entirely
  stabilizes the film. Including convective transport leads to a
  further destabilization as compared to the purely diffusive case. In
  some cases the inclusion of convective transport and the related
  widening of the range of available film configurations (it is then
  able to change its surface profile) change the stability behavior
  qualitatively.
  We furthermore present results regarding the dependence of the
  instability on several other parameters, namely, the Reynolds
  number, the Surface tension number and the ratio of the typical
  velocities of convective and diffusive transport. 
\end{abstract}

\maketitle

\section{Introduction} \mylab{intro2}
%

The main stages of structure formation and dewetting in thin films of
simple liquids are generally well understood
\cite{deGe85,BrDa89,ODB97,Thie03,Seem05,KaTh07}. Films of polymers rupture
forming holes, and evolve towards a network of liquid rims that may
decay subsequently into small droplets \cite{Reit92,ShRe96}. Moreover,
using physically/chemically heterogeneous substrates it is possible to
create ordered soft matter films \cite{Rock99,KKS00,KaSh01,BKTB02,TBBB03}.
For single component thin films there are two main pathways of
dewetting: spinodal surface instability and heterogeneous nucleation,
both driven by effective molecular interactions between the substrate
and the film surface like, e.g., van der Waals forces. Both mechanisms
may result in the rupture of the film leading to various patterns and
have been studied extensively in numerical and experimental works.
\cite{Reit93b,Xie98,TVN01,BeNe01,KaSh02,TWZD03,Beck03}

The dewetting of thin films composed of binary mixtures is more
involved.  Experiments on mixtures have reported dewetting mechanisms
that differ from the ones for simple liquids.  In particular,
Yerushalmi-Rozen \etal{} \cite{YKK99} discuss the phase-separation
induced dewetting of a polymer blend consisting of deuterated
oligomeric styrene and oligomeric ethylene-propylene.  They assume
that first there occurs a separation of the two components of the
blend.  The phase separation is followed by a dewetting process that
is characterized by the formation of holes at a dewetting front moving
inwards from the sample edge. This route of dewetting differs from the
classical ones for single component films by its short induction time
(at comparable film thickness), the properties of the front, and its
morphological characteristics.

The evolving gradients of concentration of the two components of the
binary mixture along the film surface originate surface tension
gradients that are responsible for an additional flow in the vicinity
of the decomposition/dewetting front. It is thought that this promotes
the acceleration of the formation of holes. However, no detailed
theoretical description is available at present. The relevance of
interfacial phenomena in binary mixtures has as well been reported in
spin-cast polymer blend films, where interfacial instabilities lead to
a horizontal phase separation \cite{HeJo05}.

Another element relevant to the dynamics of films of binary mixtures
is the energetical influence of the surfaces. In particular, an
energetically biased surface is rapidly enriched in the preferred
component and may become the 'source' of so-called spinodal
decomposition waves \cite{BaEs90,Jone91,KrDa93}. Jones \etal{}
\cite{Jone91}, for instance, study the spinodal decomposition of
critical mixtures of poly(ethylene-propylene) and its per-deuterated
variant in presence of a substrate with a preferential attraction for
the poly(ethylene-propylene). They find that that composition waves
origin at the surfaces and propagate into the bulk. Their wave-vectors
are oriented normal to the surface.

The interplay between phase separation, surface tension gradients, and
surface properties make the dynamics of films of mixtures extremely
rich and its theoretical description challenging.  The dynamics of a
fluid binary mixture is often described by the so called model-H,
which couples transport of the mass of one component (convective
Cahn-Hilliard equation) and momentum (Navier-Stokes-Korteweg
equations) \cite{HoHa77,AMW98}. It has been extensively studied for
various isothermal and non-isothermal settings
\cite{GuPo96,JaVi96,AMW98,VMM99,VMM98,BoBe03,BMB04,FNOG08}. None,
however, involves beside the diffuse internal interface a free
liquid-gas interface, i.e., a free surface. This implies that existing
theories can not be readily applied to thin film experiments with
polymer blends whose evolution is governed by the driving influence of
the dynamics of and at the free surface.

In an alternative approach Clarke \cite{Clar04,Clar05} constructs
and investigates a simple thin film model for the coupled evolution of
film thickness profile and mean concentration. It is, however, based
on the assumption that the films show no vertical concentration
profile. This implies that it can not be applied to the case of the
lateral stability of layered films we are interested in here.

The present series of works aims at the development and application of
a model-H for systems involving free surfaces. Note that the
incorporation of hydrodynamic flow is necessary even in the case of
extremely slow creeping flow. Otherwise no evolving surface
deflections can be described.
The first part (Ref.~\onlinecite{TMF07}) discusses the inclusion of
free evolving surfaces into model-H and determines basic stratified
film states for various types of energetic bias at the surfaces.  In
particular, the framework of phenomenological non-equilibrium
thermodynamics is used to derive a generalized model-H coupling
transport equations for momentum, density and entropy that is then
simplified for an isothermal setting, vanishing interface viscosity
and simplified internal energies. A comparison with literature results
and a variational derivation of the static limiting case clarify the
issue of defining pressure and chemical potential. Furthermore,
boundary conditions at the solid substrate and the free interface are
introduced. After non-dimensionalisation it is shown that the
dimensionless numbers entering the boundary conditions for the
Cahn-Hilliard and the Korteweg-Navier-Stokes equations are closely
related and can not by any means be chosen independently of each
other. Physically this means that the energetic bias with respect to 
decomposition at a free surface is intrinsically coupled to a solutal
Marangoni effect. Neither of the two effects can be considered
independently of the other one. Ref.~\onlinecite{TMF07} continues with
an analysis of steady base state solutions for laterally homogeneous
films of decomposing mixtures. A plethora of stratified solutions is
determined and ordered for various types of energetic bias at the
surfaces using continuation techniques and symmetry arguments.

This present second paper presents the analysis of the lateral
stability of the laterally homogeneous steady films states obtained in
Ref.~\onlinecite{TMF07}, i.e., it determines the stability of
homogeneous and layered films with respect to modulations in thickness
and/or composition along the substrate. A similar approach (however,
with other boundary conditions than used here) was recently employed to obtain
instability thresholds for long-wave instabilities \cite{FNOG08}. The paper
is organized as follows.  In Section~\ref{dimlessmod} we review the
non-dimensional model-H including appropriate boundary conditions at
the solid substrate and the free film surface. In
Section~\ref{sec-linstab2} equations are linearized for base states
corresponding to homogeneous and stratified films. We as well discuss
a realistic range of parameters for polymer blends to be used in the
calculations. The stability results for homogeneous films are
presented in Section \ref{sec-homo} for neutral and energetically
biased surfaces.
The stability of stratified films for neutral and biased surfaces is
presented in Section \ref{sec-hete}.  Finally, conclusions are given
in Section~\ref{sec:conclusions}.
%
\section{Dimensionless model} \mylab{dimlessmod}
%
We consider a film of a binary mixture on a horizontal homogeneous solid
substrate. The film has a free surface that may evolve in time. 
The system is two-dimensional and infinitely extended in the
horizontal direction. The origin of the Cartesian frame is fixed
to the substrate. The dimensionless governing equations
expressing the balance of mass and momentum, and the continuity
equations are re-derived and discussed in Ref.~\onlinecite{TMF07}. They read
\begin{equation}
\text{Ps}\left[\frac{\partial \vec{v}}{\partial t}\,+\,\vec{v}\cdot\nabla\vec{v}\right]=\,
-\nabla\cdot\left\{(\nabla c)(\nabla c)+p_{\mathrm{eff}}\tens{I} \right\}
+ \frac{\text{Ps}}{\text{Re}} \Delta\vec{v}, \mylab{e:velo}
\end{equation}
\begin{equation}
\partial_t  c + \vec{v}\cdot\nabla c \,=\, 
- \nabla \cdot\left\{ \nabla [ \Delta c - \partial_c f(c)]\right\}, \mylab{e:con}
\end{equation}
and
\bea
\nabla \cdot \vec{v} =0,
\eea
respectively, with the operators $\nabla=\left(\pt_x,\pt_z \right)$,
$\Delta=\left(\pt_x^2+\pt_z^2 \right)$, and the velocity field
$\vec{v}=\left(u,w\right)$. The composition field $c=c_1-c_2$
represents the difference of concentration of the two components of
the mixture, $p_{\mathrm{eff}}=p-(c+1)\Delta c-\left(\Delta c
\right)^2/2$ is an effective pressure comprising all diagonal terms of
the stress tensor, and $p$ is the 'normal' pressure. The local bulk
free energy is assumed to correspond to the simple quartic potential
$f(c)=(c^2-1)^2/4$.  To make the equations~(\ref{e:velo})
and~(\ref{e:con}) dimensionless, in Ref.~\onlinecite{TMF07} the scales
$l=C\sqrt{\sigma_c/E}$, $U=M\,E/l\,C^2$, $\tau=l/U=l^2\,C^2/(M E)$ and
$P=E$ are used for length, velocity, time and pressure
respectively. The length $l$ corresponds to the thickness of the
diffuse interface between the two phases of the mixture and is
determined from the coefficient $\sigma_c$ of the gradient term of the
convective Cahn-Hilliard equation (cf.~Ref.~\onlinecite{TMF07}), the
energy scale $E$ and the concentration $C$ at the binodal. The
parameter $M$ is the diffusion coefficient in the Cahn-Hilliard
equation.  With these scales, two dimensionless numbers appear in the
bulk equations~(\ref{e:velo}) and (\ref{e:con}), the pressure number
$\text{Ps}=\rho\,\text{M}^2\,E^2/\text{C}^6\,\sigma_c$ and the
Reynolds number $\text{Re}= \text{M}\,\text{E}\rho/\eta\,C^2$.  Note
that, as the diffusion constant is $D=ME$ and $U\sim D/l$, the used
Reynolds number can be seen as an inverse Schmidt number Sc$=\eta/\rho
D$.  The ratio Re/Ps=$\sigma_c C^4/\eta M E$ will turn out to be the
most important bulk parameter for the extremely slow creeping flow we
are interested in (see below Section~\ref{sec:param}). It corresponds
to the ratio of the typical velocity $U’=\sigma/\eta$ of the viscose
flow driven by the internal 'diffuse interface tension'
$\sigma=\sigma_c/l$ and the typical velocity of diffusive processes
$U\sim M\,E/l=D/l$. This implies that Re/Ps can as well be seen as a
Peclet number Pe=$U' l/D$ or as an inverse Capillary number
Ca$^{-1}=\sigma/U\eta$.  Such alternative scalings are used, for
instance, in Refs.~\onlinecite{VMM99} and \onlinecite{VMM98}.

Equations~(\ref{e:velo}) and (\ref{e:con}) are supplemented with
boundary conditions for the concentration and velocity
fields.\cite{TMF07} We first present the conditions for the
concentration field.  At the solid substrate ($z=0$) one prevents a
diffusive mass flux through the substrate and allows for an
energetic bias, i.e.,
\bea
\pt_z\left[(\pt_{xx}+ \pt_{zz})c -\pt_c f(c) \right]&=&0\\
-\pt_z c+ \text{S} \pt_c f^{-}(c) &=&0
\eea
respectively. The energy bias is $f^-(c)=\gamma_s+a^-\,c+b^-c^2/2$,
where S$\gamma_s$ is the dimensionless solid-liquid interface tension
at $c=0$. Note that it does not influence later calculations as only
$\partial_cf^-(c)$ enters.  Parameters $a^-$ and $b^-$ model
preferential adsorption of one of the species at the substrate and
changes in the mixing/demixing behaviour of the species at the substrate,
respectively.  The dimensionless parameter S$=\gamma_0/l\,E$ is the
dimensionless surface tension of the liquid-gas interface, and
$\gamma_0$ is the reference surface tension at $c=0$.
Similar conditions are applied at the (curved) free surface
($z=h(x,y,t)$), i.e.
\bea
\left[\pt_z -(\pt_x h)\pt_x \right] \left[(\pt_{xx}+ \pt_{zz})c -\pt_c f(c)
  \right]&=&0 \mylab{e:evol-c-surf-a} \\
\left[\pt_z -(\pt_x h)\pt_x \right] c + \text{S} \pt_c f^{+}(c)\sqrt{1+\left(\pt_x
  h\right)^2} &=&0 \mylab{e:evol-c-surf-b}
\eea
with $f^+(c)=1+a^+\,c+b^+c^2/2$. Parameters $a^+$ and $b^+$ quantify
preferential adsorption of one of the species and changes in the
the mixing/demixing behaviour at the free interface, respectively.

The boundary conditions for the velocity field are the no-slip and
no-penetration condition at the substrate, i.e., $v=w=0$ at $z=0$. At
the free surface the conditions result from the balance of tangential forces
\bea
&&-\left[ \pt_x c+(\pt_x h)\pt_z c \right] \left[  \pt_z c-(\pt_x h)\pt_x c  \right]\\
&&+\frac{\text{Ps}}{\text{Re}}\left[(u_z+w_x)(1-h_x^2)+2(w_z-u_x)h_x \right] \nonumber \\
&& =\text{S} \sqrt{1+h_x^2}\left[\pt_x +(\pt_x h)\pt_z\right]f^+(c) \nonumber  
\eea
and normal forces
\bea
&&-\frac{1}{1+h_x^2}\left[ \pt_z c-(\pt_x h)\pt_x c \right]^2-p_{\mathrm{eff}}\\
&&+\frac{\text{Ps}}{\text{Re}}\frac{2}{1+h_x^2}\left[u_x h_x^2+w_z-h_x(u_z+w_x) \right] \nonumber\\
&& =  \text{S} f^+(c) \pt_x \left[ \frac{\pt_x h}{(1+h_x^2)^{1/2}}  \right].  \nonumber 
\eea
Notice that the tangential gradient of $f^+(c)$   enters
the tangential force balance corresponding to a solutal Marangoni
force. Furthermore, at the free surface one has a kinematic condition
assuring that the free surface follows the velocity field
\bea
\pt _t h = w - u \pt_x h.
\mylab{e:kine}
\eea
After deriving the system of equations and boundary conditions,
Ref.~\onlinecite{TMF07} embarks on an extensive study of quiescent, vertically
homogeneous and vertically stratified base state solutions that are
all homogeneous (or translationally invariant) with respect to the
direction parallel to the substrate. The corresponding 
concentration profiles $c_0(z)$ are obtained by solving
the steady bulk equation (steady version of Eq.~(\ref{e:con}))
\bea
0=\pt_{zz}\left[\pt_{zz}c_0-\pt_c f(c)|_{c_0}\right]
\mylab{ss1}
\eea
with the boundary conditions
\bea
0=\pt_{z}\left[\pt_{zz}c_0-\pt_c f(c)|_{c_0}\right] \qquad\mbox{at}\qquad z=0,h_0
\eea
and
\bea
0=\pm \pt_z c_0+\text{S} \pt_c f^\pm(c)|_{c_0} \quad\mbox{at}\quad
z=0,\quad h_0. \mylab{e:bc-base-state}
\eea
Here we are concerned with the stability of the base state concentration profiles $c_0(z)$
with respect to perturbations along the direction parallel to the substrate.
The rational behind this approach is the experimental observation that
thin films of a decomposing mixture first stratify vertically on a
relatively short time scale and then develop on a slower time scale a
horizontal structure. Here, we want to capture the characteristics of
the latter process by the linear stability analysis.
%
\section{Linearized equations for vertically stratified and homogeneous films}
\mylab{sec-linstab2}
%
\subsubsection{General ansatz}
%
To analyze the stability of the quiescent base states with respect to
infinitesimally small perturbations, we write the general solution of
the problem in the form $\vec{v}=\vec{v_0}+\varepsilon\vec{\tilde
v_1}$, $p_{\textrm{eff}}=p_0+\varepsilon \tilde p_1$,
$c=c_0+\varepsilon \tilde c_1$, and $h=h_0+\varepsilon \tilde h_1$,
with $\vec{v_0}=0$ and $p_0=-(\partial_z c_0)^2$. The fields
$\varepsilon \vec{\tilde v_1}$, $\varepsilon \tilde p_1$, $\varepsilon
\tilde c_1$, and $\varepsilon \tilde h_1$ denote the infinitesimal
perturbations of velocity, pressure, concentration, and thickness
fields, respectively. The small parameter $\varepsilon$ will be
used to order terms in the series expansion.  
The perturbations are decomposed into a sum of normal modes
$(\vec{\tilde v_1},\tilde p_1,\tilde c_1,\tilde
h_1)=(\vec{v}_1(z),p_1(z),c_1(z),h_1)\exp(\beta t +i k x)$, where
$\beta$ is the growth rate and $k$ the lateral wavenumber.  Using
this ansatz in Eqs.~(\ref{e:velo}) to~(\ref{e:kine}) we obtain the
linearized convective Cahn-Hilliard equation
\begin{equation}
\beta  c_1 + w_1 \partial_z c_0 \,=\, 
- (\partial_{zz}-k^2) \left[(\partial_{zz}-k^2) c_1 -  c_1 \partial_{cc} f|_{c_0}
\right].
\mylab{mh-eqc-lin3}
\end{equation}
and the linearized momentum equation
\begin{eqnarray}
\beta\,\text{Ps}\, \vec{v}_1\,&=&\,
-\nabla\cdot\left\{ [(\nabla c_0)(\nabla c_1)+(\nabla c_1)(\nabla c_0)] + p_1\,\tens{I}
\right\} \nonumber \\
&& + \frac{\text{Ps}}{\text{Re}} (-k^2+\partial_{zz}) \vec{v}_1
\mylab{mh-mom-lin3}
\end{eqnarray}
where the tensor $[(\nabla c_0)(\nabla c_1)+(\nabla c_1)(\nabla
c_0)]=[(0,ik c_1 \partial_z c_0),(ik c_1 \partial_z c_0, 2(\partial_z
c_1) (\partial_z c_0))]$, i.e. splitting the velocity field in its
components we obtain
\begin{eqnarray}
\beta \text{Ps} \, u_1 \,&=&\,-ik \partial_z( c_1 \partial_z c_0)
- i k p_1\nonumber\\ 
&&+ \frac{\text{Ps}}{\text{Re}} (-k^2 + \partial_{zz}) u_1
\mylab{mh-mom-lin4a}\hspace*{0.6cm} \\
\beta \text{Ps}\, w_1\,&=&\,-[-k^2 c_1 \partial_z c_0 
+ 2\partial_z((\partial_z c_0)(\partial_z c_1))]
- \partial_z p_1\, \nonumber \\ &&+ \frac{\text{Ps}}{\text{Re}}(-k^2 +
\partial_{zz}) w_1.
\mylab{mh-mom-lin4b}
\end{eqnarray}
The incompressibility  condition leads to
\begin{equation}
ik u_1 +\partial_z w_1 \,=\,0.
\mylab{cont-lin3}
\end{equation}
The linearized boundary conditions assuring zero mass flux through the
substrate at $z=0$ and through the free surface at $z=h_0$ are both of
the form
\begin{equation}
0\,=\, \partial_z \left[  (-k^2+\partial_{zz}) c_1 - c_1 \partial_{cc} f|_{c_0}
\right].
\mylab{bc-lin1c}
\end{equation}
The linearized energy bias conditions for the concentration are
\begin{eqnarray}
-\pt_z c_1+ \text{S} c_1 \pt_{cc} f^{-}|_{c_0} &=&0 \quad\mbox{at}\quad z=0\\
\mbox{and}\quad\pt_z  c_1 + \text{S} c_1 \pt_{cc} f^{+}|_{c_0} &=&0 \quad\mbox{at}\quad z=h_0.
\mylab{bc-lin2c}
\end{eqnarray}
The non-slip condition for the velocity at the substrate becomes
\begin{equation}
u_1\,=\, w_1 \,=\,0.
\mylab{bc-linz0}
\end{equation}
and the linearized  tangential and normal force balance at the free interface are
\begin{eqnarray}
&&\frac{\text{Ps}}{\text{Re}}\,(\partial_z u_1\,+\,ik w_1)=0,
\mylab{bc-lin3c}\\
&& 2(\partial_z c_0)(\partial_z c_1) + p_1
-2\frac{\text{Ps}}{\text{Re}}\partial_z w_1=\text{S} f^+|_{c_0} k^2
h_1, 
\mylab{bc-lin4c} 
\end{eqnarray}
respectively. Finally, the kinematic condition reads
\begin{equation}
w_1 \,=\,\beta h_1.
\mylab{bc-kin3}
\end{equation}
Note that the r.h.s.\ of Eq.~(\ref{bc-lin3c}) is exactly zero
following from the boundary condition Eq.~(\ref{e:bc-base-state}) of
the base state.  Physically this means that the Korteweg and the
Marangoni stress at the free surface exactly compensate, i.e.\ all the
tangential driving force is already contained in the linearized bulk
equation. The situation would be different if we were to allow for a
dynamic surface tension different from the used static one.
%
\subsubsection{Eigenvalue problem for stratified film}
%
To carry out the stability analysis of the linearized model-H we
eliminate the pressure in the momentum equation and boundary
conditions, and write the linearized model-H as an eigenvalue problem
of the form
\bea
\partial_{zzzz}c_1&&\,=\,-(\beta  c_1 + w_1 \partial_z c_0) \nonumber\\
&&- \left[(k^4 - 2k^2 \partial_{zz}) c_1 -(\partial_{zz}-k^2)(c_1 \partial_{cc} f|_{c_0})
\right],\mylab{ds1}
\eea
and
\begin{eqnarray}
\partial_{zzzz} w_1
&&\,=\,\beta \,\text{Re}\,(\partial_{zz} - k^2) w_1 
\,+\frac{\text{Re}}{\text{Ps}}\,k^4 \,c_1 \partial_z c_0 \nonumber \\ 
&&\,-\,\frac{\text{Re}}{\text{Ps}}\, 2 k^2 \partial_z[(\partial_z
c_0)(\partial_z c_1)]\nonumber \\
&&+\frac{\text{Re}}{\text{Ps}}\,k^2\partial_{zz}(c_1\partial_z c_0) 
+ 2k^2\partial_{zz}w_1-k^4 w_1,
\mylab{ds2}
\end{eqnarray}
with the boundary conditions
\begin{equation}
w_1\,=\,\partial_z w_1\,=\,0 \qquad\mbox{at}\qquad z=0;
\mylab{bc-orso1}
\end{equation}
\begin{equation}
\frac{\text{Ps}}{\text{Re}}(\partial_{zz} w_1 + k^2 w_1)\,=\,0
\qquad\mbox{at}\qquad z=h_0;
\mylab{bc-orso2}
\end{equation}
and
\bea
&&k^2(\partial_z c_0)(\partial_z c_1) 
\,-\, k^2 \,c_1 \partial_z^2 c_0
\,+\,\frac{\text{Ps}}{\text{Re}}(\partial_{zz}-3k^2) \partial_z w_1 \nonumber \\
&&-\beta \,\text{Ps}\,\partial_z w_1 =\,\text{S} f^+|_{c_0}
k^4 h_1 \qquad\mbox{at}\qquad z=h_0; \mylab{bc-orso3-het}
\eea
where $h_1=w_1/\beta$. Furthermore, 
\begin{equation}
 \partial_z \left[ (-k^2+\partial_{zz}) c_1 - c_1 \partial_{cc} f|_{c_0}
 \right]=0 \quad\mbox{at}\quad z=0,h_0;
\mylab{bc-lin1c-2}
\end{equation}
\begin{eqnarray}
-\pt_z c_1+ \text{S} c_1 \pt_{cc} f^{-}|_{c_0} &=&0
\quad\mbox{at}\quad z=0;\mylab{bc-lin2c-2-a} \\
\pt_z  c_1 + \text{S} c_1 \pt_{cc} f^{+}|_{c_0}  &=&0 \quad\mbox{at}\quad z=h_0.
\mylab{bc-lin2c-2}
\end{eqnarray}
%
\subsubsection{Eigenvalue problem for  homogeneous films} \mylab{sec-linstab1}
%
For homogeneous films the concentration does not depend on $z$, i.e., the base state 
is uniform along both, the $x-$ and $z$-direction. In consequence, the eigenvalue problem 
reduces to the bulk equations
\begin{equation}
\partial_{zzzz}c_1\,=\,-\beta  c_1 
-(k^4 - 2k^2 \partial_{zz}) c_1  -(\partial_{zz}-k^2)(c_1 \partial_{cc} f|_{c_0}),
\mylab{ds1-homo}
\end{equation}
and
\begin{eqnarray}
\!\!\partial_{zzzz} w_1 \!=\! \beta \,
\text{Re}\,(\partial_{zz} - k^2) w_1  + 2k^2\partial_{zz}w_1-k^4 w_1,
\mylab{ds2-homo}
\end{eqnarray}
with the boundary conditions
\begin{equation}
w_1\,=\,\partial_z w_1\,=\,0 \mbox{ at } z=0;
\mylab{bc-orso1-homo}
\end{equation}
\begin{equation}
\frac{\text{Ps}}{\text{Re}}(\partial_{zz} w_1 + k^2 w_1)\,=\,0
\qquad\mbox{at}\qquad z=h_0; \mylab{bc-orso2-homo}
\end{equation}
\begin{eqnarray}
&&\frac{\text{Ps}}{\text{Re}}(\partial_{zz}-k^2) \partial_z w_1-\beta \,\text{Ps}\,\partial_z w_1
\,-\,\frac{\text{Ps}}{\text{Re}}\,2k^2\partial_z w_1 \nonumber \\ 
&&\,=\,\text{S}f^+|_{c_0} k^4 w_1/\beta  \quad\mbox{at}\quad z=h_0; \mylab{bc-orso3}
\end{eqnarray}
\begin{equation} 
\partial_z \left[  (-k^2+\partial_{zz}) c_1 -c_1\partial_{cc} f|_{c_0} \right]
\,=\,0 \qquad\mbox{at}\qquad z=0,h_0;
\mylab{bc-lin2}
\end{equation} 
\begin{eqnarray}
-\pt_z c_1+ \text{S} c_1 \pt_{cc} f^{-}|_{c_0} &=&0 \quad\mbox{at}\quad z=0; \mylab{bc-lin3-homo-a}\\
\mbox{and}\quad
\pt_z  c_1 + \text{S} c_1 \pt_{cc} f^{+}|_{c_0}  &=&0 \quad\mbox{at}\quad z=h_0. \mylab{bc-lin3-homo}
\end{eqnarray}
Inspection of the equations shows that the perturbations in the
concentration and in the velocities do decouple entirely, i.e., the
linear stability problem for the concentration field in a homogeneous
film reduces to the one resulting from the Cahn-Hilliard equation in a
slab \cite{Kenz01}. The decoupled velocity perturbations are all
damped out.
%
\subsection{Numerical technique}
\mylab{sec:num}
%
In the present work the calculations of the base states and their
linear stability are carried out using numerical continuation
techniques bundled in the package AUTO\cite{AUTO97,DKK91,DKK91b}.

Continuation techniques allow to obtain solutions of a problem for a
given set of control parameters by 'extrapolation' from known
solutions that are nearby in the parameter space. In particular, the
set of equations for the steady state (\ref{ss1}) \textit{and} the for
real linear perturbations (\ref{ds1})-(\ref{ds2}) can be written as a
10-dimensional dynamical system
$\vec{y}'(z)=\vec{f}(\vec{y}(z),\lambda)$, where
$\vec{y}=(c_0,c_{0z},c_1,c_{1z},c_{1zz},c_{1zzz},w_1,w_{1z},w_{1zz},w_{1zzz})$
and $\lambda$ denotes a set of control parameters (in our case Re,
Ps$/$Re, S, $\beta$, $k$, $\bar{c}$).  This system of ordinary
differential equations together with the boundary conditions at $z=0$
and $z=h$ (\ref{bc-orso1})-(\ref{bc-lin2c-2}) and one integral
condition (mass conservation) is discretized in space and the
resulting algebraic system is solved iteratively.  The package
AUTO\cite{AUTO97} uses the method of orthogonal collocation to
discretize solutions, where the solution is approximated by piecewise
polynomials with two collocation points per mesh interval. The mesh is
adaptive to equidistribute the discretization error. Starting from
known solutions, AUTO tries to find nearby solutions to the
discretized system by using a combination of Newton and Chord
iterative methods. Once the solution has converged, AUTO proceeds
along the solution branch by a small step in the parameter space
defined by the free continuation parameters and restarts the
iteration.  Boundary conditions and integral conditions require
additional free parameters which are determined simultaneously and are
part of the solution of the differential equation.  The package AUTO
is limited to the continuation of ordinary differential equations, and
has been successfully applied to other thin film problems like, e.g.,
the determination of dispersion relations for transverse instabilities
of advancing liquid fronts and ridges \cite{ThKn03}; the determination
of steady and stationary thickness profiles and their linear stability
for sliding \cite{Thie01,Thie02,ThKn04}, running \cite{JBT05} or
depinning \cite{ThKn06} droplets; the analysis of steady film profiles
for epitaxial growth \cite{Thie08}; and the analysis of surface waves
in falling film problems \cite{Sche05,TVK06,TGV09}.

The main difficulty is usually to provide a starting solution for the
continuations. Here, we follow different sequences of steps starting
from trivial analytically known solutions.  For $n=0$ branches (for
definition see below) we (i) increase the thickness starting from a
homogeneous solution of small thickness, (ii) increase the energetic
bias at the surfaces if needed, (iii) look for branching points when
increasing the linear growth rate for some fixed wave number, (iv)
calculate the dispersion relation and its maximum, (v) and follow the
maximum in a multi-parameter continuation. For $n\ne 0$ branches we
(i) start with some fixed thickness, increase the energetic bias at
the surface if needed and identify branching points where non-trivial
solutions emerge (ii) increase the thickness following the various
emerging branches, (iii)-(v) as above. In this way the linear stability 
of the various stratified films can be determined in a rather 
effective manner.
%
\subsection{Parameter range}
\mylab{sec:param}
%
Due to the large number of dimensionless parameters needed to describe
a film of a binary mixture with a free surface, we proceed to discuss
the stability results in a restricted range of parameters that is of
particular experimental interest. For instance, the polymer blend
Polystyrene/Polyvinylmethylether (PS/PVME) has been widely used in
thin film experiments, without\cite{Scha01,MoTesis04} and
with\cite{MoTesis04} the presence of an external electric field. We
will use this mixture as a reference for our calculations. The density
of polystyrene at $T=170^o\,C$ is $\rho\approx 0.987 \cdot
10^3$kg/m$^3$, its viscosity is $\eta=4062$kg/ms, and its surface
tension is $\gamma_0=0.03$ N/m.  The surface tension of
Polyvinylmethylether is $\gamma_0=0.021$ N/m.  A linear estimate of
the variation of the surface tension of the mixture with the
concentration gives $d\gamma/dc=0.018$ N/m.\cite{MoTesis04}

To calculate the coefficients of the free energy
$a$ and $b$,\cite{note1} and the coefficient of the gradient
term in the Cahn-Hilliard equation $\sigma_c$ (cf.~Ref.~\onlinecite{TMF07} for a
more detailed discussion of the coefficients), we use the Flory-Huggins
model \cite{ScBi85,GeKr03}. In this model one supposes that the
monomers of the chains of the two polymers composing the binary
mixture occupy different sites on a square lattice.  For polystyrene
molecules with typically $1000$ monomers, and a lattice spacing of
$\sim 10^{-10}\,$m the random phase
approximation\cite{St97b,GeKr03,FlBu96} gives $\sigma_c\sim 10^{-16}$
at $T=170^o\,C$.\cite{note3}
This finally allows us to estimate $a=-\mathcal{O}(1)$
and $b=\mathcal{O}(1)$. 
The scale of the concentration field $C$ must be $\le \mathcal{O}(1)$
and we consider $C\sim 0.5$, consistent with $C=\sqrt{a/b}$.

The diffusion coefficients for polymers measured in experiments range
from $10^{-17}$ to $10^{-19}\,$cm$^2$/s.
\cite{BuBu99,QiBo99,CoKr98,GrKr86} We choose here the value obtained
by Reiter and Steiner \cite{ReSt91} of $D\sim 10^{-17}\,$cm$^2$/s
because their case is close to ours. The mobility coefficient of the
Cahn-Hilliard equation $M$ is calculated according to the relation
$M=D/E$ with the characteristic energy per unit volume $E=4a^2/b$ and
the characteristic diffusivity $D$. We find $M\sim
10^{-20}\,$m$^3$s/kg. The obtained parameters correspond to a
characteristic time scale of a few hundredths of a second
and a characteristic length scale of some nanometers. 
From the estimates of the physical parameters we obtain the
dimensionless parameters Re$\simeq 10^{-17}$ and Ps$\simeq 10^{-15}$.
Therefore, we set Re$=$Ps$=0$ but keep the ratio Re$/$Ps as an
important parameter of order one. Normally, we will keep Re$/$Ps$=1$
fixed but we investigate its influence below in Section~\ref{sec:pars}.
These values correspond to the fact that polymers
flow in an extremely slow creeping flow as expected. However, as
explained before that flow has to be taken into account in order to
explain and describe the evolution of the film profile. In
consequence, the only free parameter that enters the bulk equations
(\ref{mh-eqc-lin3}) and (\ref{mh-mom-lin3}) is the ratio Re/Ps.
Next we discuss the dimensionless numbers that enter the boundary
conditions.
For the surface tension number S, we obtain S$\simeq 10^3$. However, it
depends linearly on the interfacial tension and therefore on
the type of polymers of the binary mixture. Therefore we will allow it to vary.
Finally, $a^+$, $a^-$, $b^+$, and $b^-$ depend on the nature of the
surfaces and therefore we will treat them as free surface parameters.

From the original large number of parameters we, finally retain as
free parameters for realistic binary mixtures the time scale ratio
Re/Ps important for the bulk flow, and the numbers related to surface
properties $a^\pm$, $b^\pm$ and surface tension number S. The
remaining dimensionless numbers discussed above or in
Ref.~\onlinecite{TMF07} are either very small, considered as zero for
all purposes, or equal to one because of the scaling used.
%
%
%
\section{Linear stability of homogeneous films} \mylab{sec-homo}
%
Homogeneous base states are characterized by a concentration $c_0$
constant across the sample and a quiescent fluid $\vec{v_0}=0$.
Arbitrary concentrations $c_0$ are possible for neutral surfaces
$a^\pm=b^\pm=0$. An energetic bias at the surfaces leads to the
restriction $c_0=-a^+/b^+=-a^-/b^-$ (see Ref.~\onlinecite{TMF07})
whenever $b^\pm \ne 0$.

Inspecting Eqs.~(\ref{ds1-homo}) to (\ref{bc-lin3-homo}) we note that
the evolution of the perturbation of the concentration field $c_1$ is
independent of the perturbation of the velocity field $w_1$, i.e., the
perturbation fields $c_1$ and $w_1$ are neither coupled in the bulk
equations nor in the boundary conditions. This decoupling of $w_1$ and
$c_1$ results in $w_1=0$ and implies that hydrodynamics has no
influence on the evolution of the homogeneous film in the linear
stage.  A further consequence is the lack of surface deflection
defined as $h_1=w_1/\beta$.

The decoupling of velocity and concentration fields implies that all
unstable modes are purely diffusive and correspond to the ones that
can be obtained in a model based solely on the Cahn-Hilliard
equation. We stress again that no surface deflection can develop.  In
the following we study the diffusive modes to put the present results
in the context of results obtained in the literature.
%
\subsection{Neutral surfaces $a^\pm=b^\pm=0$}
\mylab{sec-homneu}
%
Neutral surfaces correspond to the simplest possible configuration for
a binary film. The bounding substrate and the free surface do not
prefer attachment by any component. They do neither influence mixing
or demixing phenomena. They only confine the film passively. Thereby,
however, they restrict the possible instability modes as compared to
the bulk.
\begin{figure}[t]
(a) \includegraphics[width=0.5 \hsize,angle=0]{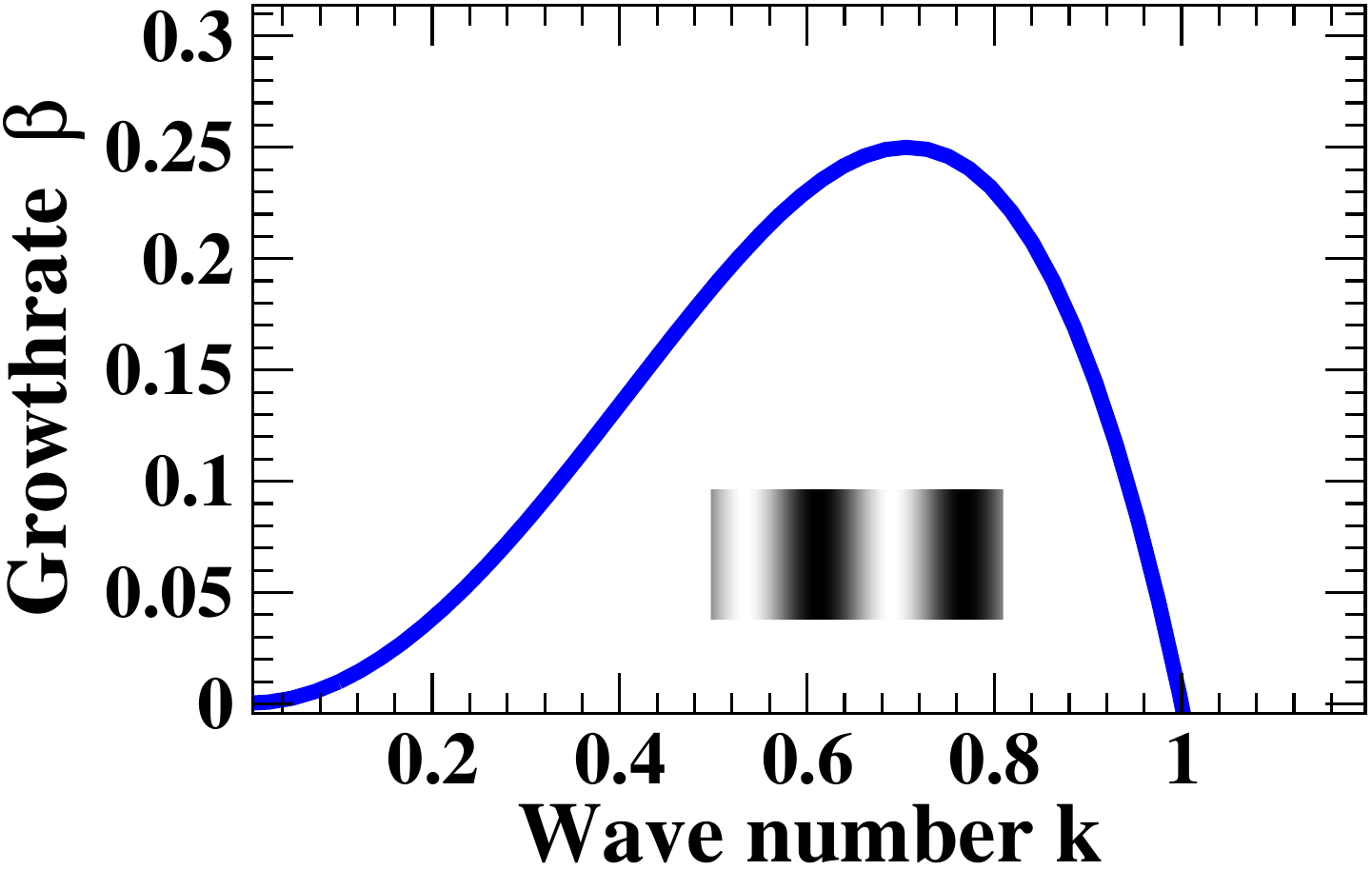}\\
(b) \includegraphics[width=0.5 \hsize,angle=0]{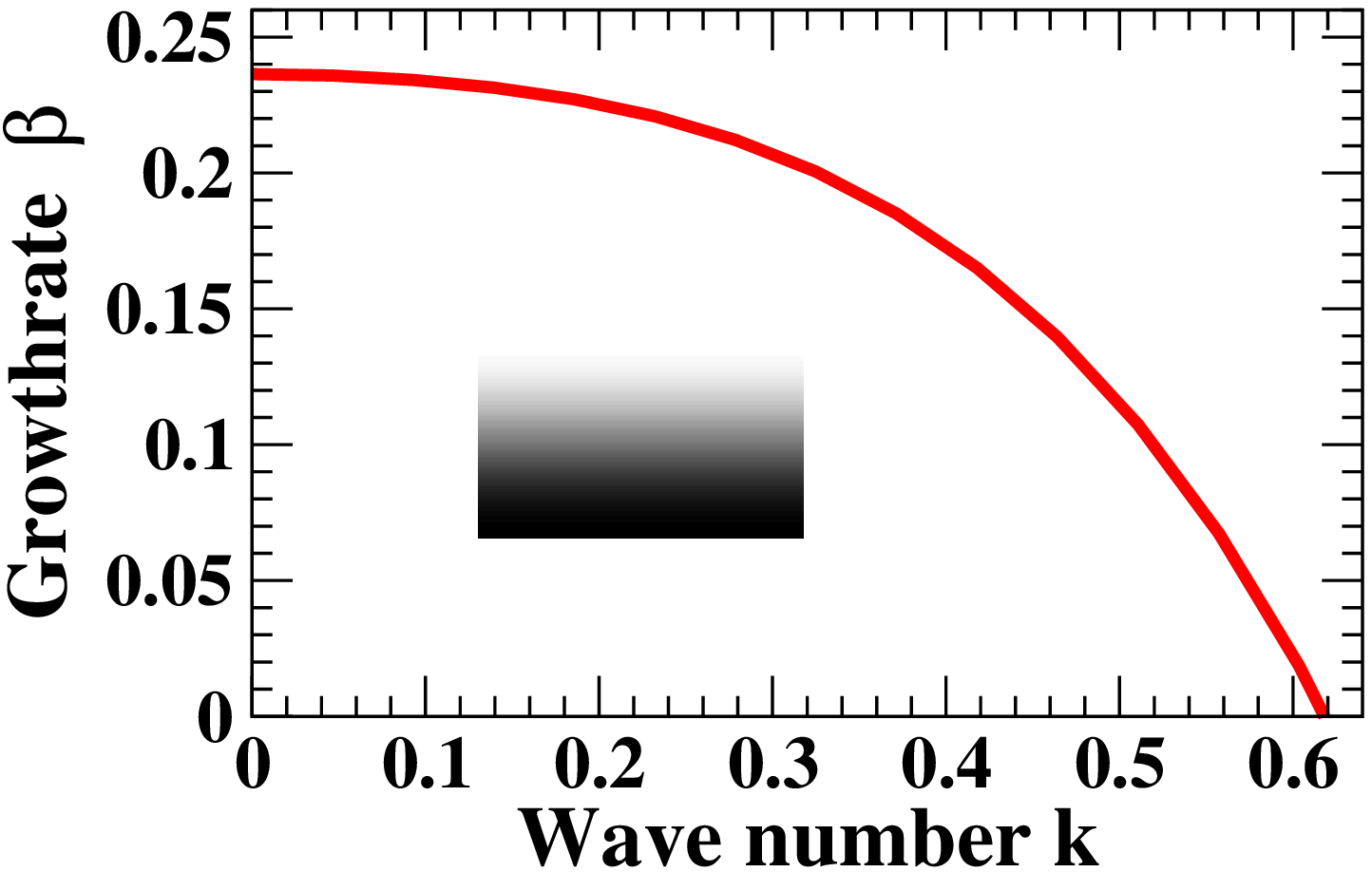}\\
(c) \includegraphics[width=0.5 \hsize,angle=0]{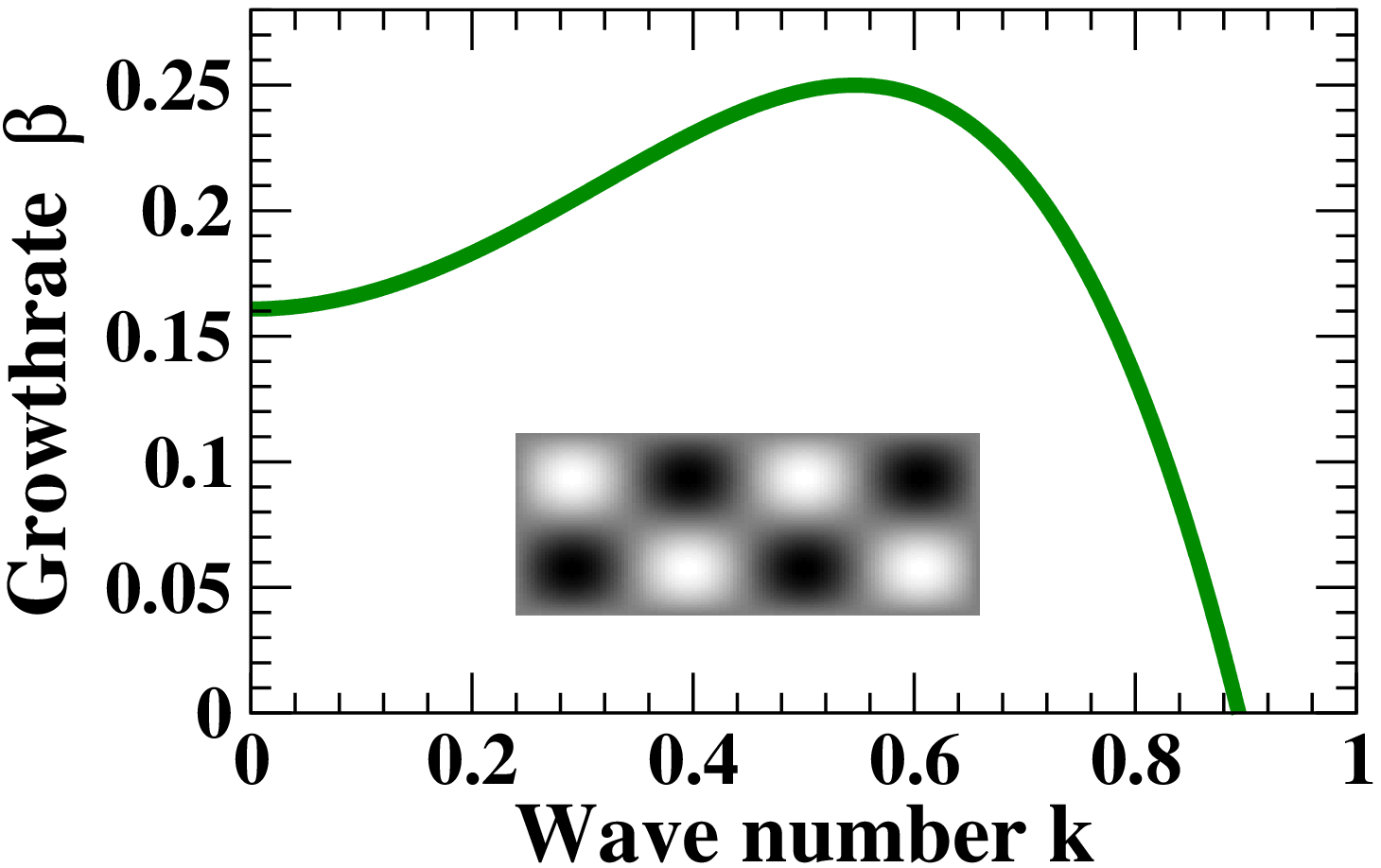}\\
\caption{ (color online) Selected dispersion relations for a {\em
    homogeneous} film of a critical mixture $c_0=0$ with energetically
  neutral surfaces $a^\pm=b^\pm=0$ and thicknesses (a) $h=3$
  (horizontal mode), (b) $h=4$ (vertical mode), and (c) $h=7$ (mixed
  mode). The insets show sketches of the evolving concentration
  patterns for the corresponding modes.  The dispersion relations
  discussed below for homogeneous films with energetically biased
  surfaces and as well for stratified films have similar shapes 
and will not be shown.\\
  \mylab{f:homo}}
\end{figure}
From Eq.~(\ref{ds1-homo}) we obtain the dispersion relation for neutral
surfaces
\bea
 \beta=-q^2\left(q^2+\pt_{cc}f(c)|_{c_0}  \right)\mylab{e:disQ}
\eea
with $q^2=k^2+k_z^2$ and the vertical eigenfunctions are of the form
$\sim \exp{\left(ik_zz\right)}$. For off-critical mixtures (i.e.,
$c_0\neq0$) $\pt_{cc}f(c)|_{c_0}=3c_0^2-1$, i.e., spinodal
decomposition only occurs for $|c_0|<1/\sqrt{3}$. Here, we entirely
focus on critical mixtures ($c_0=0$) with $\pt_{cc}f(c)|_{c_0}=-1$.
As the system is infinitely extended in $x$-direction the horizontal
wavenumber $k$ is continuous. However, the vertical wavenumber can
only take discrete values $k_z=2\pi m/h$ where $m=0,1/2,1,\dots$ is a
mode number.  The values 0, 1/2 and 1 correspond to a vertically
homogeneous mode, a vertical two-layer mode and a vertical three-layer
sandwich mode, respectively.

The $m=0$ mode is characterized by a purely horizontal variation of
the concentration field $c_1$ (cf.\ Fig.~\ref{f:homo}(a)) without any
vertical structure ($k_z=0$). For neutral surfaces
it is present for all film thicknesses. The corresponding dispersion
relation $\beta_{0}=-k^2\left(k^2-1 \right)$ is shown in
Fig.~\ref{f:homo}(a).  The maximum growth rate is
$\beta^{m=0}_{\mathrm{max}}=1/4$ at $k_{\mathrm{max}}=1/\sqrt{2}$. Due
to mass conservation $\beta=0$ for $k=0$.  The mode corresponds to the
long-wave {\em bulk mode} of the Cahn-Hilliard equation.\cite{Lang92,MST96}

The first mode with a vertical structure is the $m=1/2$ mode. It is
unstable for $h>\pi$. For $h\le\sqrt{2}\pi$ and $h>\sqrt{2}\pi$ the
mode has the dispersion relations presented in Figs.~\ref{f:homo}(b)
and \ref{f:homo}(c), respectively.
For $h\le\sqrt{2}\pi$ the mode shows its maximum growth rate
$\beta^{m=1/2}_{\mathrm{max}}$ at zero horizontal wavenumber $k=0$,
i.e., it is purely vertical. Note that for
neutral surfaces $\beta^{m=1/2}_{\mathrm{max}}$ is always smaller than
$\beta^{m=0}_{\mathrm{max}}$ (equal at $h=\sqrt{2}\pi$, see
Fig.~\ref{f:homo-max}(a)). For $h>\sqrt{2}\pi$ the $m=1/2$ mode has a
dispersion relation shown in Fig.~\ref{f:homo}(c) that we call 'mixed
mode' because one has a vertical as well as a horizontal structuring
(checkerboard-like) of the unstable film because $m>0$ and
$k_{\mathrm{max}}\neq0$. Note as well that $\beta>0$ at $k=0$ and
$\beta_{\mathrm{max}}=1/4$ as for the $m=0$ mode. As $h$ increases the
fastest growing horizontal wavenumber of the mixed mode becomes larger
(Fig.~\ref{f:homo-max}(b)) and approaches the one of the horizontal
mode whereas its growth rate equals $\beta^{m=0}_{\mathrm{max}}=1/4$.
The minimum of the dispersion relation at $k=0$ becomes deeper as
$h$ increases (thin dashed line in Fig. \ref{f:homo-max}(a)).

\begin{figure}[t]
(a)\includegraphics[width=0.7 \hsize,angle=0]{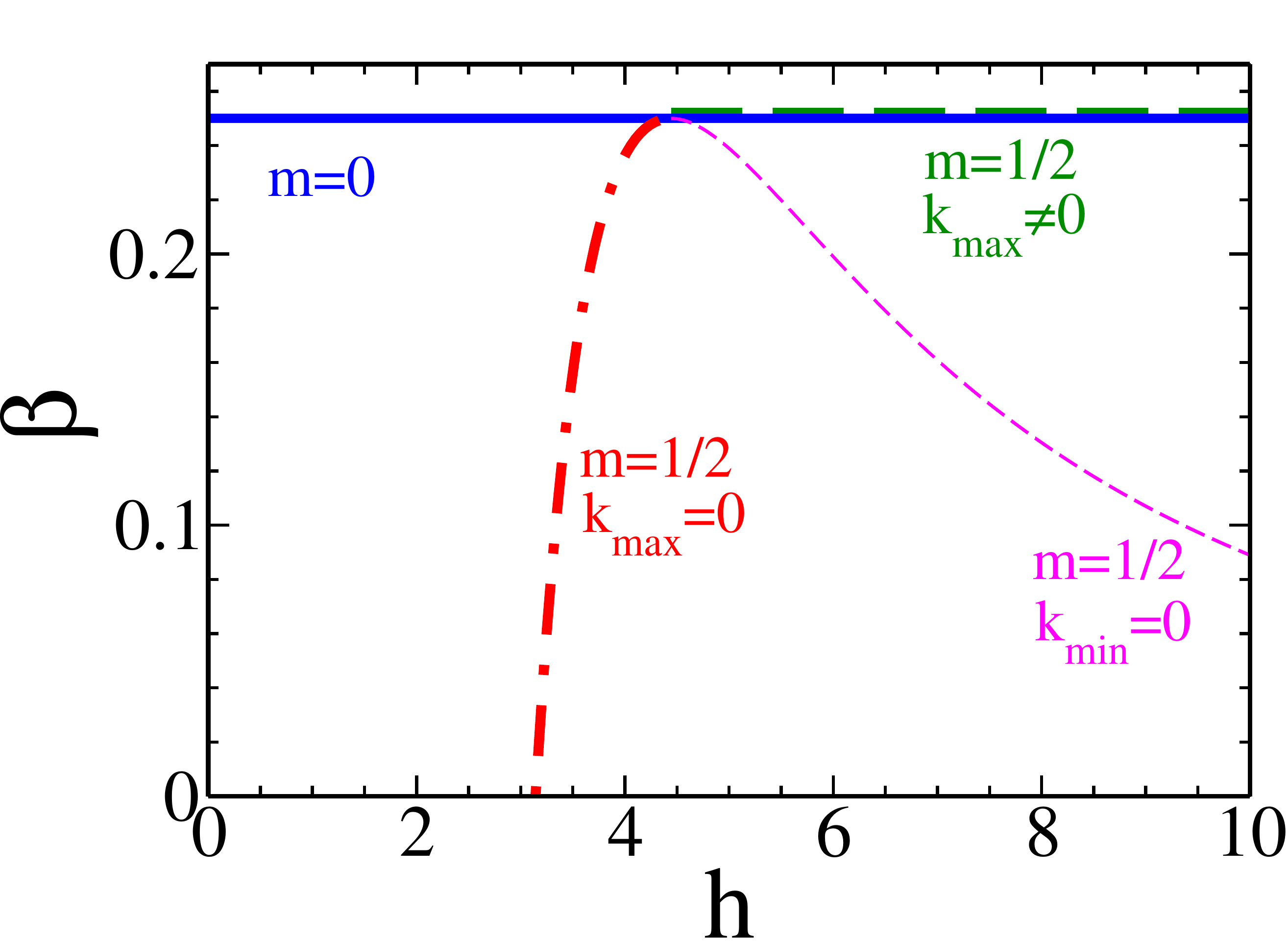}
(b)\includegraphics[width=0.7 \hsize,angle=0]{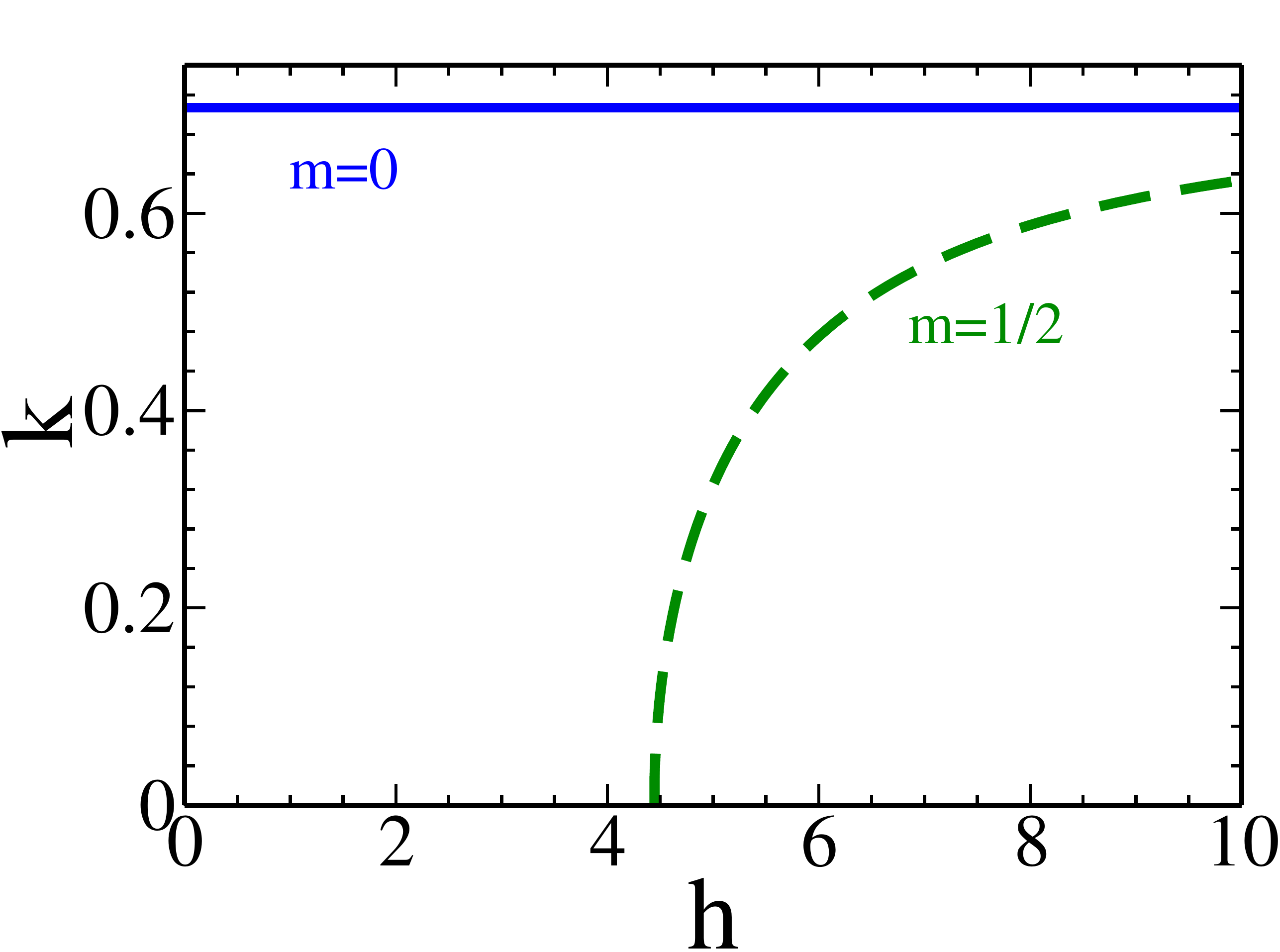}
\caption{ (color online) Maximum growth rate (a) and corresponding
  wavenumber (b) as a function of the thickness for a {\em
    homogeneous} film of critical composition $c_0=0$ and {\em neutral
    surfaces} $a^\pm=0$ and $b^\pm=0$.  Solid line: mode $m=0$
  (Fig.~\ref{f:homo}(a)). Dot-dashed line: mode $m=1/2$ with
  $k_{\mathrm{max}}=0$ (Fig.~\ref{f:homo}(b)). Dashed line: mode $m=1/2$ with
  $k_{\mathrm{max}}\ne 0$ (Fig.~\ref{f:homo}(c)). Thin dashed-line: growth rate
  $\beta$ of the mixed mode $m=1/2$ at the minimum of dispersion
  relation at $k_{\mathrm{min}}=0$.  \mylab{f:homo-max}}
\end{figure}

Higher order modes $m\ge1$ destabilize successively at $h= 2
m\pi$. The resulting instability correspond to purely vertical
modes with $\beta_{\mathrm{max}}\le1/4$ for $h\le 2\sqrt{2} m\pi$, and
to mixed modes with $\beta_{\mathrm{max}}=1/4$ for $h > 2\sqrt{2}
m\pi$.

Note, finally, that the dispersion relations that will be discussed
below for homogeneous films with energetically biased surfaces and for
stratified films have similar shapes as the ones in
Figs.~\ref{f:homo}(a) to \ref{f:homo}(c).  However, they do not fall
into the classification of modes used in this section.
%
\subsection{Energetically biased surfaces $a^\pm=0$, $b^\pm\ne0$}
%
The relevance of energetically biased surfaces in the mixing/demixing
dynamics has been recognized in the design of
materials based on polymer blends \cite{GeKr03}. A biased surface can
cause the formation of spinodal composition waves in a direction
normal to the surfaces \cite{BaEs90,Jone91} giving rise to the
so-called surface directed spinodal decomposition \cite{Jone91}.

According to the expression of the surface free energies
$f^+(c)=1+a^+c+b^+c^2/2$ and $f^-(c)=\gamma_s+a^-c+b^-c^2/2$ and the
boundary conditions (\ref{bc-orso3}), (\ref{bc-lin3-homo-a}), and
(\ref{bc-lin3-homo}), the influence of the surfaces enters the
stability problem through the coefficients $a^\pm$ that account for a
preferential adsorption of one of the components at the respective
surface, and the coefficients $b^\pm$ that describe mixing/demixing
properties at the surfaces that differ from the bulk.

An energetic bias at the surfaces leads to the restriction
$c_0=-a^+/b^+=-a^-/b^-$ for the mean concentration.\cite{TMF07} It is
very unlikely that the general case can be easily realized in an
experiment.  Therefore we restrict our analysis for the
homogeneous film to the case $b^\pm\neq0$ with $a^\pm=0$. Experimental
realization seems as well questionable, however, the case is more
generic as a critical mixture yields a valid base state for any $b^+$
and $b^-$. The case was studied in depth in Ref.~\onlinecite{FMD98} in
the context of purely diffusive decomposition of a binary mixture in a
gap between two solid plates. This allows us to use the case as a
benchmark for our numerical procedure by comparing our results with
the literature.

We introduce a number of archetypal cases for $b^\pm$: the symmetric
case with $b^+=b^-$ where both surfaces of the film have  identical
bias; the antisymmetric case $b^+=-b^-$, where the two surfaces have
opposite bias; and the asymmetric case where either $b^+$ or $b^-$ is zero.

Fig.~\ref{f:homo-active} presents typical results for the two most
dangerous transverse instability modes for the symmetric
($b^+=b^-=1$, top row), asymmetric ($b^+=1,\,b^-=0$, middle row), and
antisymmetric ($b^+=-b^-=1$, bottom row) case. A selection of the
corresponding eigenmodes $c_1(z)$ is given in
Fig.~\ref{f:homo-active-c1}(a) to (c), respectively.

In all cases one of the two most dangerous modes has a dispersion
relation similar to Fig.~\ref{f:homo}(a), i.e.,
$\beta_{\mathrm{max}}\to0$ for $k_{\mathrm{max}}\to0$. It is
nevertheless not a purely 'horizontal mode' as the eigenfunction $c_1$
shows a clear vertical structure (see Fig.~\ref{f:homo-active-c1}). It
dominates for symmetric bias (Fig.~\ref{f:homo-active}(a)).  The other
mode has normally a dispersion relation similar to
Fig.~\ref{f:homo}(c), i.e., $\beta_{\mathrm{max}}\neq0$ at
$k_{\mathrm{max}}=0$. However, for $4.8<h<6$ in the symmetric case
$\beta_{\mathrm{max}}$ occurs at $k_{\mathrm{max}}=0$ (analogous to
Fig.~\ref{f:homo}(b)). This mode dominates for asymmetric and
antisymmetric bias (Fig.~\ref{f:homo-active}(b) and
Fig.~\ref{f:homo-active}(c), respectively).

Inspecting Fig.~\ref{f:homo-active} one notes that for symmetric and
asymmetric bias the dependency on film thickness is
similar to the one for the $m=1/2$ mode in the case of
neutral surfaces (Fig.~\ref{f:homo-max}): All modes stabilize at a
thickness $h_c$, wavenumber and growth rate increase with
increasing $h>h_c$ and for large $h$ both $\beta_{\mathrm{max}}$
approach 1/4.  The convergence at large $h$ is expected as the
relative influence of the walls decreases with increasing $h$. Below
$h_c$ the film is linearly stable. 

For antisymmetric surface bias (bottom row of
Fig.~\ref{f:homo-active}) one of the modes shows a similar behavior as
the above modes. The other one, however, becomes progressively more
unstable as $b^-$ decreases (as compared to the asymmetric case). The
growth rate converges for large thicknesses to a value much larger
than the bulk value of 1/4. The corresponding eigenmode given in
Fig.~\ref{f:homo-active-c1}(c) indicates that the film remains nearly
homogeneous at the top where $b^+=1$ suppresses the demixing. At the
bottom demixing is enforced by $b^-=-1$ and becomes much stronger than
bulk demixing.


\begin{figure}[t]
\includegraphics[width=0.4 \hsize,angle=0]{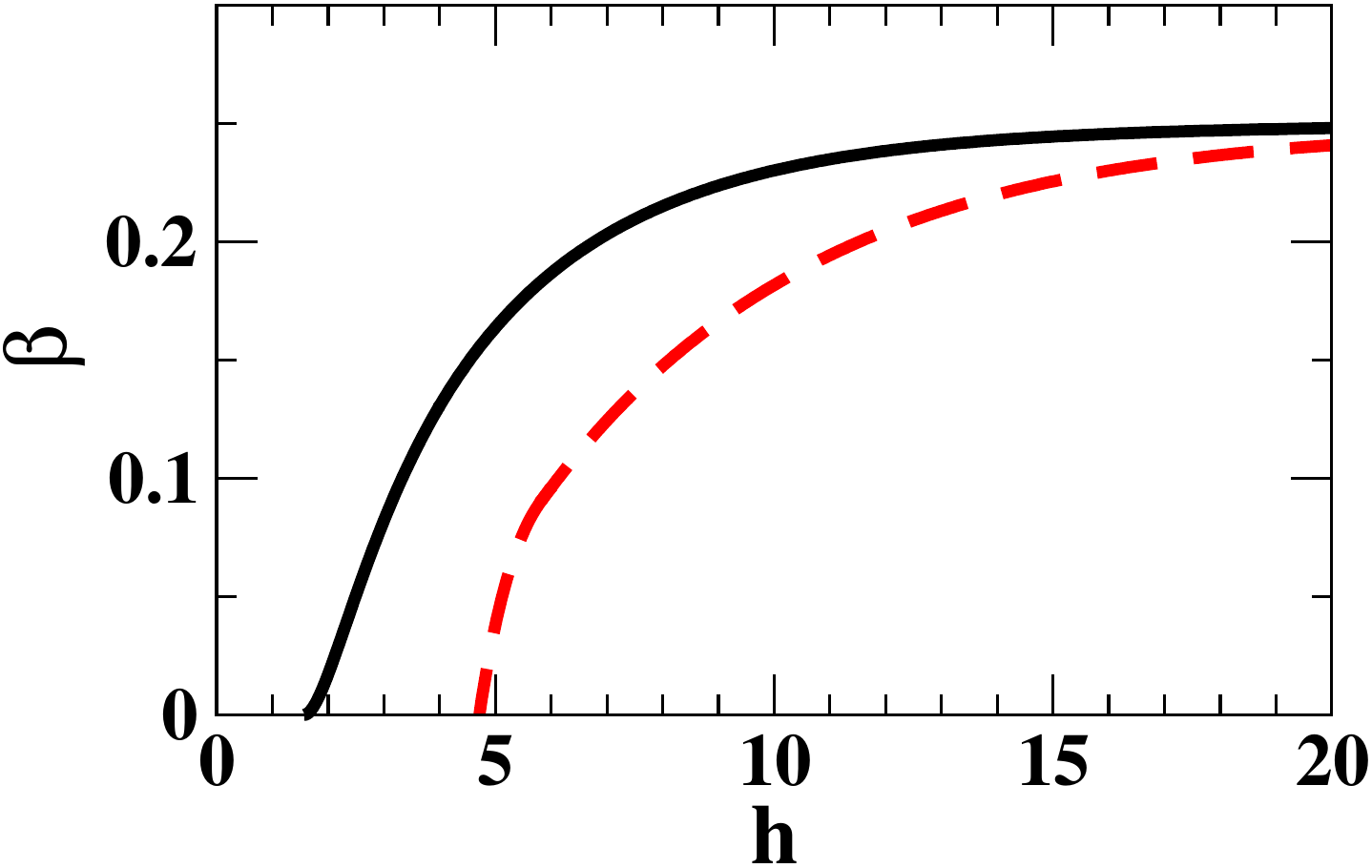}
\hspace*{0.0cm} \includegraphics[width=0.4 \hsize,angle=0]{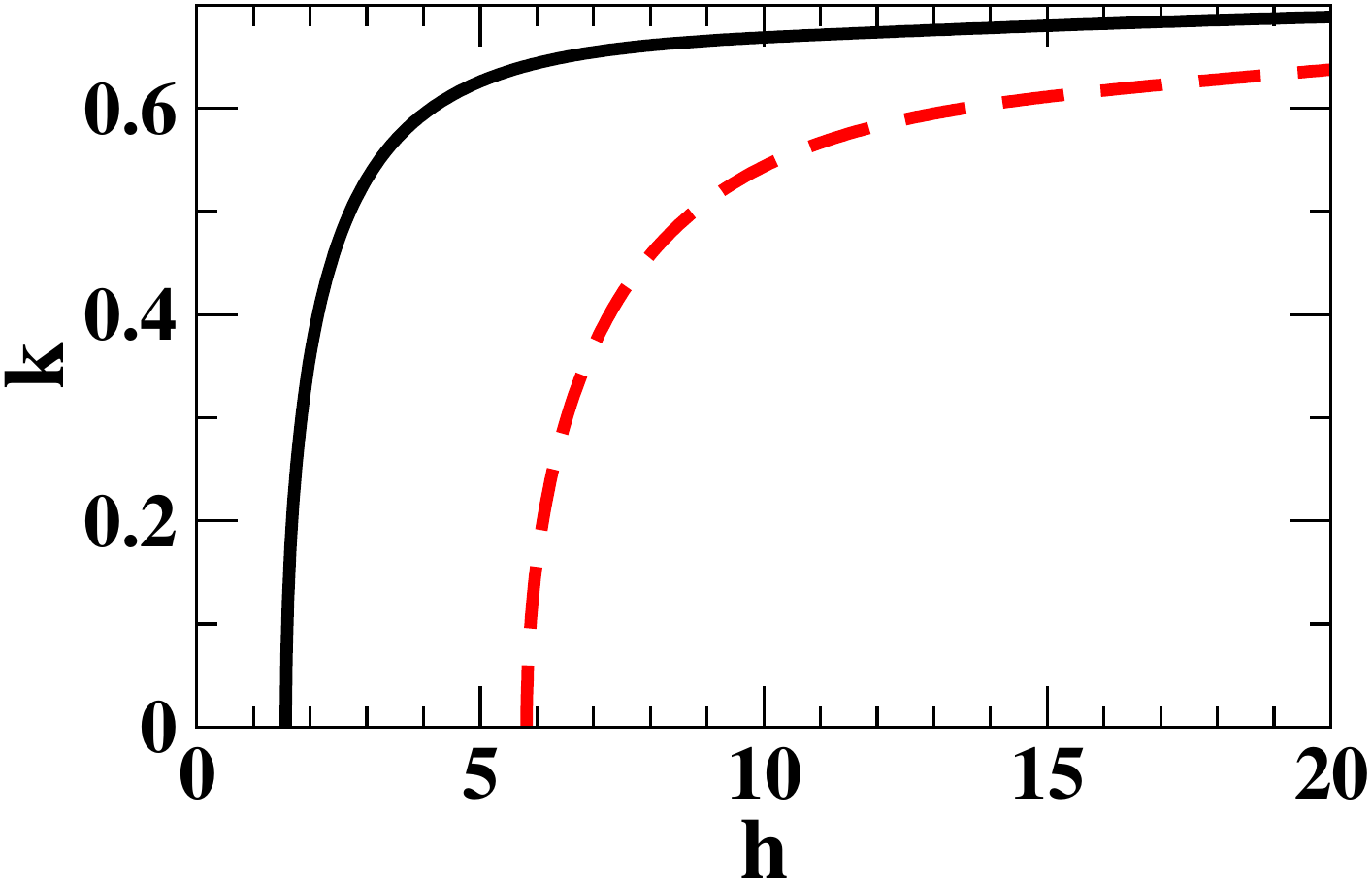}
\includegraphics[width=0.4 \hsize,angle=0]{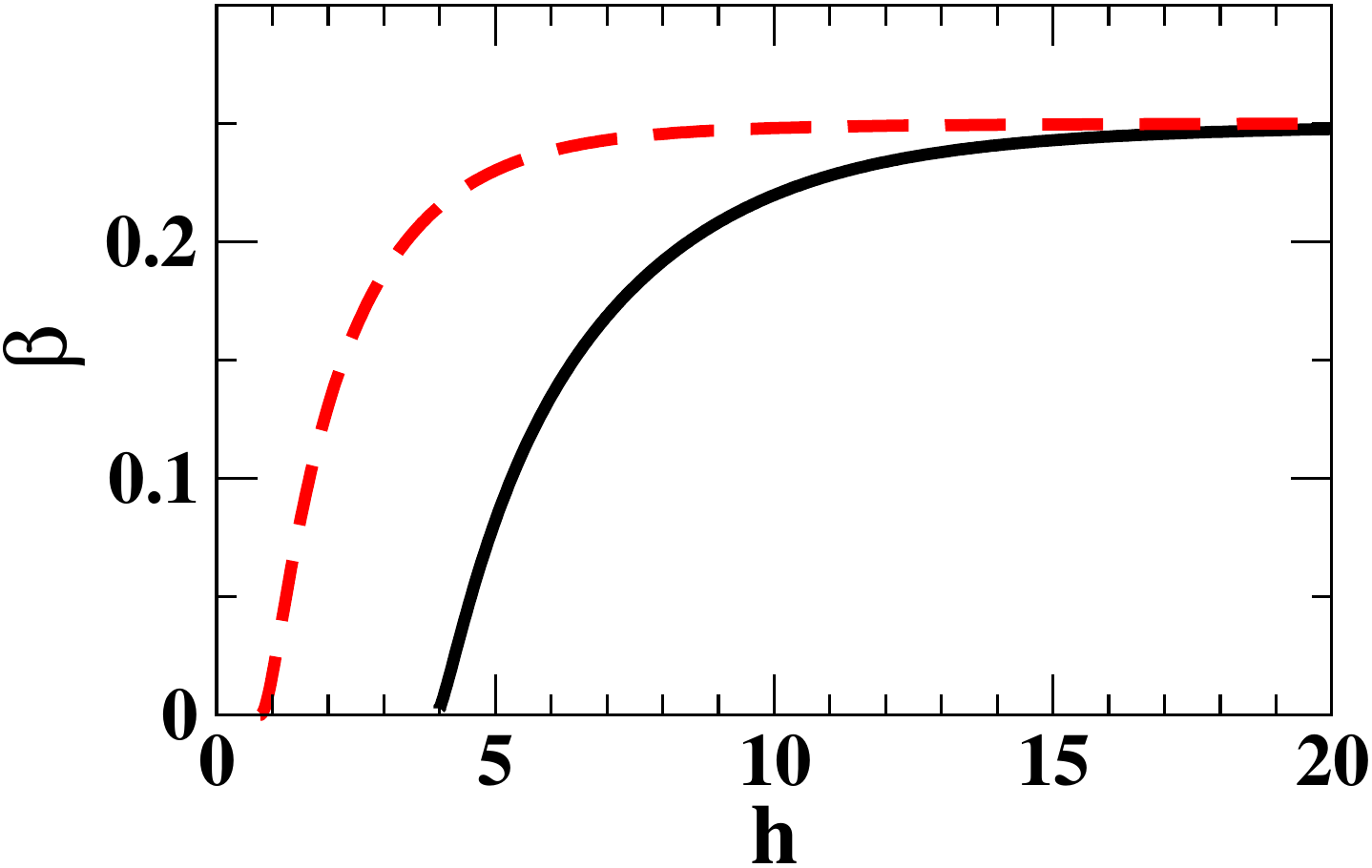}
\hspace*{0.0cm} \includegraphics[width=0.4 \hsize,angle=0]{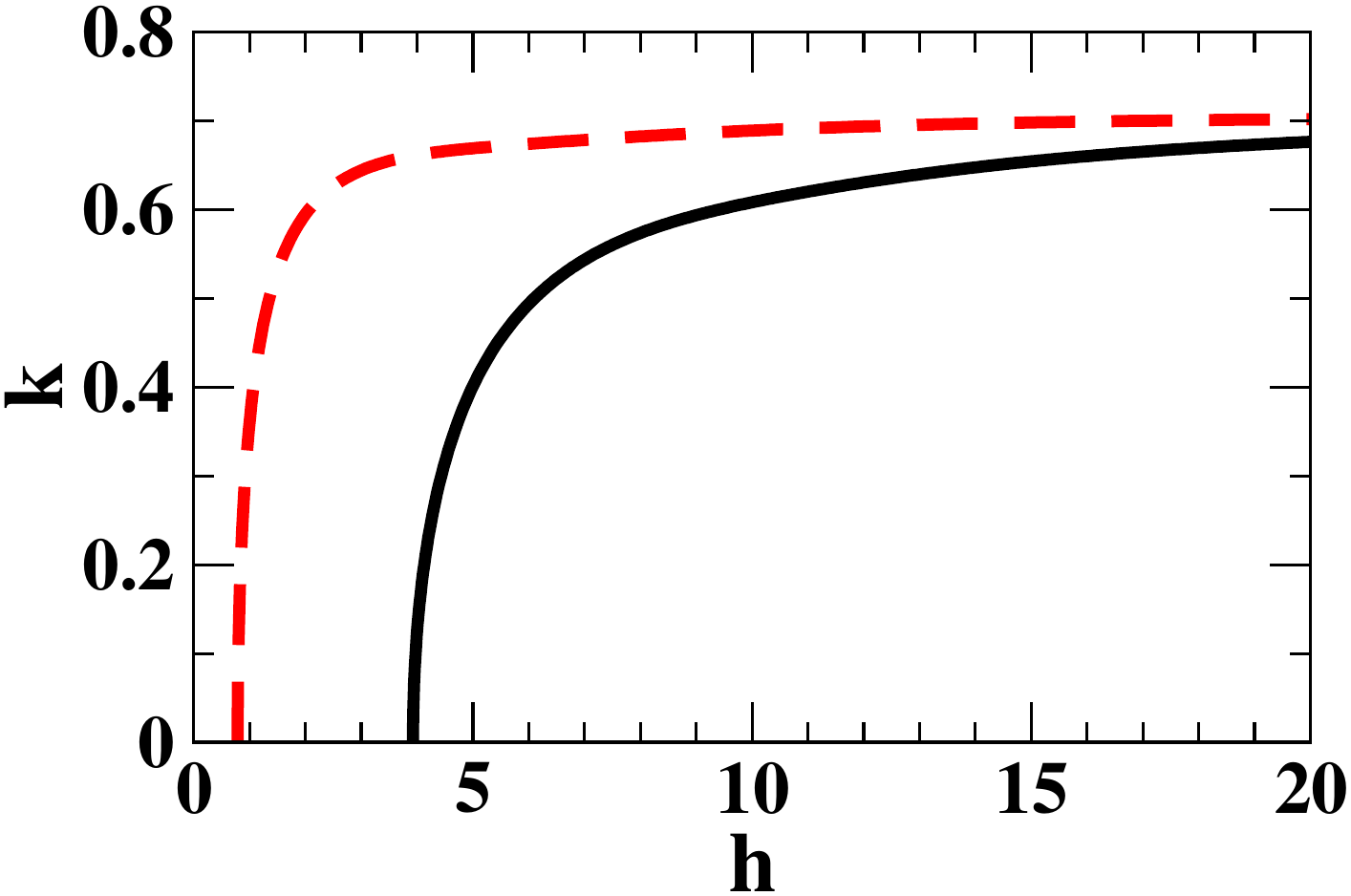}
\includegraphics[width=0.4 \hsize,angle=0]{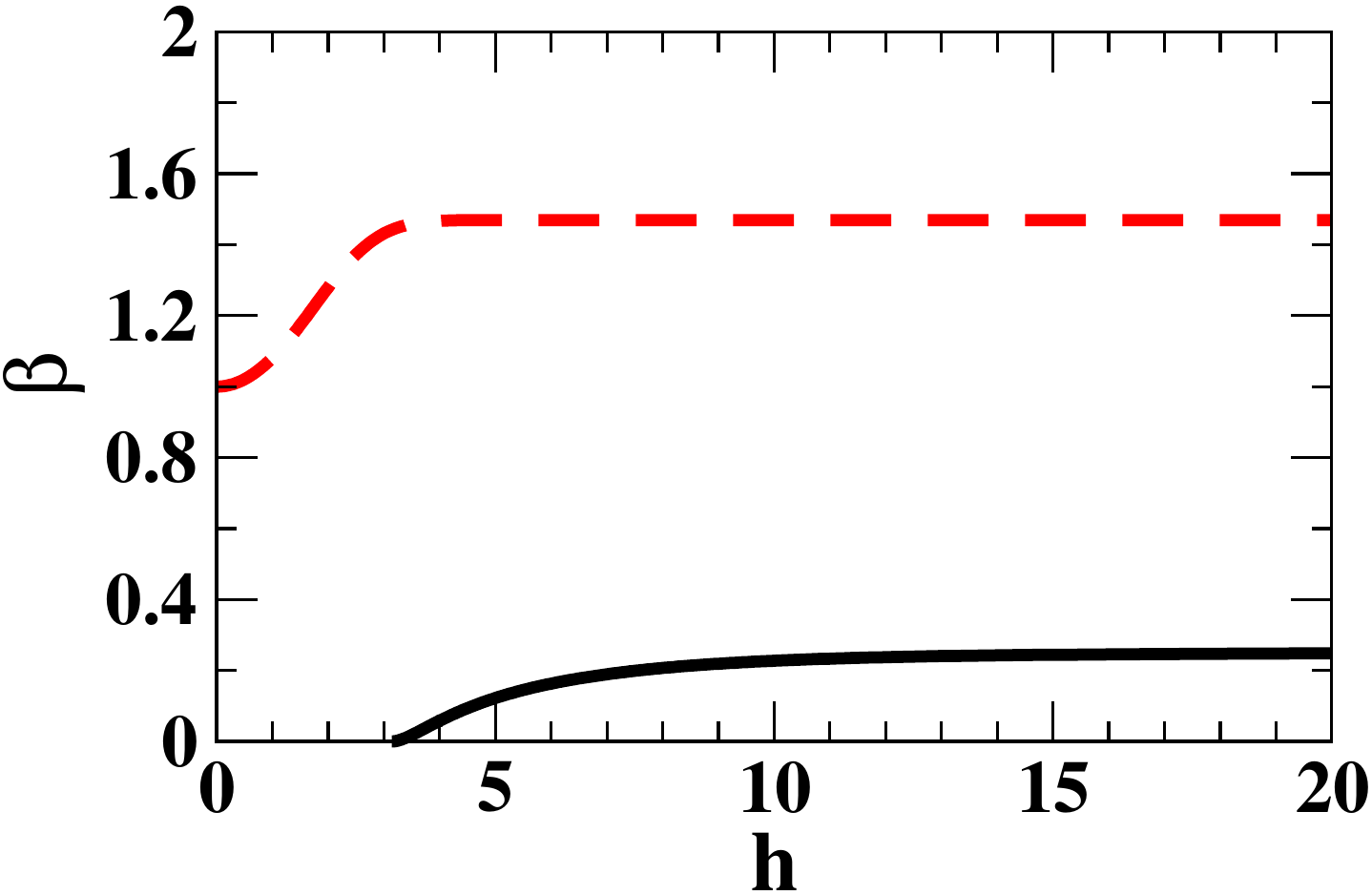}
\hspace*{0.0cm} \includegraphics[width=0.4 \hsize,angle=0]{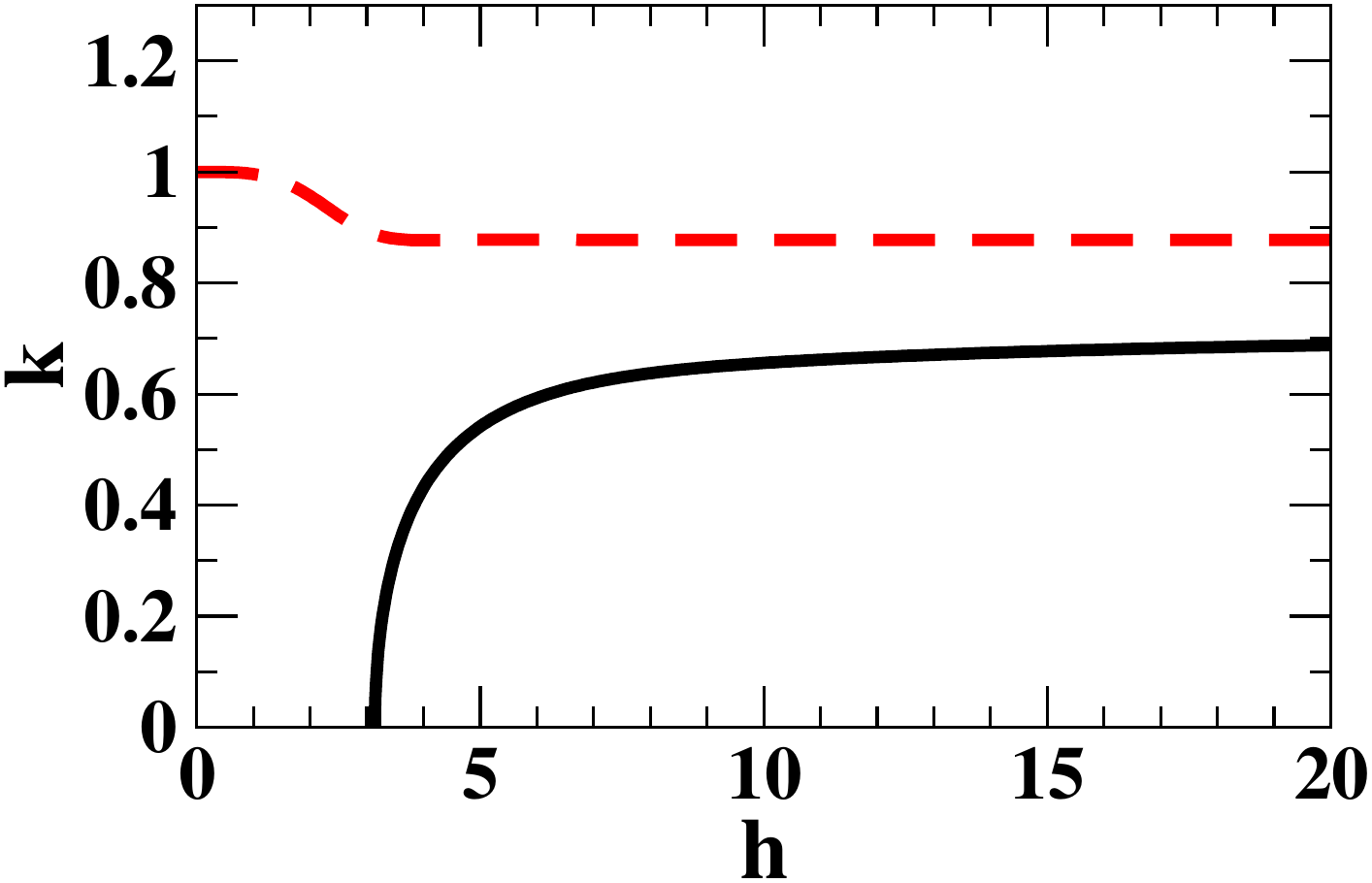}
\caption{ (color online) Maximum growth rate (left) and corresponding wavenumber
  (right) as a function of the thickness for an {\em homogeneous} film
  $c_0=0$ with {\em biased surfaces}.  Top row: symmetric bias with
  $b^\pm=1$ disfavoring demixing at both surfaces; middle row:
  asymmetric bias with $b^+=1$ and $b^-=0$ disfavoring demixing at the
  top surface; bottom row: antisymmetric bias with $b^+=1$ and
  $b^-=-1$ disfavoring (favoring) demixing at the top (bottom)
  surface.  All solid lines correspond to maxima of dispersion
  relations similar to the one  illustrated in Fig.~\ref{f:homo}(a),
  whereas dashed lines normally correspond to relations similar to the
  one in Fig.~\ref{f:homo}(c).  \mylab{f:homo-active} }
\end{figure}
\begin{figure}[t]
\includegraphics[width=0.7 \hsize,angle=0]{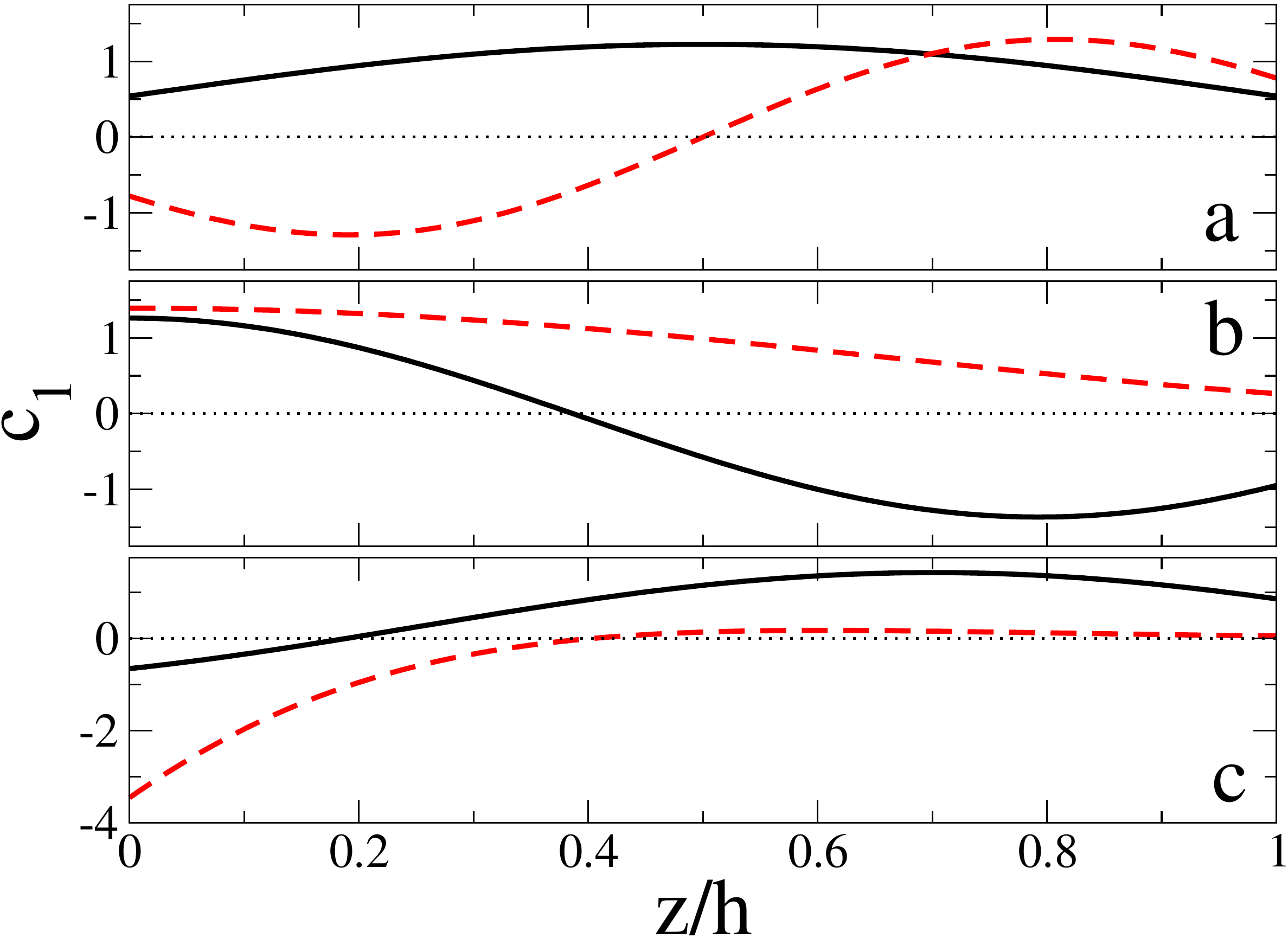}
\caption{(color online) Selection of eigenmodes $c_1(z)$ for fixed
  thickness $h_0=5$  corresponding to the instability modes presented in
  Fig.~\ref{f:homo-active}: (a) symmetric bias with $b^\pm=1$; (b)
  asymmetric bias with $b^+=1$ and $b^-=0$; (c) antisymmetric bias
  with $b^+=1$ and $b^-=-1$. Solid lines correspond to maxima of dispersion
  relations similar to the one  illustrated in Fig.~\ref{f:homo}(a),
  whereas dashed lines correspond to relations similar to the
  one in Fig.~\ref{f:homo}(c).  \mylab{f:homo-active-c1} }
\end{figure}

%
\section{Results for stratified base states} \mylab{sec-hete}
%
Non-trivial quiescent horizontally homogeneous base states of model-H
are vertically non-uniform solutions, i.e., layered films. They are
characterized by $z$-dependent concentration profiles $c_0=c_0(z)$ and
$\vec{v_0}=0$. In part I\cite{TMF07}, we discuss for different
energetic biases at the surfaces various types of stratified film
states in dependence of film thickness $h$. Several branches of
stratified solutions may co-exist. They can be ordered by the mode
type of the related linear instability mode, i.e., the instability
mode that destabilizes the trivial homogeneous solution at the
bifurcation of the respective branch of stratified states.  For
details see Ref.~\onlinecite{TMF07}.

Although the pattern of branching is intriguing and interesting from
the point of view of bifurcation theory, only a subset of branches has
a large importance for experimental systems. They can be identified
calculating their energy (cf.~Figs.~4, 8 and 10 of
Ref.~\onlinecite{TMF07}). The figures show that depending on parameter
values the stratified films of either type $n=0$ or $n=1/2$ correspond
to the energetic minimum. For weak surface bias, these solutions
correspond roughly to a weak vertical stratification for $n=0$ and to
strongly stratified two-layer configurations for $n=1/2$. For stronger
bias they might as well correspond to a 3-layer sandwich
configuration.  In the following, we restrict our attention to the
lateral stability of the solutions that are energetically most
favorable.

\subsection{Stability without hydrodynamics}

To isolate the effect of diffusive transport from the one of
hydrodynamic convective motion on the stability of the binary mixture
we first study the stability of stratified films without fluid motion,
i.e.\ we set $w_1=0$ in Eq.~(\ref{ds1}) and only solve (\ref{ds1}),
(\ref{bc-lin1c-2}), (\ref{bc-lin2c-2-a}) and (\ref{bc-lin2c-2})
together with the equations for the steady states. This implies as
well that the free surface remains flat.

\subsubsection{Neutral surfaces}

\begin{figure}[t]
(a)\includegraphics[width=0.35\hsize,angle=0]{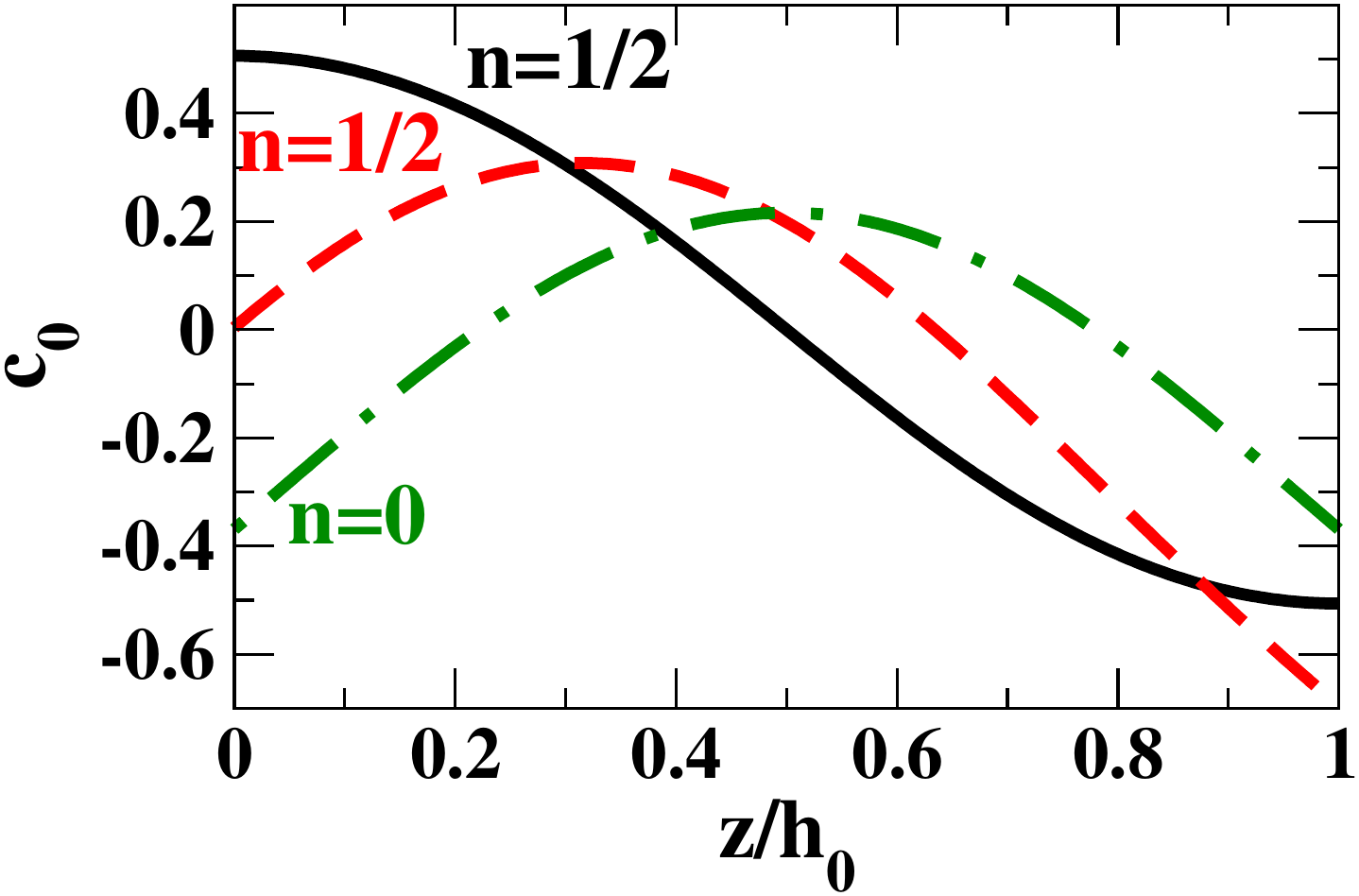}
\includegraphics[width=0.35\hsize,angle=0]{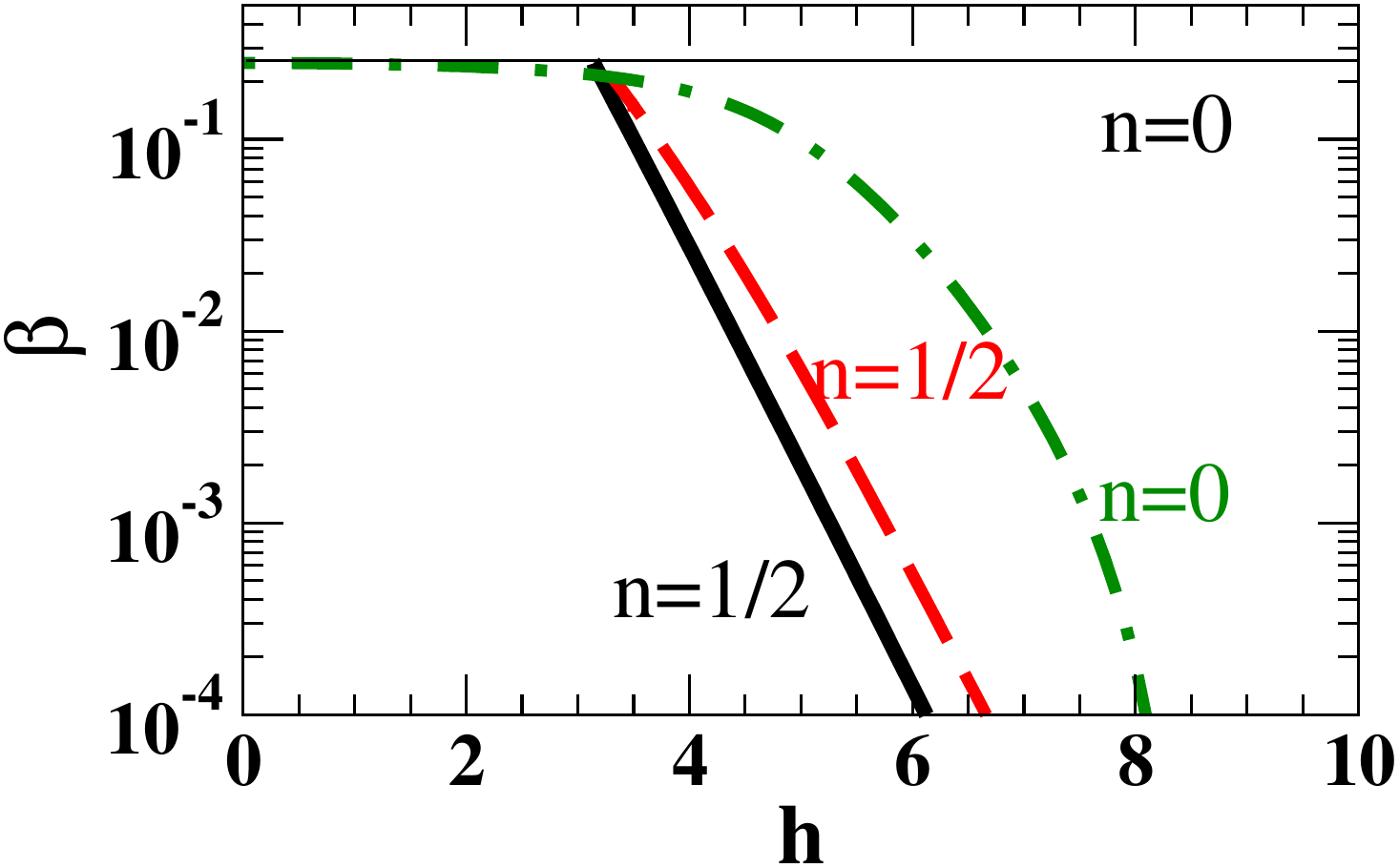} (b)\\
(c)\includegraphics[width=0.35\hsize,angle=0]{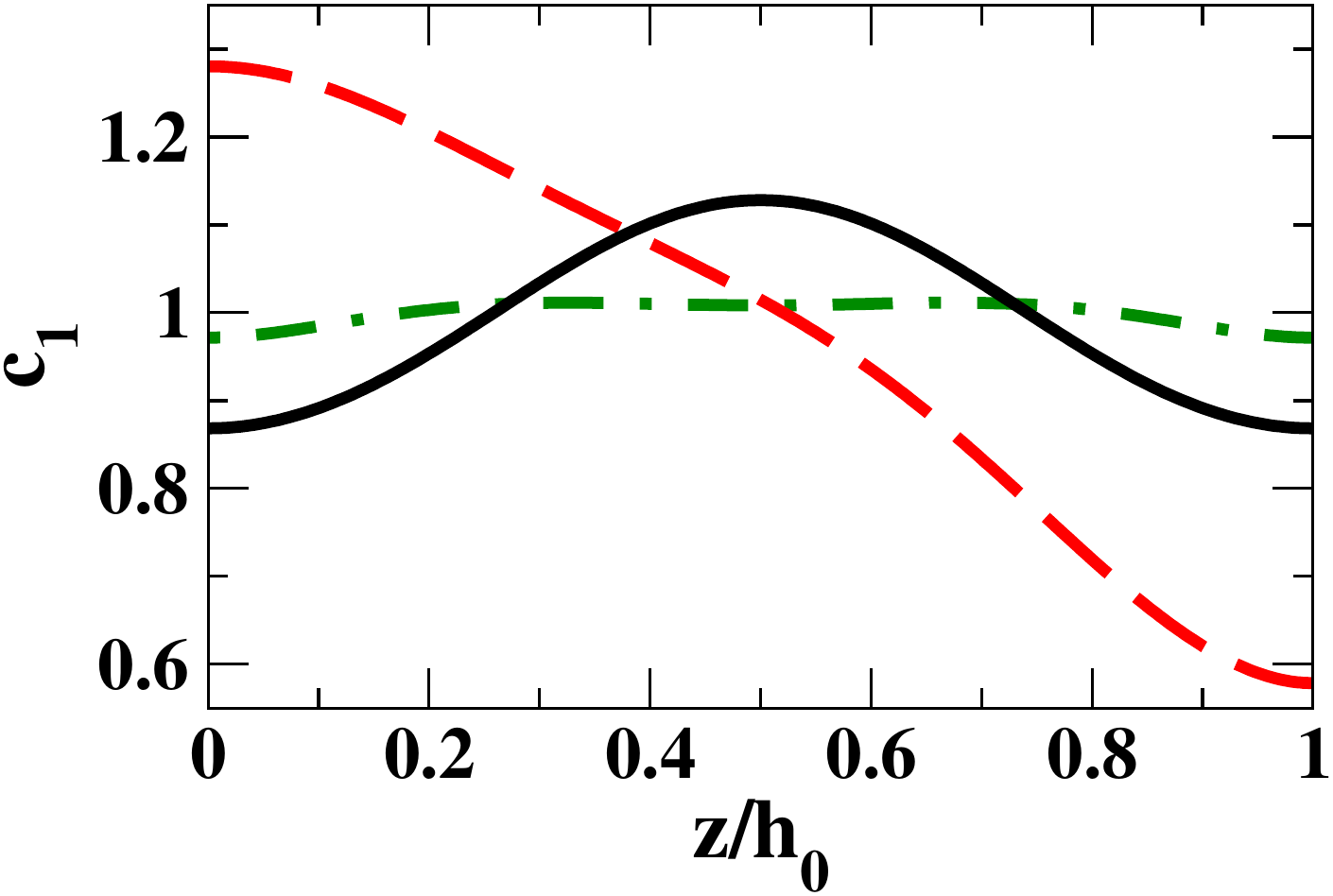}
\includegraphics[width=0.35\hsize,angle=0]{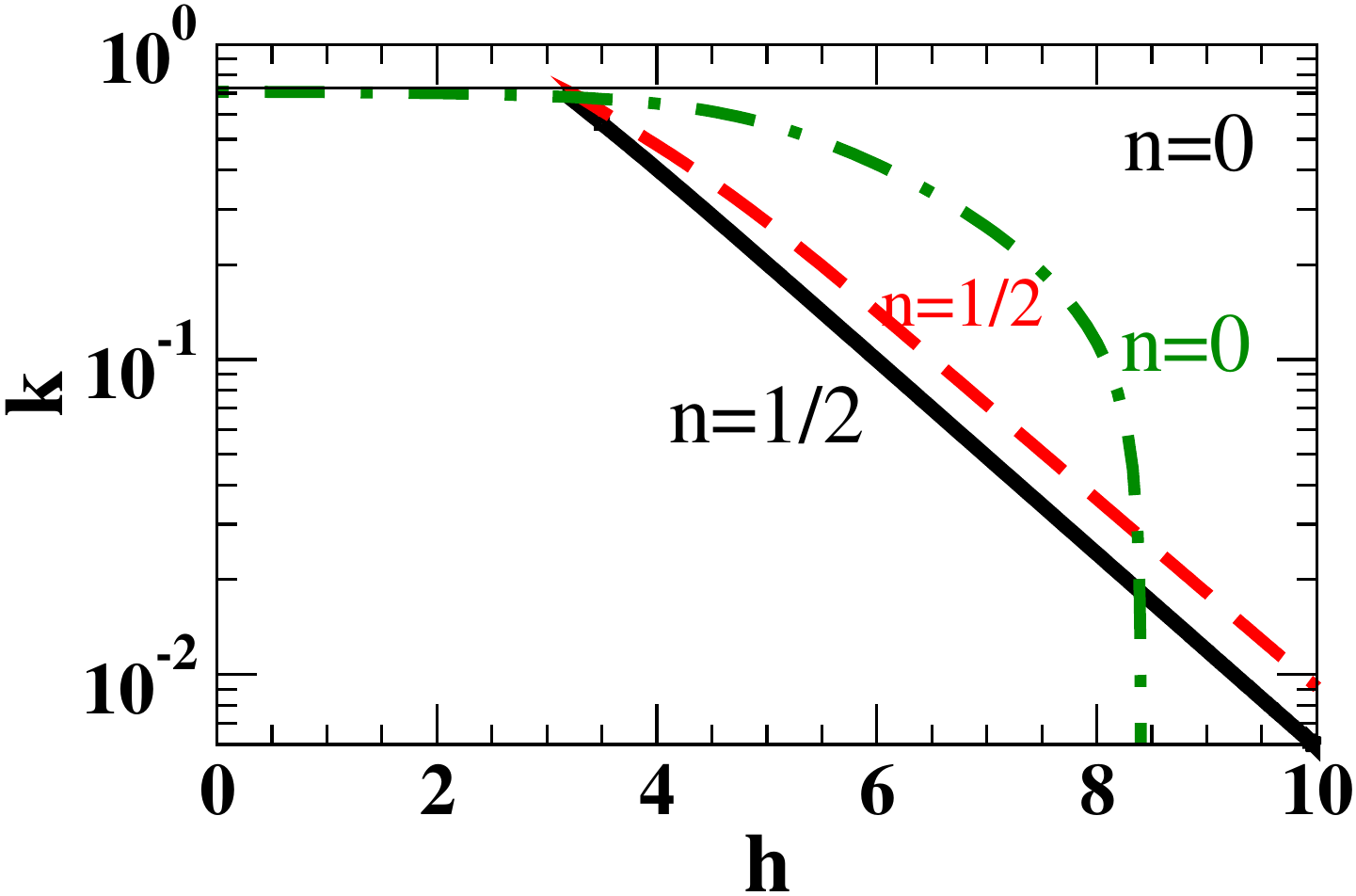} (d)
\caption{ (color online) Characteristics of the horizontal instability
  modes in the case of purely {\em diffusive transport} for {\em
    neutral} ($a^\pm=b^\pm=0$, solid lines in (b) and (d)) and {\em
    symmetrically biased} ($a^\pm=0.5$, $ b^\pm=0$, dashed and
  dot-dashed lines in (b) and (d)) surfaces.  Panels correspond to (a)
  concentration profiles $c_0(z)$ of the stratified base states at
  $h_0=3.5$ for neutral surfaces ($n=1/2$ branch, solid line) and
  symmetric bias ($n=0$ branch, dot-dashed line and $n=1/2$ branch,
  dashed line); (c) perturbation profiles $c_1(z)$ for $h=3.5$ and
  line styles as in panel (a); (b) maximum growth rate and (d)
  associated wavenumber as a function of the film thickness for $n=0$
  and $n=1/2$ base states (cf.~Figs.~1 and~4 of
  Ref.~\onlinecite{TMF07}). In all cases S$=1$, Re$=0$, Ps/Re$=1$.
  \mylab{f:strati-nohydro-a0}
  }
\end{figure}

First, we discuss the case of neutral surfaces summarized in
Fig.~\ref{f:strati-nohydro-a0}.  The solid line in panel (a) gives the
concentration profile of the base state for neutral surfaces for a
film thickness not much above the branching of the $n=1/2$ branch from
the trivial homogeneous state.  The concentration decreases
monotonically with vertical position corresponding to a layered film
with component 1 enriched near the substrate and component 2 enriched
near the free surface (remember that $c=c_1-c_2$). As the film is
quite thin, the diffuse interface between the two phases nearly spans
all the layer.  The solid line in panel~(c) gives the perturbation
mode $c_1(z)$.

Figs.~\ref{f:strati-nohydro-a0}(b) and~(d) show the maximal growth
rate and the associated wavenumber, respectively, as a function of the
film thickness. Solid lines refer to results for the energetically
preferable base state solution. We distinguish two regimes: (i) for
$0<h<\pi$ the base state corresponds to the homogeneous film, the most
dangerous mode is the horizontal mode discussed above in
Section~\ref{sec-homneu}, i.e., the growth rate equals 1/4 and the
horizontal wavenumber is constant as well (cf.~Fig.~\ref{f:homo-max});
and (ii) for $h> \pi$ the energetically preferable base state
corresponds to the two-layer film shown in
Fig.~\ref{f:strati-nohydro-a0}(a) as solid line. It is unstable
w.r.t.~lateral perturbations, however, the growth rate and wavenumber
both decay exponentially with increasing film thickness.  The
exponential decay indicates that in practical terms a film above
20-50\,nm might appear to be stable w.r.t.~lateral concentration
perturbations when only diffusive transport is taken into account.
The corresponding dispersion relations (not shown) are similar to
Fig.~\ref{f:homo}(a), i.e., $\beta_{\mathrm{max}}\to0$ for
$k_{\mathrm{max}}\to0$. In the neutral case the eigenmodes $c_1(z)$
show an up-down symmetry (cf.~solid line in
Fig.~\ref{f:strati-nohydro-a0}(c)). The
mean concentration will develop a horizontal variation as $\int
c_1(z) dz$ strongly deviates from zero. 
The variation is strongest along the diffuse interface.
The thin solid lines in Figs.~\ref{f:strati-nohydro-a0}(b) and (d)
indicate the behavior for the (energetically unfavorable)
homogeneous $n=0$ state for $h>\pi$.
%
%
\subsubsection{Symmetrically biased surfaces}
\mylab{s:ebs}
Results for symmetrically biased surfaces with $a^\pm=0.5$ are given
as dashed and dot-dashed lines in Fig.~\ref{f:strati-nohydro-a0}.  The
energetical bias strongly influences the base states (see Figs.~3 to~5
of Ref.~\onlinecite{TMF07}). All branches known from neutral surfaces
are modified already for weak bias.  Their number reduces with
increasing bias via a sequence of bifurcations. For strong bias
normally only a 3-layer sandwich film survives.

Even for small thicknesses $h<\pi$ the composition of the base state
is not uniform any more, i.e., the homogeneous solutions on the $n=0$
branches are modified and become (weakly) stratified. The
corresponding norm $||\delta h||$ increases with $h$ and $a^+$. The
strongly stratified $n=1/2$ solution branch bifurcates before from
the (modified) $n=0$ branch, however, the bifurcation is shifted from
$h=\pi$ to $h\sim 3.3$ (see Fig.~\ref{f:strati-nohydro-a0}).  The
bifurcated $n=1/2$ branch is always the energetically favorable one.

The concentration profiles given in Fig.~\ref{f:strati-nohydro-a0}(a)
(dashed line for $n=1/2$ branch and dot-dashed line for $n=0$ branch)
indicate that the preference for component 2 at the top and at the
bottom surface causes component 1 to concentrate within the film
giving rise to a local concentration maximum at $z/h_0 \sim 0.3$ for
the $n=1/2$ solution and $z/h_0 = 0.5$ for the $n=0$ solution. This
corresponds to the creation of a 3-layer sandwich structure similar to
the $n=1$ solution for unbiased surfaces.\cite{TMF07}

At small thicknesses below $\approx\pi$, the maximal growth rate as
well as the associated wavenumber for the $n=0$ branch now decrease
slightly with $h$ (Fig.~\ref{f:strati-nohydro-a0}(b) and (d)). For
larger $h$ growth rate and wavenumber decay faster than exponentially,
until for $h \approx 8.3$ the $n=0$ branch gains stability with
respect to lateral perturbations. This is, however, practically of no
importance as for $h>3.3$ the $n=0$ branch is not the energetically
favorable one. The energetically favorable one is the $n=1/2$ branch
that behaves similar as in the case of neutral surfaces.

The eigenmodes $c_1$ are given in Fig.~\ref{f:strati-nohydro-a0}(c).
For the $n=1/2$ branch the mean concentration $\bar{c_1}$ is smaller
than in the case of neutral boundaries, i.e., the perturbation is not
predominantly lateral.  The horizontal variation is smallest at the
top and strongest along the bottom
(cf.~Fig.~\ref{f:strati-nohydro-a0}(c)). The $n=0$ branch is unstable
w.r.t.~a nearly purely lateral mode as $c_1(z)\approx1$.
The direct comparison of the neutral and the
biased case in Fig.~\ref{f:strati-nohydro-a0} shows, however, that the
symmetrically biased surfaces actually decrease the lateral instability
of the layered ($n=1/2$) branch.  Note finally, that the
$n=0$ branch is much less unstable for symmetrical bias
than for neutral surfaces.
\subsubsection{Antisymmetrically and asymmetrically biased surfaces}
Stability results for antisymmetric surfaces with $a^+=-a^-=0.2$ in
the case of purely diffusive transport are presented in
Figs.~\ref{f:diff-conv-anti}(b) and (d) using dashed lines (notice
that dashed lines are superposed to solid lines). Panel (a)
and (c) present selected base state ($c_0(z)$) and perturbation
($c_1(z)$) profiles. The figures give results for the $n=1/2$ branch
only as it is the energetically favorable one (cf.~Figs.~6 to 8 of
Ref.~\onlinecite{TMF07}).

The boundary conditions for antisymmetric surfaces favor a bi-layer
structure of the film, i.e., they give rise to a monotonous change of
$c_0(z)$ with $z$ (Fig.~\ref{f:diff-conv-anti}(a)). Growth rate and
wavenumber of the most dangerous lateral mode both decrease with
increasing film thickness until the film becomes laterally stable at
about $h=4.5$. This implies that an antisymmetric bias has a much
stronger stabilizing effect than a symmetric one (compare
Figs.~\ref{f:strati-nohydro-a0} and \ref{f:diff-conv-anti}).

The corresponding dispersion relations (not shown) are similar to
Fig.~\ref{f:homo}(a). The eigenmodes $c_1(z)$ for the $n=1/2$
stratified film shows an up-down symmetry. It indicates that the
concentration will develop a horizontal variation that is strongest
along the diffuse interface and weakest at the surfaces
(cf.~Fig.~\ref{f:diff-conv-anti}(c)). 

\begin{figure}[t!]
(a)\includegraphics[width=0.35\hsize]{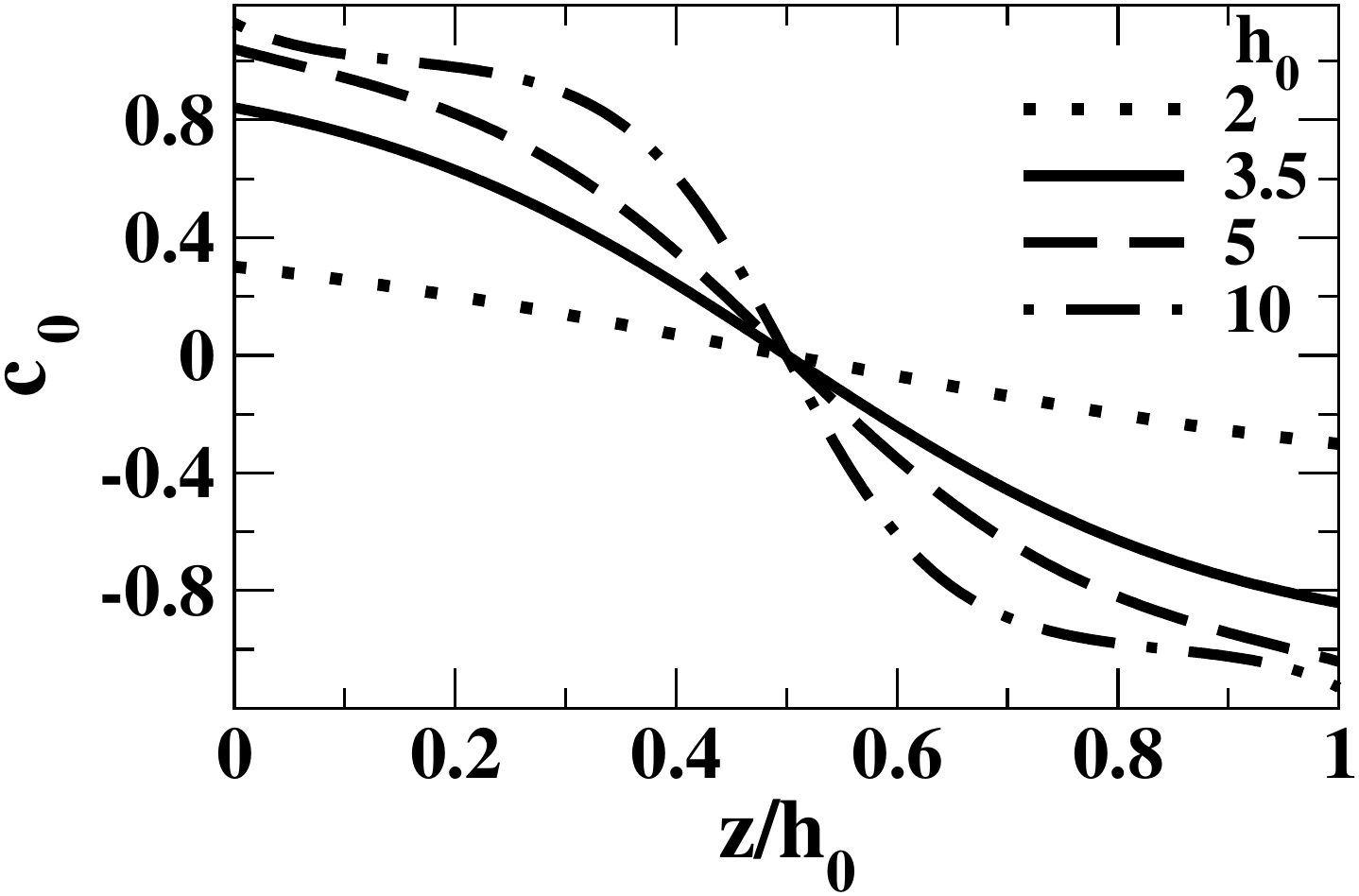}
\includegraphics[width=0.35\hsize]{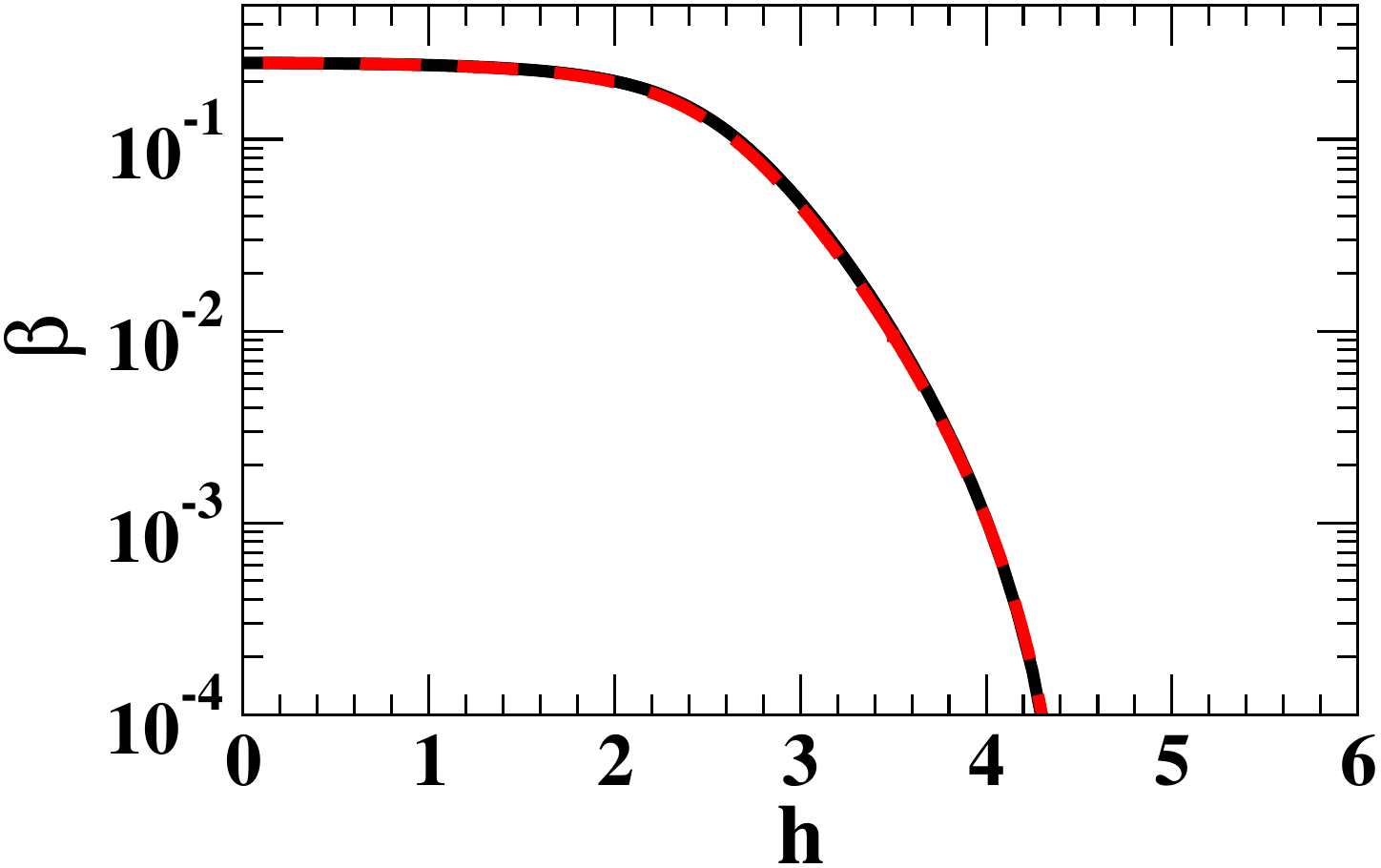}(b)\\
(c)\includegraphics[width=0.35\hsize]{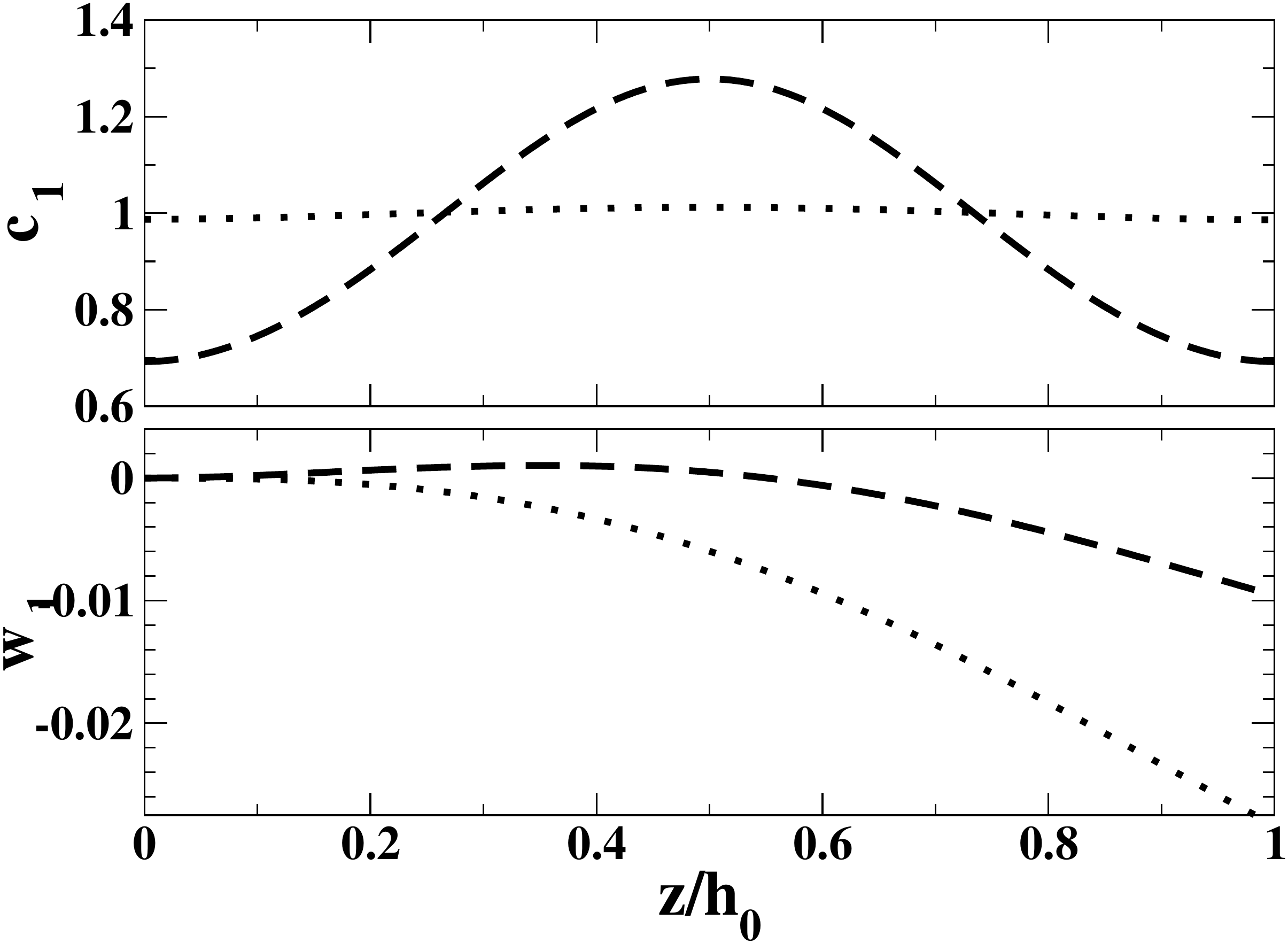}
\includegraphics[width=0.35\hsize]{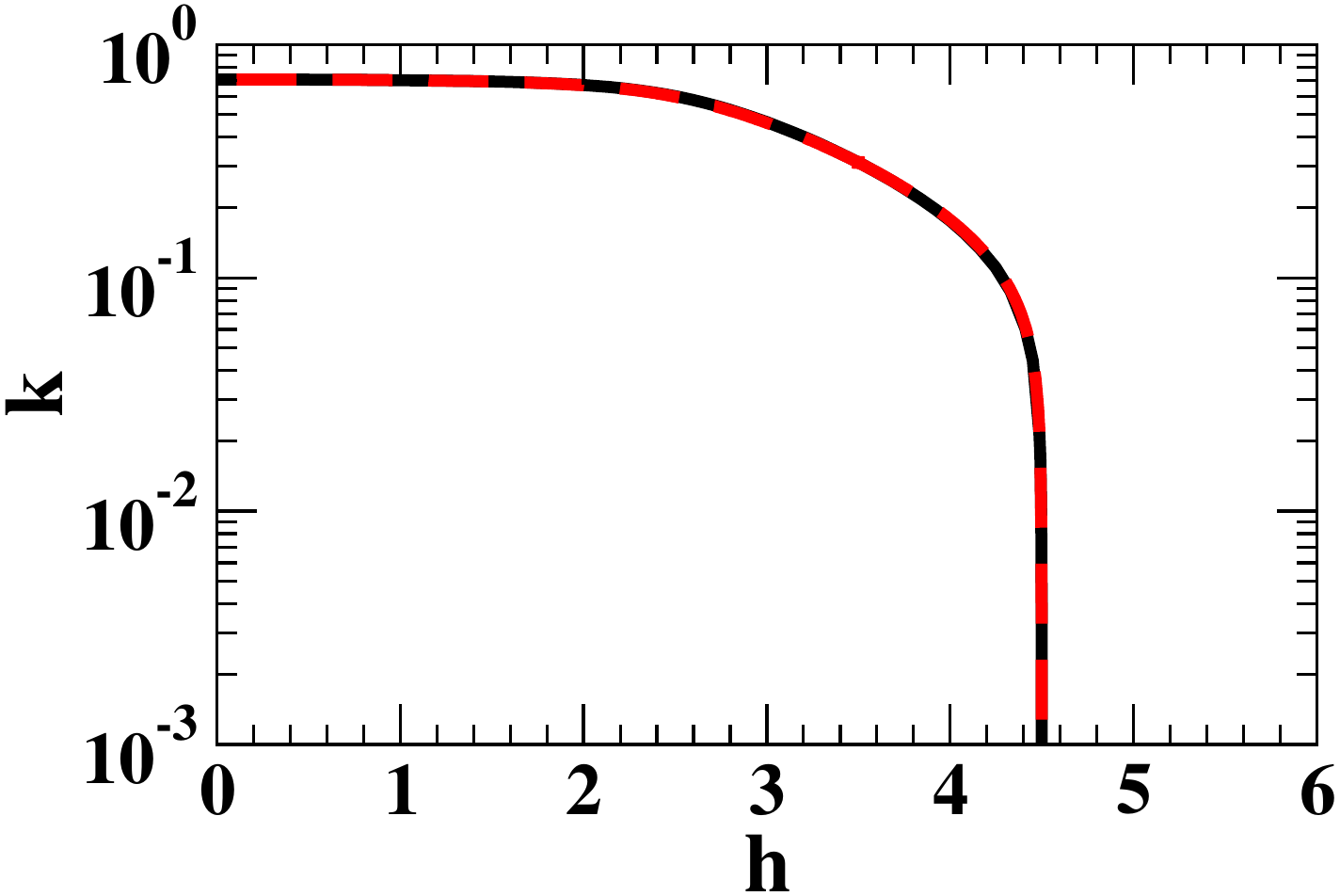}(d)
\caption{(color online) Characteristics of the horizontal instability
  modes for {\em antisymmetrically
    biased} surfaces with $a^+=0.2$, $a^-=-0.2$ and $b^\pm=0$. Panel
  (a) gives base state concentration profiles for solutions on the
  energetically favorable $n=1/2$ branch (see Figs.~6 and 8 of
  Ref.~\onlinecite{TMF07}) for selected film thickness as indicated in
  the legend. Panel (c) presents the corresponding perturbation modes
  $c_1(z)$ and $w_1(z)$ for $h=3.5$.  Note that the $c_1$ curves with and
  without convection lie on top of each other.  Panels (b) and
  (d) present characteristics of the horizontal instability modes for
  purely {\em diffusive transport} (dashed lines) and for coupled
  transport by {\em diffusion and convection} (solid lines). Note,
  that solid and dashed lines coincide for the chosen axis scales.
  (b) Maximum growth rate and (d) associated wavenumber as a function
  of the thickness for the branch $n=1/2$. Parameters: S$=1$, Re$=0$
  and Ps/Re$=1$. \mylab{f:diff-conv-anti}}
\end{figure}

The stability behavior for asymmetric surface bias is shown with
dashed lines in Fig.~\ref{f:diff-conv-asym}(b) and (d) for $a^+=0.2$
and $a^-=0$. The corresponding  base state profiles $c_0(z)$ and 
eigenmodes $c_1(z)$ are given in
Fig.~\ref{f:diff-conv-asym}(a) and (c), respectively.
Only results for the energetically favorable $n=1/2$
branch (cf.~Figs.~9 and 10 of Ref.~\onlinecite{TMF07}) are shown. The
growth rate and wavenumber of the most dangerous mode decrease with
increasing thickness $h$ until the film becomes laterally stable at
about $h=5.6$. The stabilizing effect is weaker than in the
antisymmetric case, but much stronger than in the symmetric one
(compare Figs.~\ref{f:strati-nohydro-a0} to \ref{f:diff-conv-asym}).
Dispersion relations (not shown) are similar to Fig.~\ref{f:homo}(a).

Before we investigate the influence of convective motion in the next
section we summarize our findings for the purely diffusive case: For
homogeneous films, we have found that energetically biased surfaces
(no linear bias $a^\pm=0$, purely quadratic bias $b^\pm\neq0$) have a
strongly stabilizing effect for small thicknesses (with the exception
of an antisymmetric bias). The influence of the boundaries becomes
weaker for larger film thicknesses.  For stratified films, an energy
bias at the surfaces (purely linear bias $a^\pm\neq0$, no quadratic
bias $b^\pm=0$) stabilizes the layered film against lateral
perturbations for asymmetric and antisymmetric biases. In both cases
the $n=1/2$ stratified films are stable above some critical film
thickness.  A symmetric bias, however, does only slightly stabilize
the layered films as compared to the case of neutral surfaces.  For
symmetric and neutral surfaces the maximal growth rate decreases
exponentially with increasing film thickness, but no critical film
thickness was found.
%
\subsection{Stability with hydrodynamics}
%
After having studied stability in the case of purely diffusive
transport we now allow as well for transport by convection, i.e., we
introduce the perturbations of the velocity fields back into the
model.  Note that the base states are identicl to the ones in the purely
diffusive case. However, the possible convective motion of the fluid
mixture may alter their lateral stability, and allow for an evolving
deflection of the free surface, and, in consequence, lead to films
that show lateral modulations of composition \textit{and} surface
profile.

The case of a homogeneous film is discussed above in
Section~\ref{sec-homo}: The linear perturbations of the concentration
and the velocity fields are decoupled implying that hydrodynamics has
no influence on the evolution of the homogeneous film in the linear
stage. No surface deflection can occur.

\begin{figure}[t]
(a)\includegraphics[width=0.45\hsize]{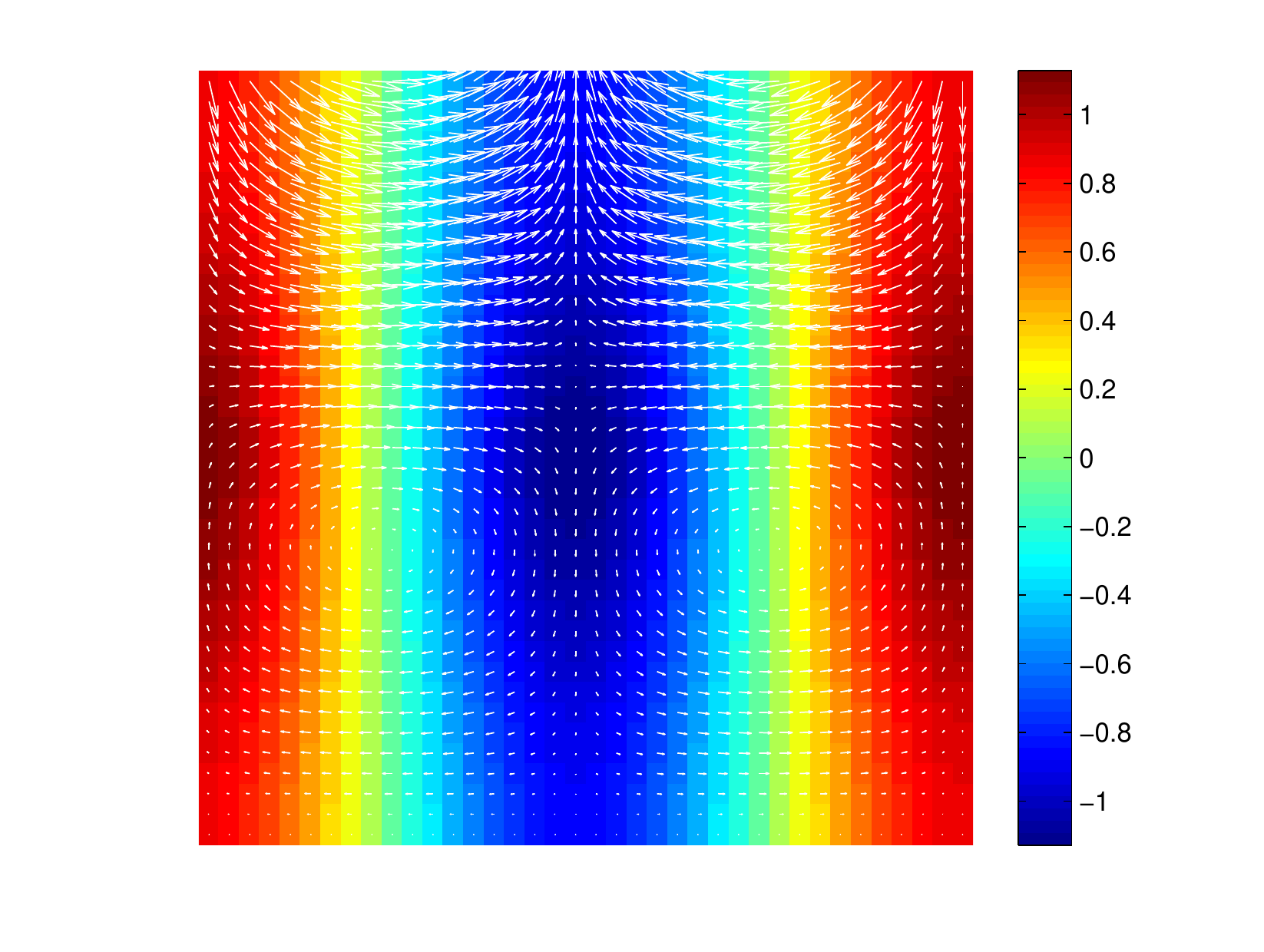}
\includegraphics[width=0.45\hsize]{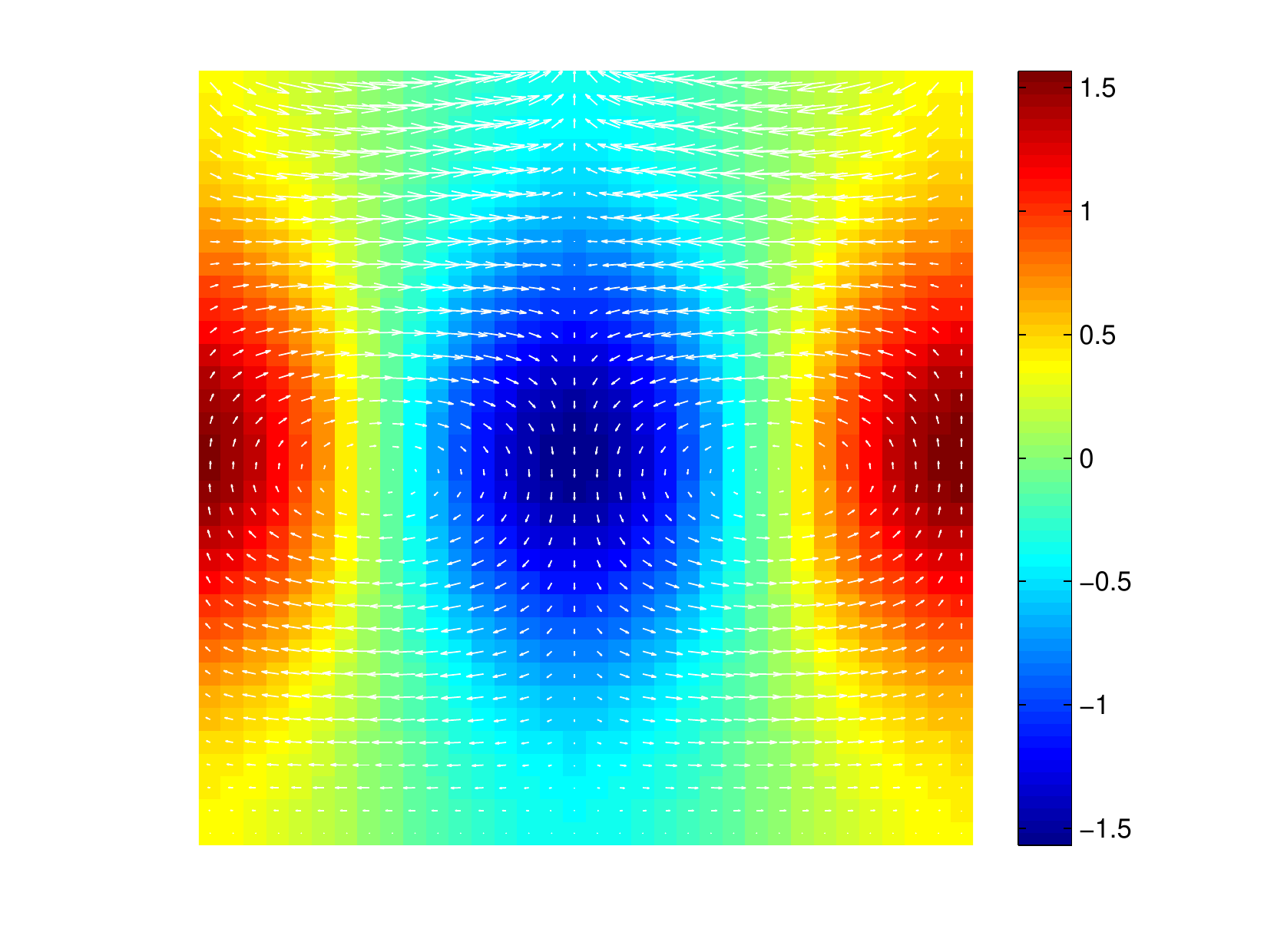}(b)\\
(c)\includegraphics[width=0.45\hsize]{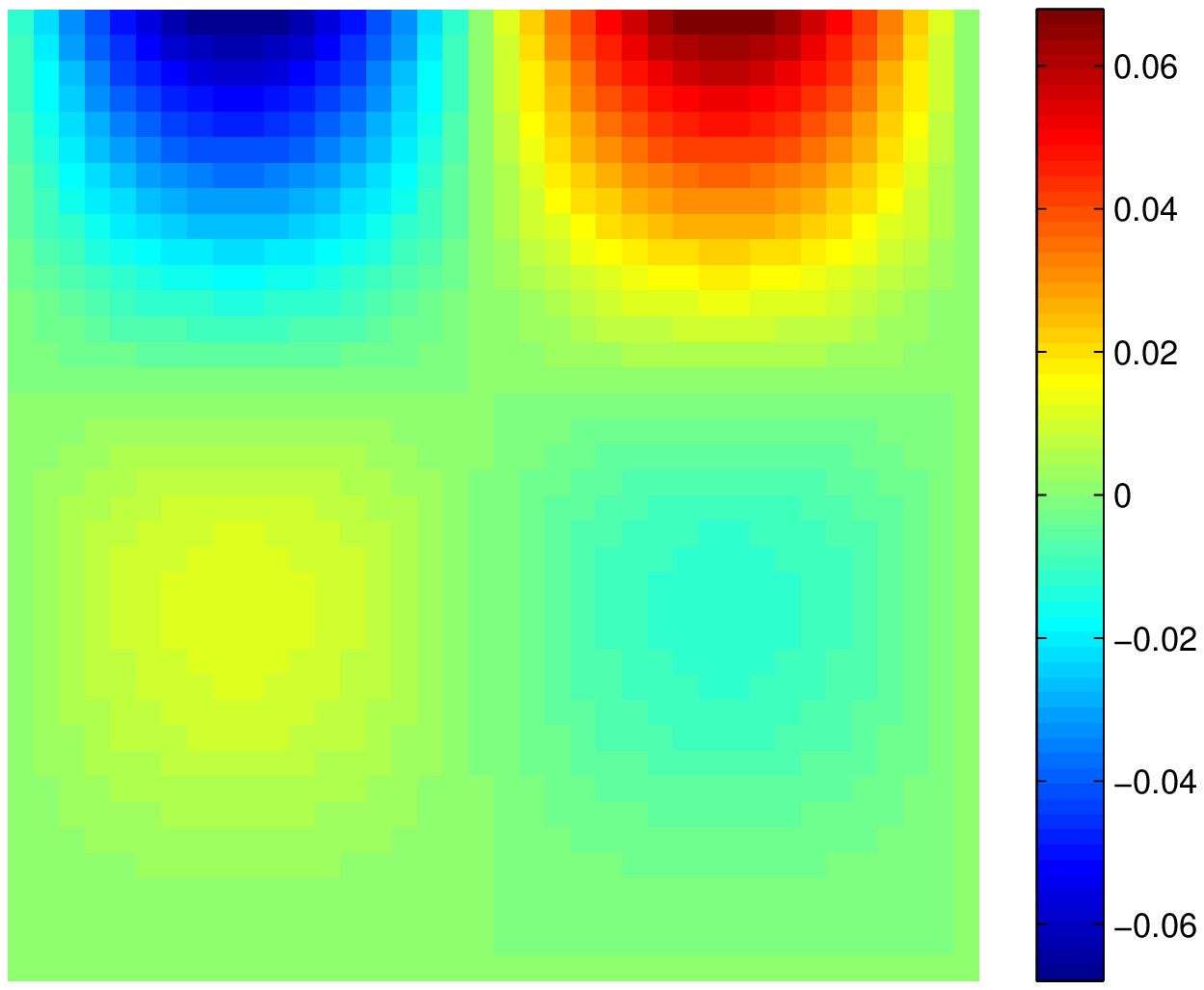}
\includegraphics[width=0.45\hsize]{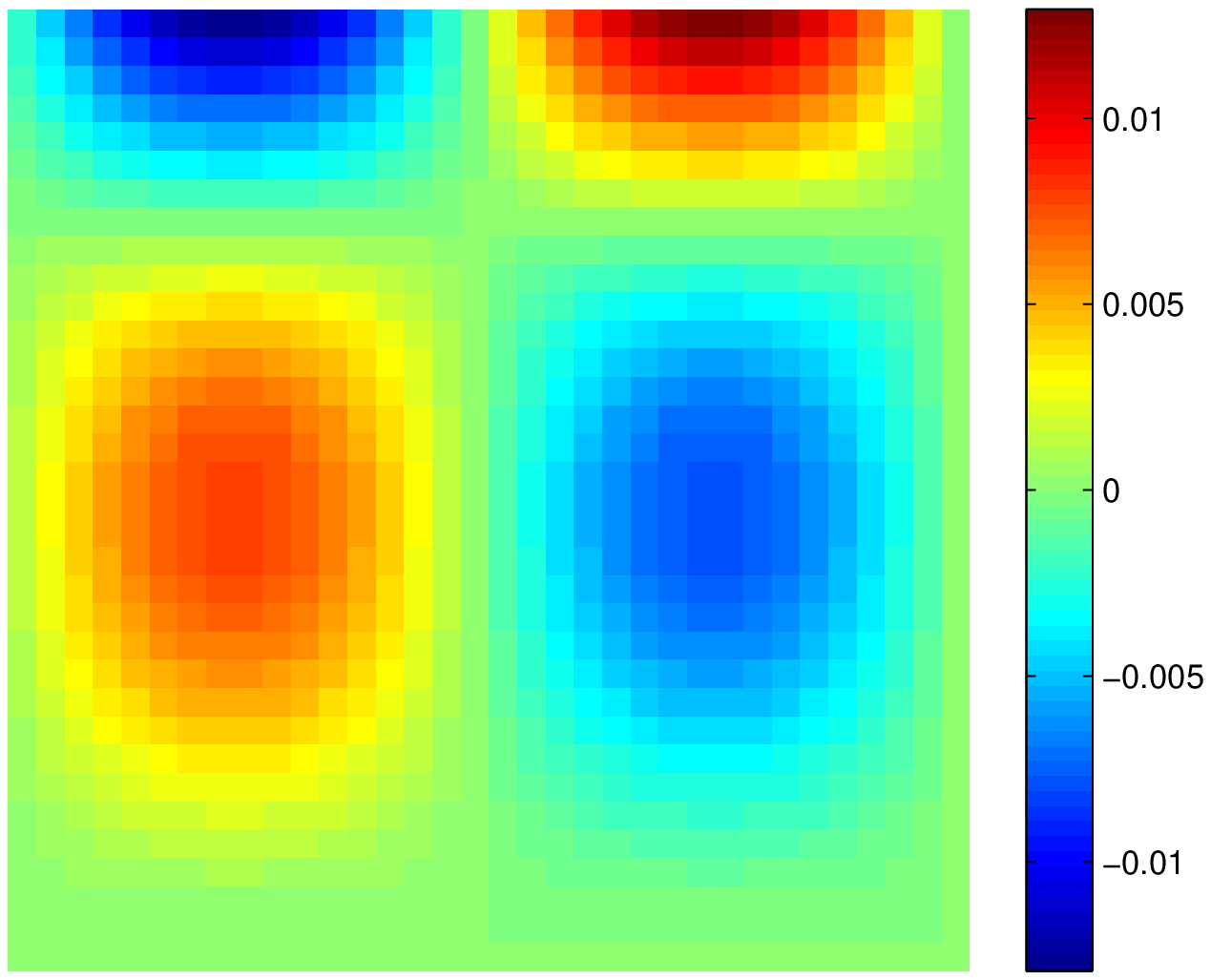}(d)\\
\caption{(color online). Strength of the perturbation fields for
  neutral surfaces ($a^\pm=0$). Panels (a) and (b) give the
  concentration field $c_1(x,z)$ with a superposed vectorial
  representation of the velocity field $(v_1,w_1)$ (white
  arrows). Panels (c) and (d) give the corresponding stream functions
  $\psi_1(x,z)$. The film thickness is (a,c) $h=3.5$ and (b,d)
  $h=5$. The color bars give the corresponding field
  'strength'. Remaining parameters are S$=1$, Re$=0$ and
  Ps/Re$=1$. Horizontal wave numbers are $k=0.558$ ($h=3.5$) and
  $k=0.202$ ($h=5$). The lateral and vertical size of each image
  correspond to the lateral period $2\pi/k$ and the film thickness
  $h$, respectively. \mylab{f:korte:neutral}}
\end{figure}

More detailed considerations are needed for vertically stratified
films as there the perturbations of concentration and velocity are
coupled in the linearized bulk equations and boundary conditions. As
above we distinguish the cases of neutral and energetically biased
surfaces.  For neutral surfaces the vertical gradient of the base
state concentration ($\partial_z c_0|_{0,h}$) is zero at both surfaces
and increases or decreases into the bulk. This implies that the flow
has to be driven by the internal diffuse interface between the two
components [cf.~momentum equation Eq.~(\ref{e:velo})]. Corresponding
flow and concentration perturbation fields are illustrated for
different film thicknesses in Fig.~\ref{f:korte:neutral}. We will call
this type of forcing ``bulk Korteweg driving''.

\begin{figure}[t]
(a)\includegraphics[width=0.45\hsize]{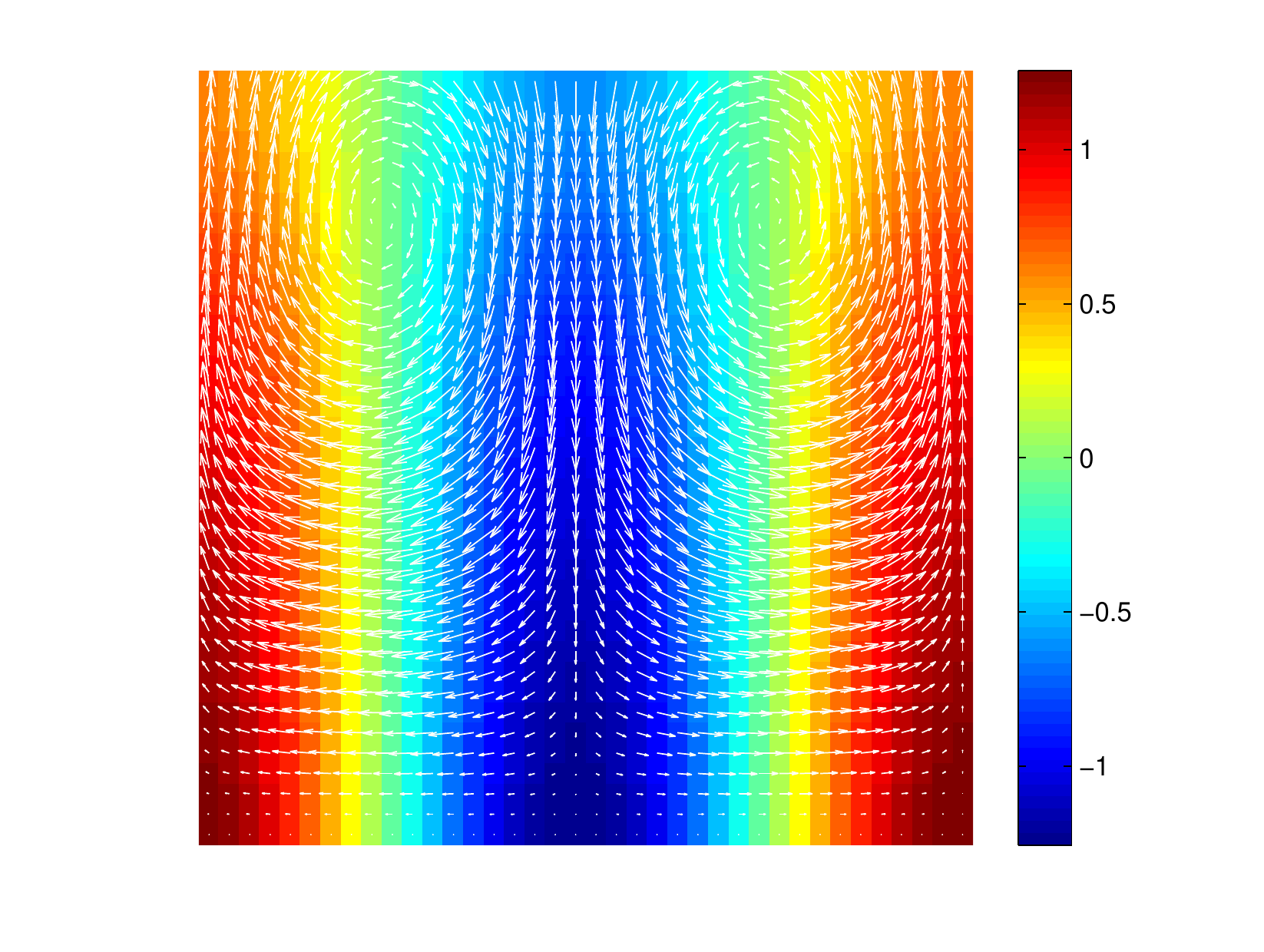}
\includegraphics[width=0.45\hsize]{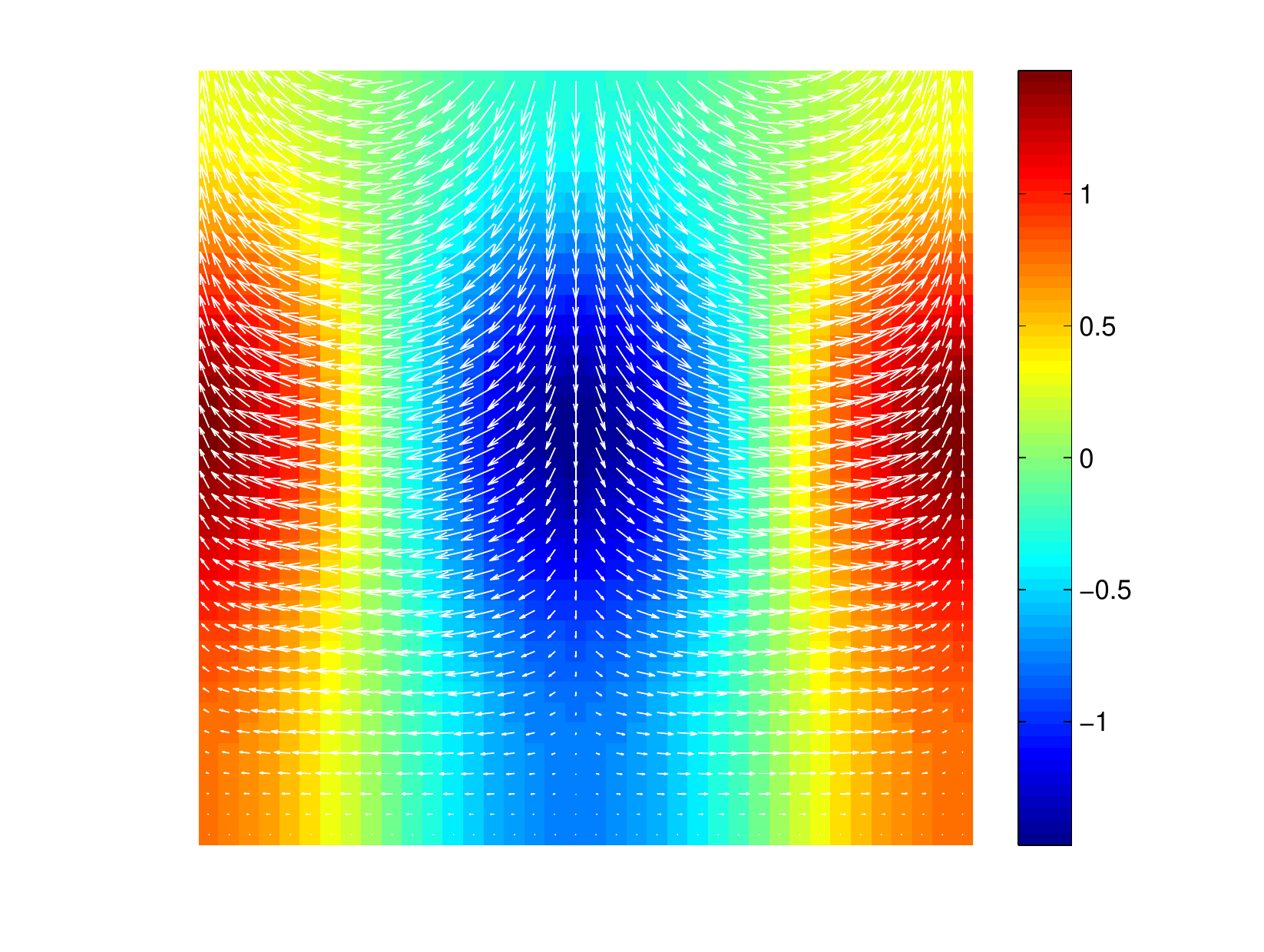}(b)\\
(c)\includegraphics[width=0.45\hsize]{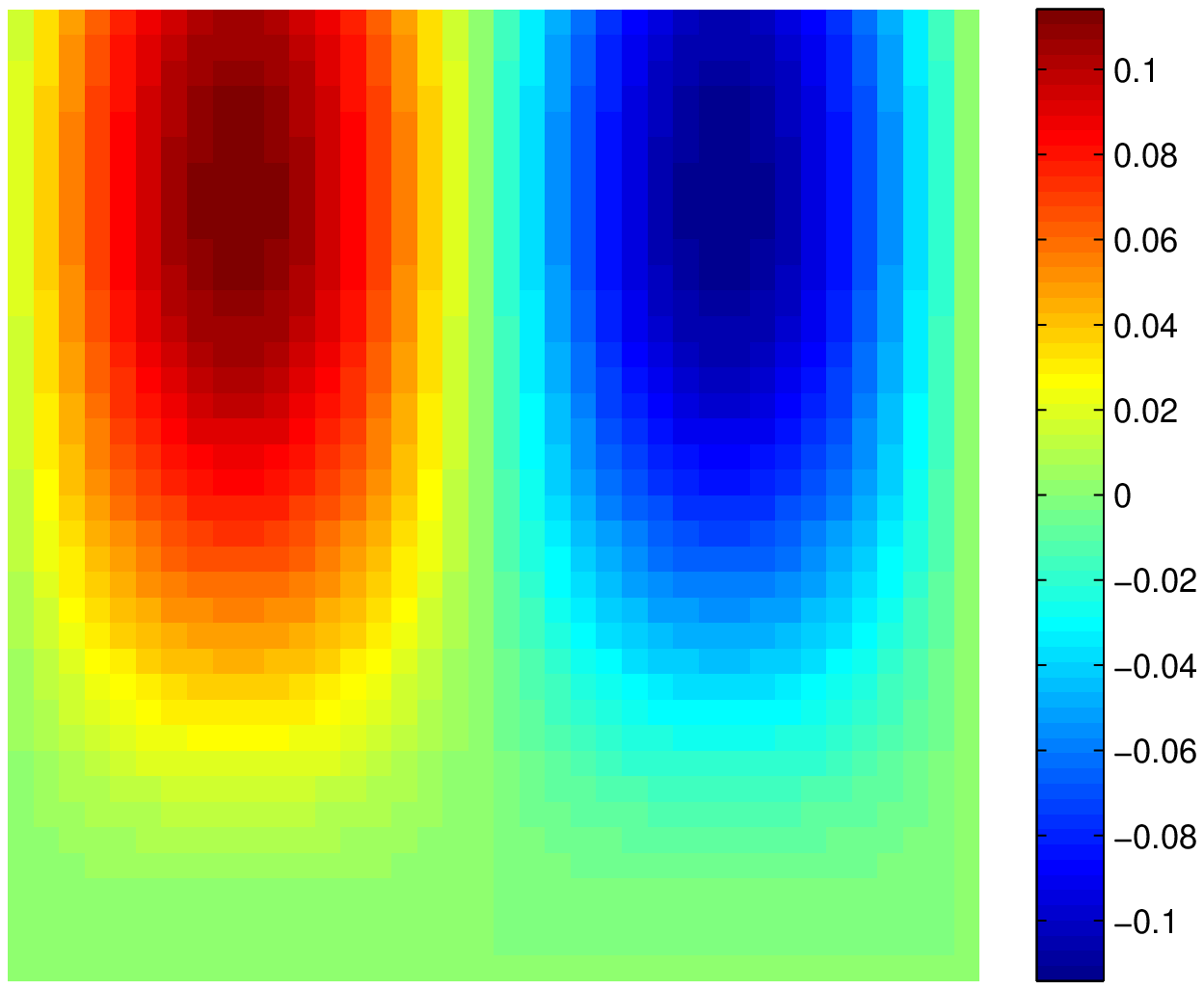}
\includegraphics[width=0.45\hsize]{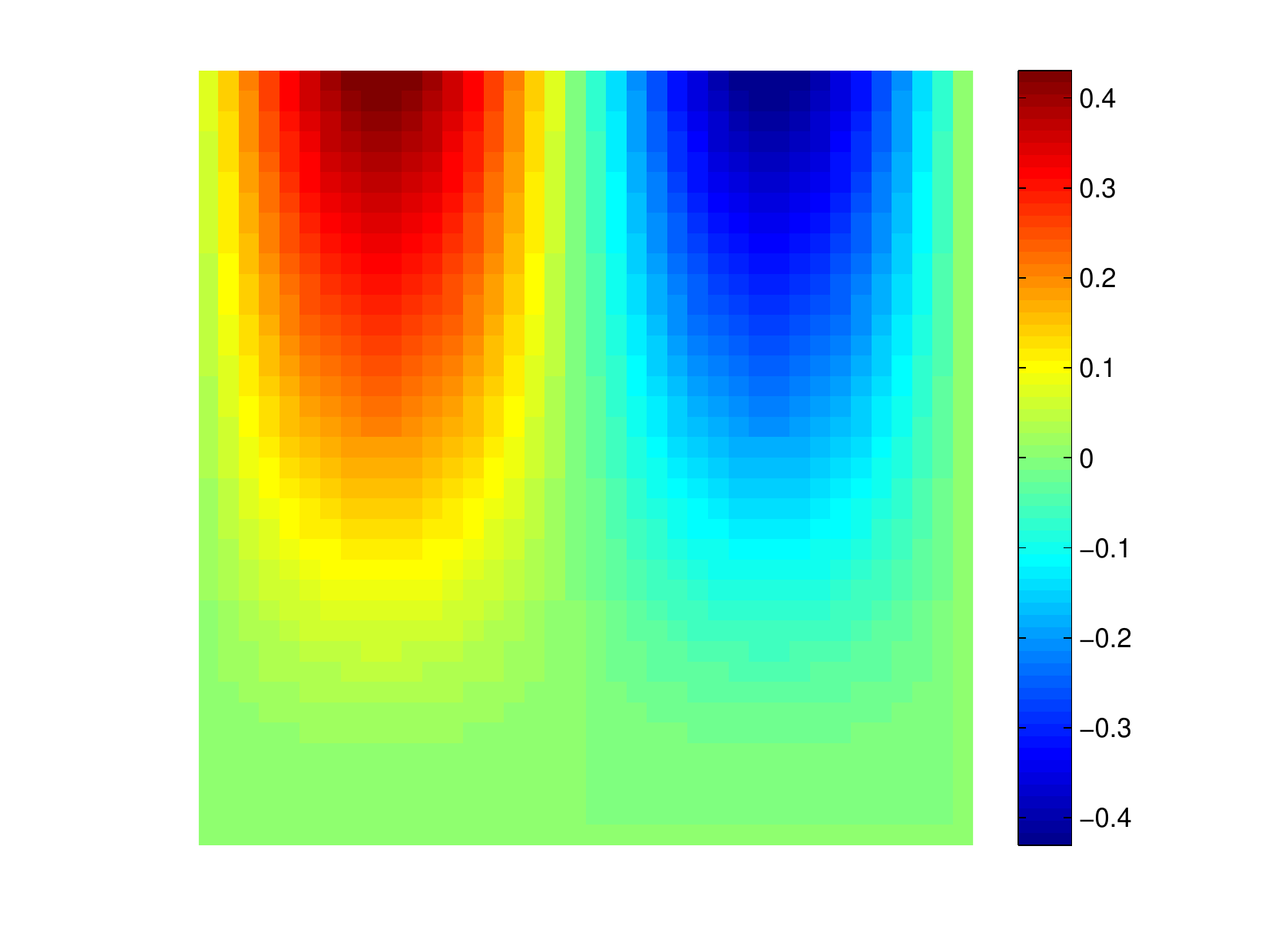}(d)\\
\caption{(color online). Presented are the perturbation fields for
  symmetric surface bias ($a^\pm=0.5$). Panels (a) and (b) give the
  concentration field $c_1(x,z)$ with a superposed vectorial
  representation of the velocity field $(v_1,w_1)$ (white
  arrows). Panels (c) and (d) give the corresponding stream functions
  $\psi_1(x,z)$. The film thickness is (a,c) $h=3.5$ and (b,d)
  $h=5$. The color bars give the corresponding field
  'strength'. Remaining parameters are S$=1$, Re$=0$ and
  Ps/Re$=1$. Horizontal wave numbers are
  $k=0.594$ ($h=3.5$) and $k=0.213$ ($h=5$).  Both solutions are on
  the $n=1/2^b$ branch. The lateral and vertical size of each image
  correspond to the lateral period $2\pi/k$ and the film thickness
  $h$, respectively. \mylab{f:korte:sym}}
\end{figure}

For energetically biased surfaces the $\partial_z c_0|_{0,h}$ are not
zero, they might even have their extrema at the surfaces
[cf.~Fig.~\ref{f:strati-nohydro-a0}(a)]. This implies that the driving
Korteweg forces can be  localized near the surfaces. Liquid
motion, however, is suppressed at the solid substrate due to the
no-slip condition, but can be strong at the free surface.
Corresponding flow and concentration perturbation fields are
illustrated for different film thicknesses in
Fig.~\ref{f:korte:sym}. This ``surface Korteweg driving'' corresponds
to the classical Marangoni driving at a free surface for one-component
systems.

Note, that strictly speaking the flow is only driven by the Korteweg
term in the bulk equations due to the exact cancellation of the
Marangoni and Korteweg terms in the tangential stress boundary
condition for the perturbations. As one is, however, still able to
distinguish a driving at the free surface and a driving at the diffuse
interface we call the former either ``surface Korteweg driving'' or
``Marangoni driving'' and the latter ``bulk Korteweg driving''.
In the following we discuss the individual cases in more detail.

\begin{figure}[t!]
(a)\includegraphics[width=0.4\hsize]{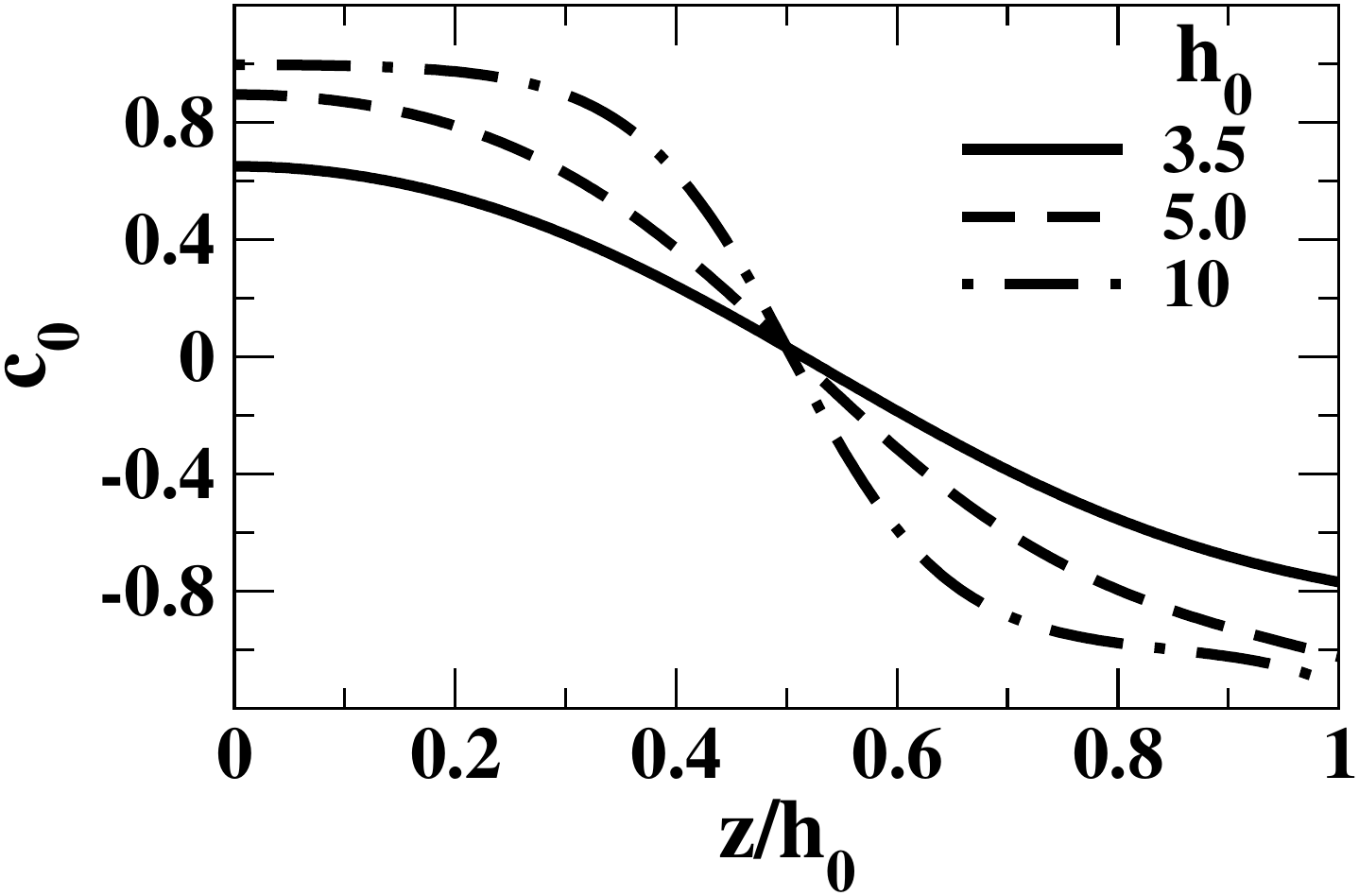}
\includegraphics[width=0.4\hsize]{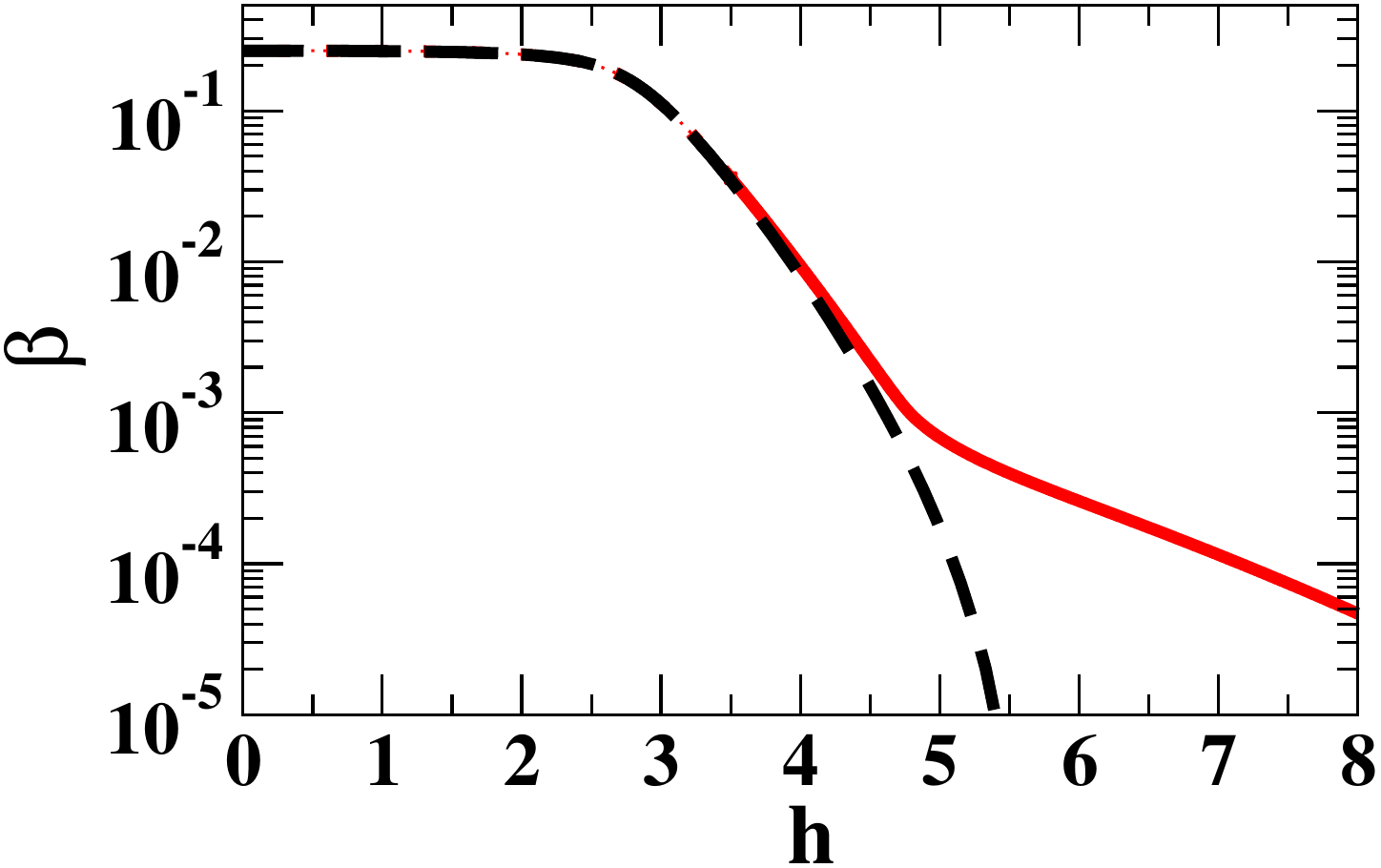}(b)\\
(c)\includegraphics[width=0.4\hsize]{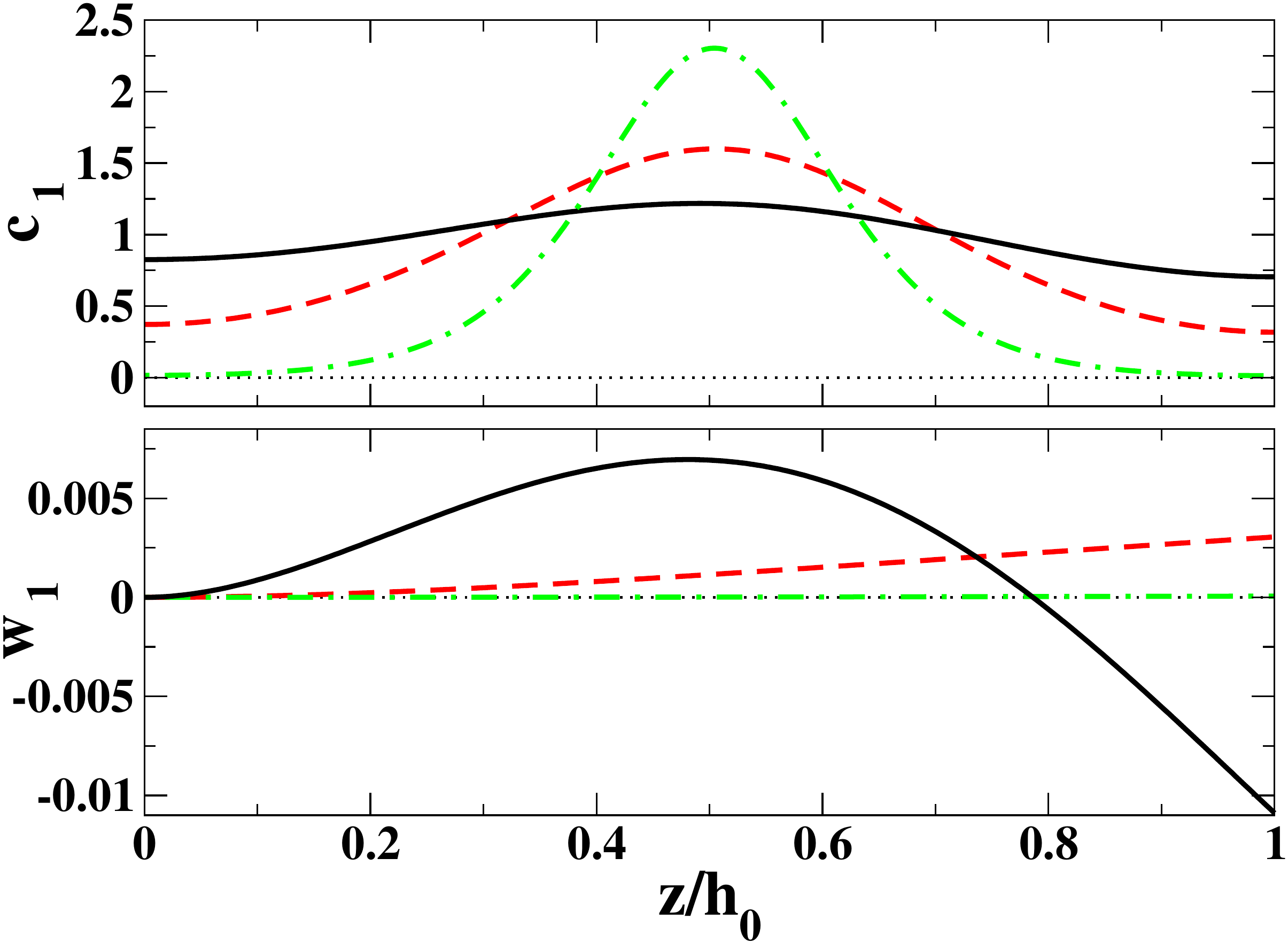}
\includegraphics[width=0.4\hsize]{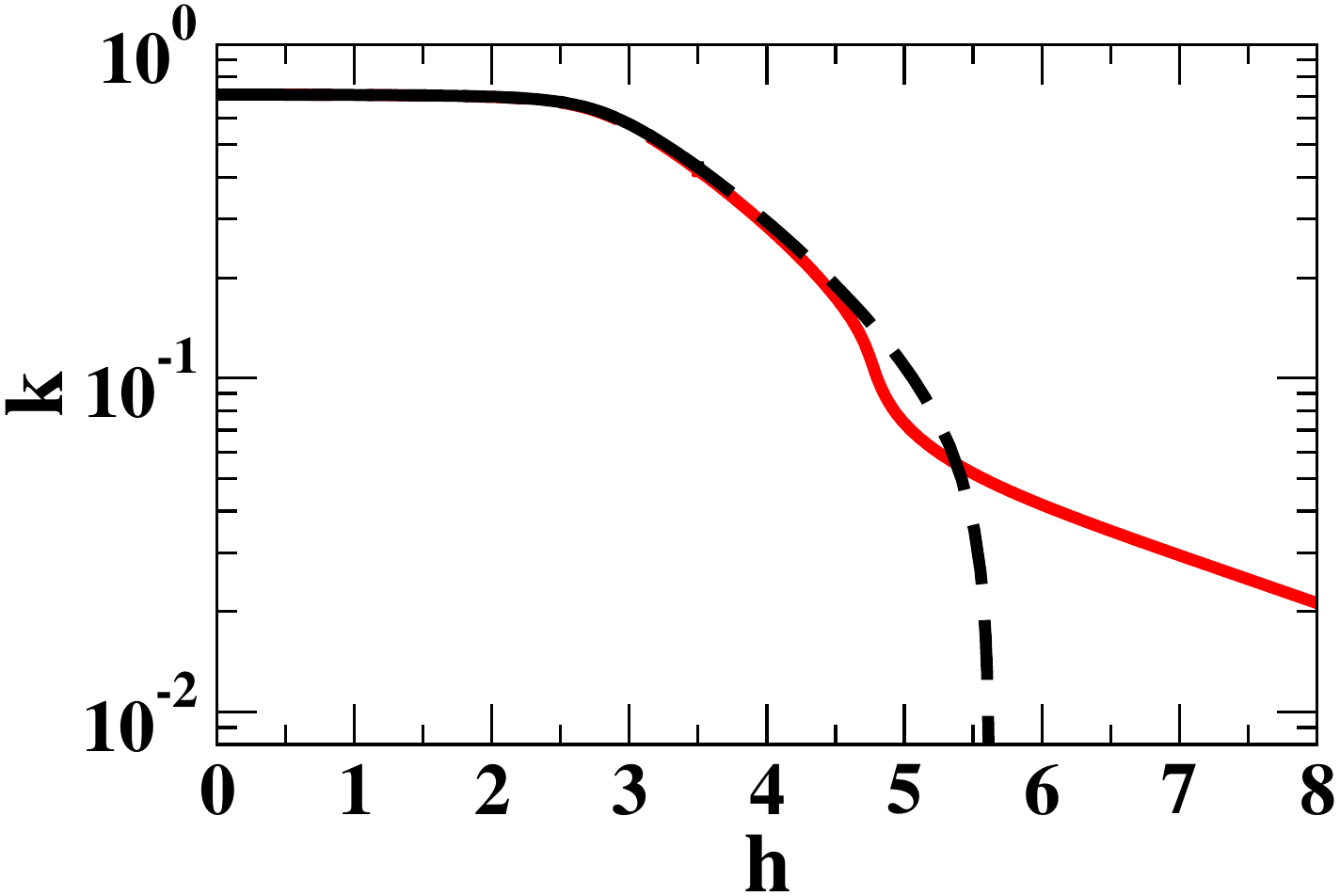}(d)
\caption{ (color online) Results for {\em asymmetrically
    biased} surfaces with $a^+=0.2$ and $a^-=b^\pm=0$. Panel (a) gives
  base state concentration profiles on the energetically favorable
  $n=1/2$ branch (see Figs.~9 and 10 of Ref.~\onlinecite{TMF07}) for
  selected film thickness as indicated in the legend. Panel (c)
  presents the corresponding perturbation modes $c_1(z)$ and
  $w_1$. Note that the $c_1$ curves with and without convection can
  not be distinguished by eye. Panels (b) and (d) present for for this
  $n=1/2$ branch the maximal growth rate and the associated
  wavenumber, respectively, of the horizontal instability modes for
  purely {\em diffusive transport} (dashed lines) and for coupled
  transport by {\em diffusion and convection} (solid lines). The
  remaining parameters are S$=1$, Re$=0$ and Ps/Re$=1$.
  \mylab{f:diff-conv-asym}}
\end{figure}

\begin{figure}[t]
(a)\includegraphics[width=0.7\hsize,angle=0]{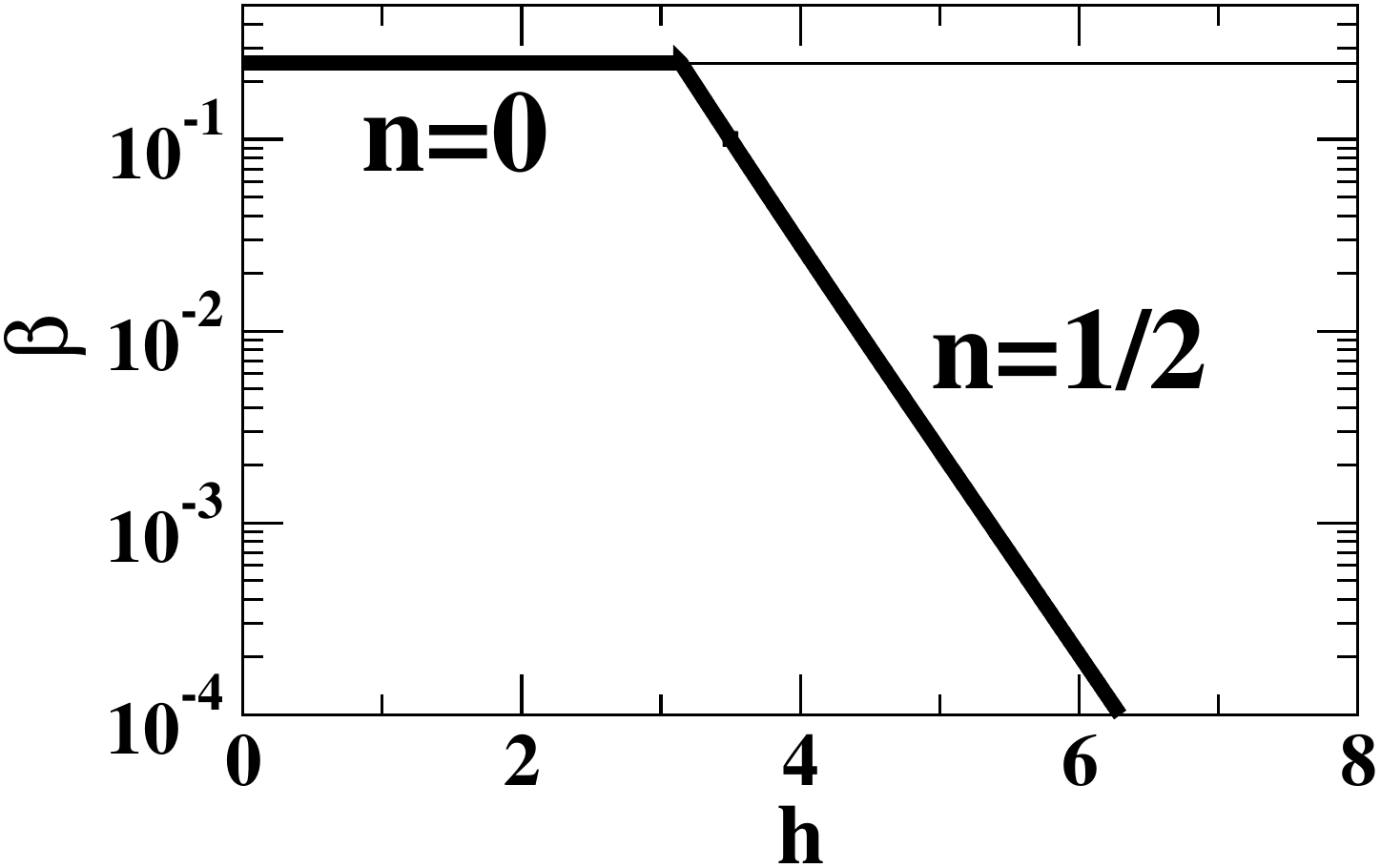}\\  
(b)\includegraphics[width=0.7\hsize,angle=0]{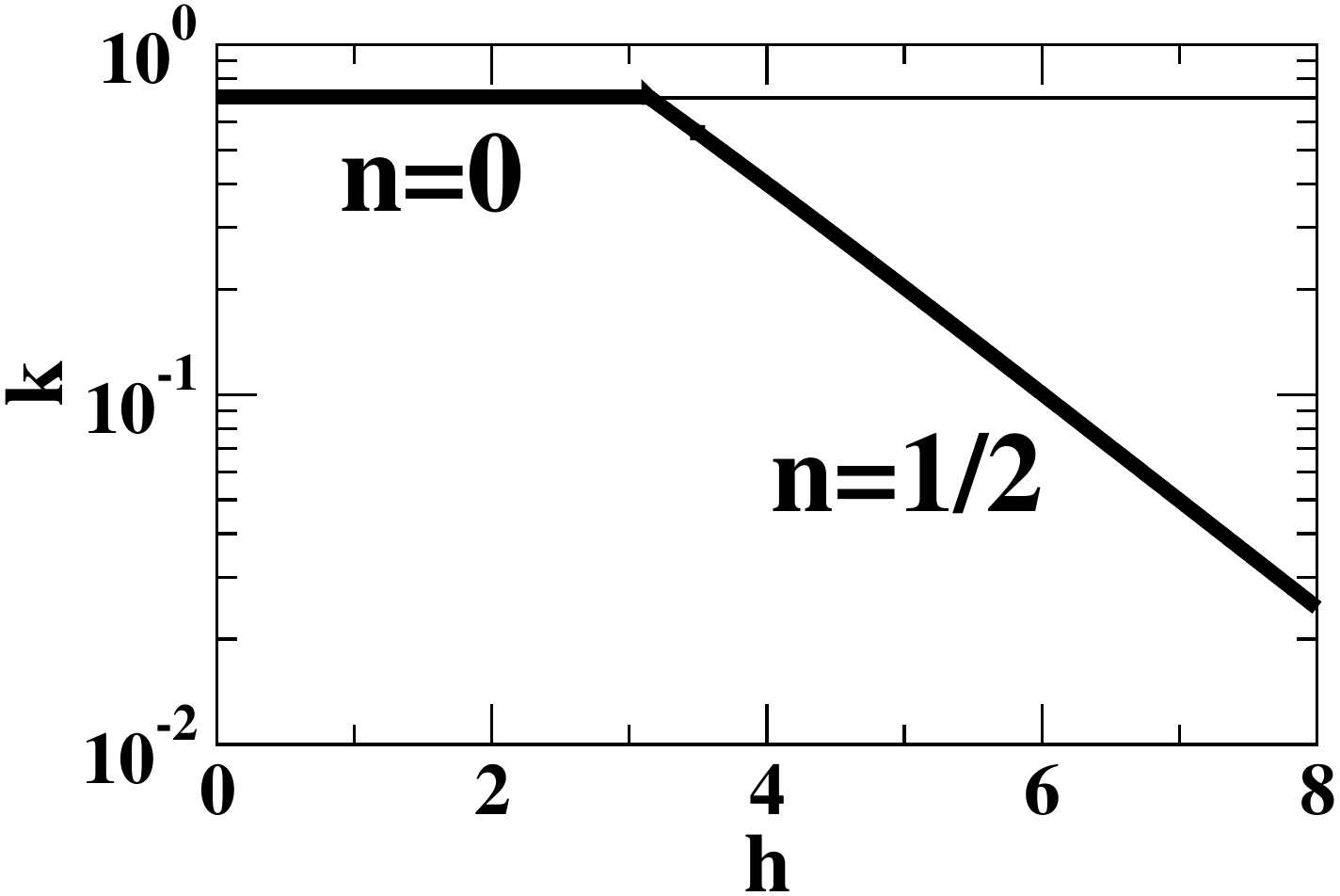}  
\caption{Characteristics of the horizontal instability modes in the
  case of coupled transport by {\em diffusion and convection} for {\em
    neutral surfaces} ( $a^\pm=b^\pm=0$).  (a) Maximum growth rate,
  and (b) associated wavenumber as a function of the film thickness
  for $n=0$ and $n=1/2$ base states (see Figs.~1 and 4 of
  Ref.~\onlinecite{TMF07}). Remaining parameters are S$=1$, Re$=0$ and
  Ps/Re$=1$. \mylab{f:strati-hydro} }
\end{figure}

\subsubsection{Neutral surfaces}\mylab{sec-hydro-ns}
%
The Marangoni number is expressed as
Ma$=a^+\,$S$=a^+\,\gamma_0/lE$. Therefore an upper neutral surface
($a^+=0$) means that gradients of concentration along the free surface
do not give rise to gradients of surface tension, i.e., do not drive a
flow. Fig.~\ref{f:strati-hydro} shows the maximum growth rate and
associated wavenumber as a function of the film thickness. A first
comparison to Fig.~\ref{f:strati-nohydro-a0} seems to indicates that
convective transport does only cause small changes. However, a careful
analysis shows that the instability is accelerated by up to $\sim
50\%$ (as measured at $h=6$: purely diffusive
$\beta=1.4\times10^{-4}$; with convection
$\beta=2.1\times10^{-4}$). This is difficult to discern in the figures
as the growth rate decreases exponentially with increasing film
thickness.

The effect of convective transport on the growth rate is,
however, a result of a rather dramatic change of the velocity
field inside the film. By definition there is no such field in the
purely diffusive case, whereas now with the incorporation of
convective transport 'convection rolls' appear driven by the bulk
Korteweg term in Eq.~(\ref{ds2}). This new dynamics
inside the film allows for a surface deflection to evolve.
Figs.~\ref{f:korte:neutral}(c,d) illustrate the convective motion by
showing the stream function corresponding to the perturbation velocity
  field $(v_1,w_1)$ in a vertical cut through the film for (c) $h=3.5$ and
(d) $h=5$. In the latter case one can clearly appreciate how the
convective cells are driven from within the bulk. At each lateral
position there is a well visible upper cell and a less visible lower
cell that rotate in opposite direction.  For $h=3.5$ the lower cell is
very weak as it is strongly 'damped' by the no-slip condition at the
substrate. The strength of the vertical flow at the upper surface is
proportional to the evolving local surface deflection. 
Figs.~\ref{f:korte:neutral}(a,b) give the corresponding perturbation
fields for the concentration $c_1$ and as well indicate the velocity
field as superposed white arrows. For $h=3.5$ the field $c_1$ shows
nearly vertical stripes, i.e., it is almost a purely horizontal
mode. For $h=5$ the lateral modulation in the concentration field is
less developed at the surfaces. It is strongest
in the region of the diffuse interface. 
%
\subsubsection{Symmetrically biased surfaces}\mylab{sec-hydro-sbs}
%
\begin{figure}[t!]
(a)\includegraphics[width=0.7\hsize]{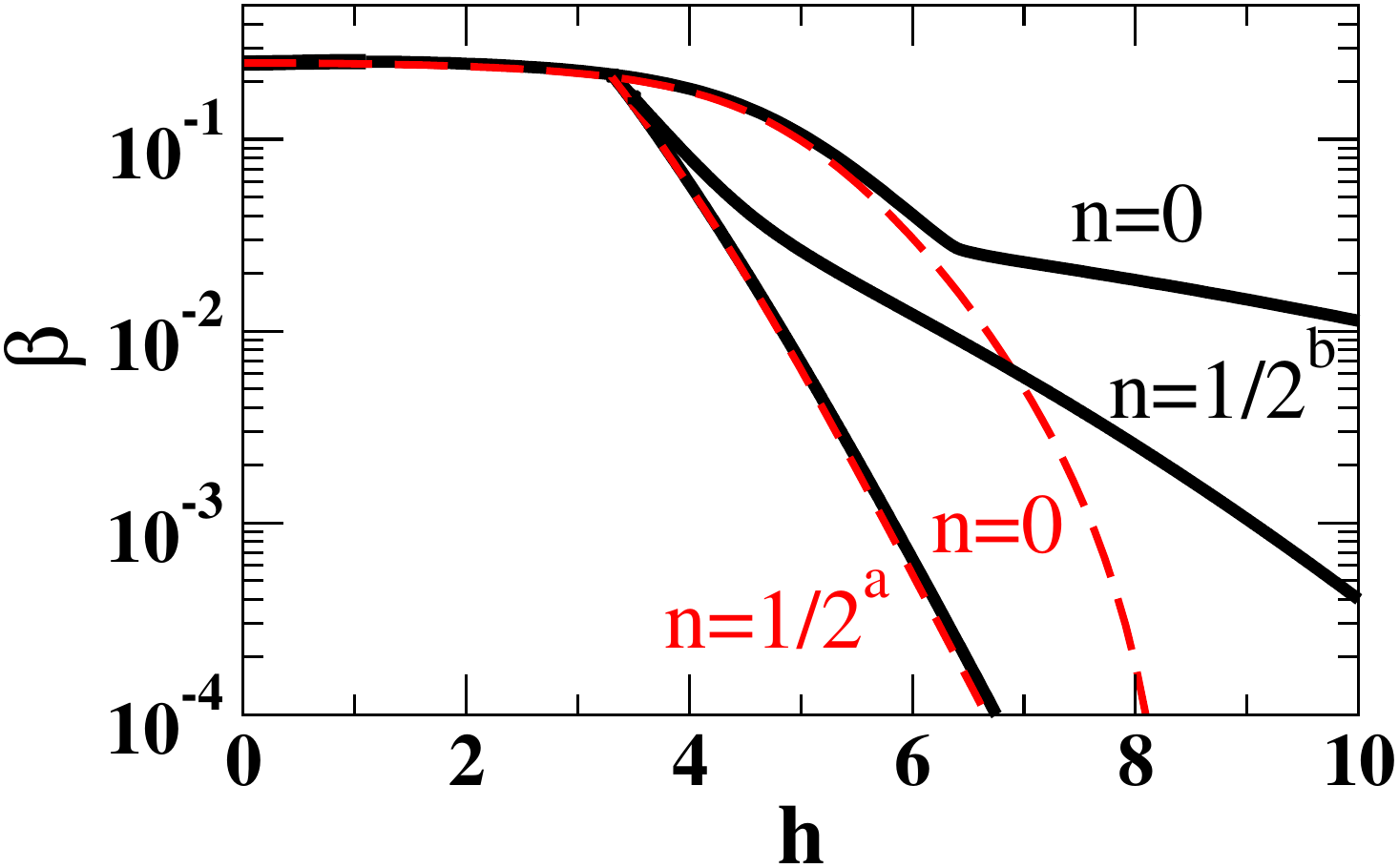}
(b)\includegraphics[width=0.7\hsize]{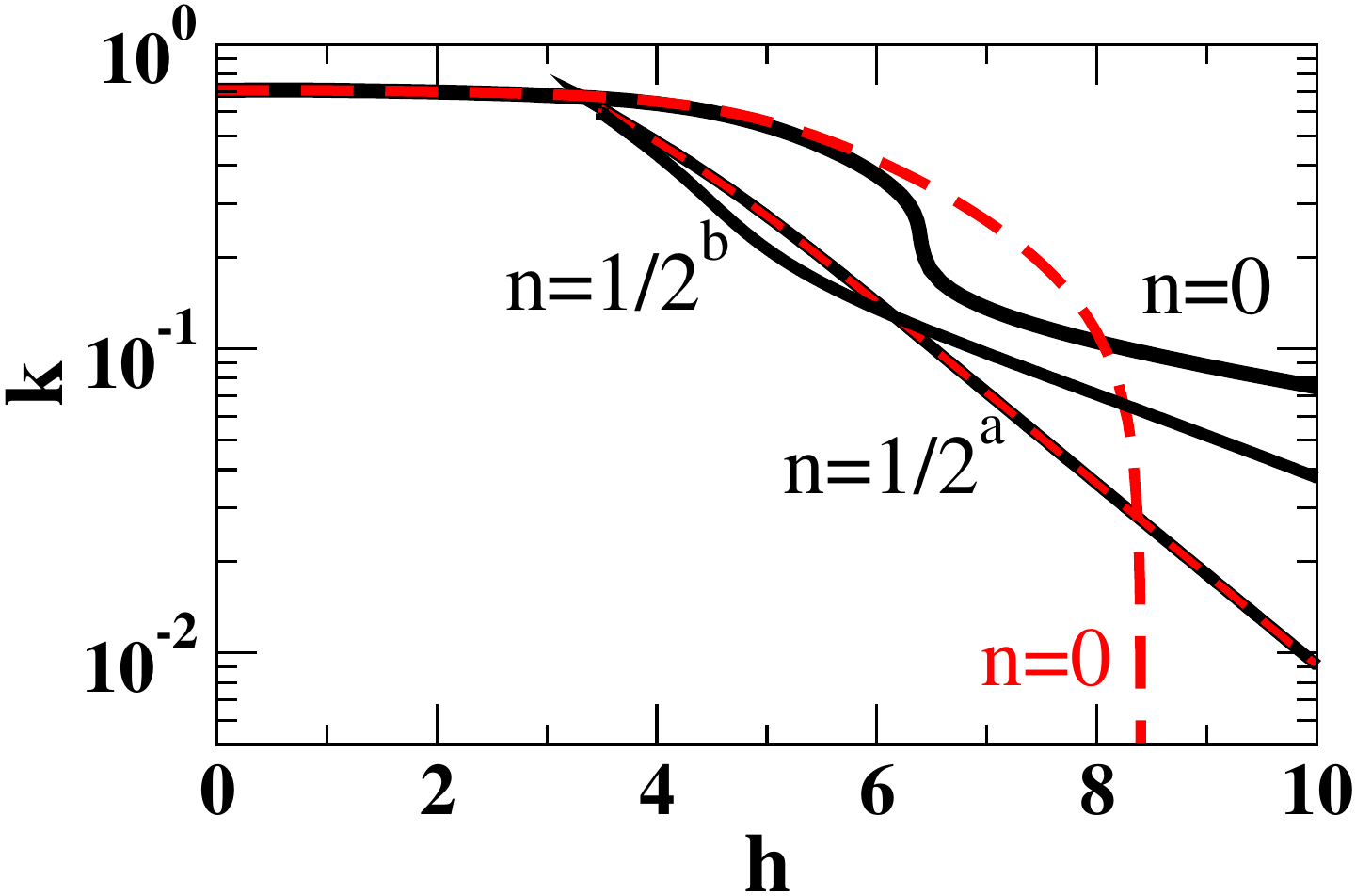}
\caption{ (color online) Characteristics of the horizontal instability
  modes for purely {\em diffusive transport}  (Dashed lines) and for
  coupled transport by {\em diffusion and convection} (solid
  lines). Surfaces are {\em symmetrically biased} with
  $a^\pm=0.5$. Panel  (a) shows the maximal growth rate and (b) the associated
  wavenumber as a function of the thickness for $n=0$ and $n=1/2$ base
  states (see Figs.~3 and 4 of Ref.~\onlinecite{TMF07}). Remaining
  parameters are S$=1$, Re$=0$ and Ps/Re$=1$. \mylab{f:diff-conv} }
\end{figure}

Fig.~\ref{f:diff-conv} gives the main stability result for symmetrical
bias (solid lines) with $a^\pm=0.5$, and S$=1$, therefore Ma$=0.5$ in
comparison to the case of purely diffusive transport (dashed lines).
There are two important differences: (i) Up to $h\approx6$ the $n=0$
branch is with convection only slightly more unstable. In contrast,
for $h>6$ the decrease in the growth rate slows strongly down, i.e.,
with further increasing thickness the branch becomes orders of
magnitude more unstable than without convection.  This is, however,
only of secondary importance as we expect the branch to be
experimentally not of great importance.  It represents the
energetically favorable solution only below $h\approx3.3$. The
behavior is nevertheless interesting and we will later on encounter
similar results for relevant branches.

(ii) Remarkable is the splitting in two of the $n=1/2$ branch. It is
a consequence of the breaking of the up-down symmetry by the
hydrodynamic boundary conditions (no-slip vs.~free
surface). Symmetries and resulting multiplicities of branches of base
states are discussed in \onlinecite{TMF07}.  In particular,
for symmetric bias the $n=1/2$ branch represents two solutions related
by the symmetry $c_0(z)\rightarrow -c_0(h/2-z)$. The two solutions
behave identically when increasing the energy bias.  Without
hydrodynamics they have as well identical stability properties.
However, as convective transport is included the symmetry is broken
and the two solutions acquire different stability properties (marked as
branches $1/2^a$ and $1/2^b$ in Fig.~\ref{f:diff-conv}). Both branches
are more unstable with convection than without, however, whereas for
the branch $n=1/2^b$ the difference reaches one order of magnitude
already at $h\approx5$, the branch $n=1/2^a$ is only slightly more
unstable. 

The velocity field inside the film is quite different from the one for
neutral surfaces given above in Fig.~\ref{f:korte:neutral}. For
symmetrical bias the surface tension depends on concentration and a
Marangoni driving at the free surface is possible. Remember, however,
that the 'classical' solutal Marangoni effect manifests itself in the
linear analysis through a surface Korteweg driving active in a diffuse
region in the vicinity of the sharp free surface.  The corresponding
perturbations of concentration and velocity fields are illustrated in
Fig.~\ref{f:korte:sym} for the $n=1/2^b$ branch at $h=3.5$ (left) and
$h=5$ (right). In the representation of the velocity field as
stream function $\psi_1(x,z)$ one can clearly discern 'convection rolls'
driven by the free surface, i.e., at each lateral position there is
only one convection cell that is concentrated near the free surface.
As the driving occurs where the film is hydrodynamically most mobile
the destabilizing effect is much stronger when caused by the Marangoni
mode (surface Korteweg mode) than when caused by the bulk Korteweg
mode.  In consequence, for larger thickness the roll reaches less
deeply into the layer.
Figs.~\ref{f:korte:sym}(a,b) give the corresponding perturbation
fields for the concentration $c_1$ and as well indicate the velocity
field as arrows. For $h=3.5$ the field $c_1$ shows again a rather
horizontal mode. It has, however, a stronger vertical element than in
the neutral case above. Here, the lateral modulation is strongest at
the substrate and weakest at the free surface.  For $h=5$ there is
nearly no lateral modulation at the free surface. The modulation is
strongest around the diffuse interface and is less developed at the
substrate.

\begin{figure}[tp!]
(a) \includegraphics[width=0.7\hsize]{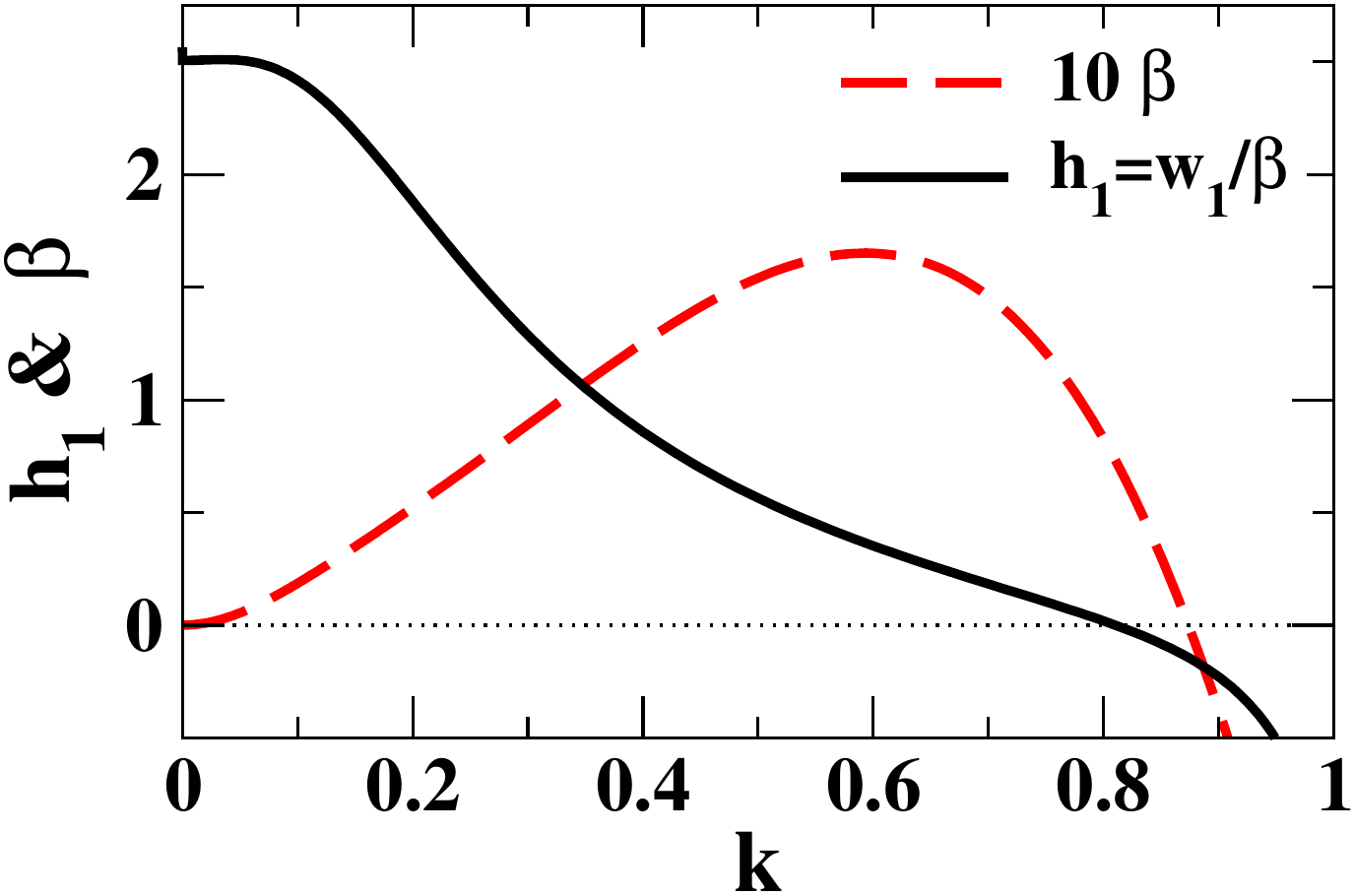}\\
(b) \includegraphics[width=0.7\hsize]{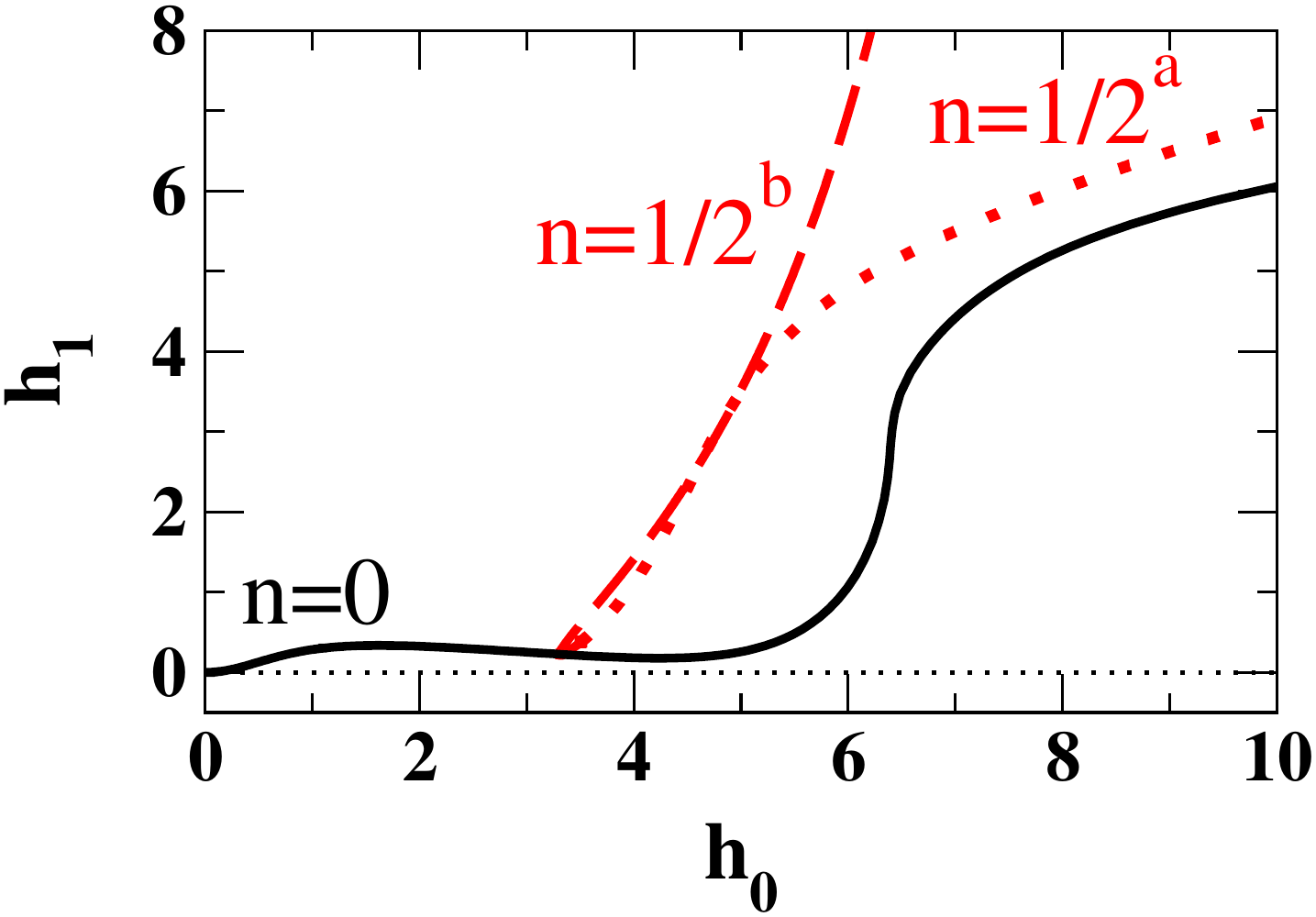}
\caption{(color online) Panel (a) shows the  relative strength  of the
  surface deflection $h_1 = w_1/\beta$ (solid line) and 
  growth rate (dispersion relation, dashed line) as a function of the
  wavenumber for a fixed thickness $h=3.5$ for the branch $n=1/2^b$.  (b) The deflection
  $h_1$ at the maximum of the dispersion relation is given as a
  function of the film thickness for the $n=0$ branch (solid line) and
  the branches $n=1/2^a$ (dotted line) and $n=1/2^b$ (dashed line) for
  {\em symmetrically biased} surfaces with $a^\pm=0.5$ and
  $b^\pm=0$. The remaining parameters
  are S$=1$, Re$=0$, Ps/Re$=1$.\\
  \mylab{f:defle-sym} }
\end{figure}

As in the neutral case the convection allows for a surface deflection
to evolve.  Fig.~\ref{f:defle-sym}(a) shows -- using $h=3.5$ as
example -- the relative strength of the surface deflection $h_1 =
w_1/\beta$ as a function of the lateral wavenumber where $w_1$ is
the value at the vertical velocity component at the free
surface. Fig.~\ref{f:defle-sym}(a) gives as well the corresponding
dispersion relation $\beta(k)$ to facilitate the identification of the
physically most relevant value of $h_1$ at the maximal growth
rate. Inspecting the full range of $k$ one finds that the maximal
deflection occurs at small but finite $k$. From the maximum value
$h_1$ decreases monotonically until reaching negative values at large
enough $k$. Note that the mode of maximal deflection does not
correspond to the most dangerous mode. We add two remarks regarding
the interpretation of the $h_1(k)$ curve: (i) The fact that
$h_1(k=0)\neq0$ does not contradict material conservation as the $k=0$
mode is marginally stable ($\beta=0$) and not unstable.  (ii) The
deflection amplitude $h_1$ does not correspond to an absolute
amplitude value, it rather represents the relative importance and the
relative sign of the surface deflection as compared to the
perturbation of the concentration field (cf.~Section~\ref{sec-linstab2}).

The accompanying Fig.~\ref{f:defle-sym}(b) presents the dependence of
the surface deflection of the most dangerous linear mode on film
thickness for all branches discussed above. For the $n=0$ branch in
the range below $h\approx3.3$, where it is energetically favorable the
deflection first increases from zero to $\approx 0.35$, then decreases
again. Above $h\approx3.3$, $h_1$ continues to decrease till
$\approx0.17$ at $h\approx4.2$ before increasing strongly at about
$h\approx6$ in accordance with the change in the
$\beta_{\mathrm{max}}(h)$ dependence discussed in connection with
Fig.~\ref{f:diff-conv}.  For the strongly stratified $n=1/2$ branches
the deflection increases quickly with film thickness above their
bifurcation from the $n=0$ branch indicating the importance of the
surface deflection for the lateral instability modes. All these
findings strengthen the above conclusion that at small thicknesses
diffusion dominates even if convection is possible. Then there occurs
a rather sharp transition over a well defined $h$-range (for the
present parameter set at about $h=6$) where the convective influence
becomes important.
%
%
\subsubsection{Antisymmetrically and asymmetrically biased surfaces}
%
Results for antisymmetrical bias with $a^+=-a^-=0.2$ are included
above in Fig.~\ref{f:diff-conv-anti} for the energetically preferred
branch $n=1/2$. Contrary to symmetrical bias, one finds that cases
without and with convection are nearly identical. For instance, at
$h=3.5$ the growth rate with convection is only two per cent larger
than without hydrodynamics and the $c_1$ perturbation fields can
visually not be distinguished (cf.~Fig.~\ref{f:diff-conv-anti}(c)).
In both cases the films stabilize w.r.t.~lateral instability modes at
$h\sim 4.5$. The following consideration might, however, help to
understand the finding: The film stabilizes at about $h=4.5$. Below
this rather small thickness the results with and without convection
are neither well distinguished in the other studied cases. That means
the antisymmetrical bias exercises such a strong stabilizing influence
that films are already stable 'before' (in terms of film thickness)
convective motion becomes relevant. The velocity field $w_1$
(cf.~Fig.~\ref{f:diff-conv-anti}(c)) changes non-monotonously with $z$
indicating a pair of convection rolls as in
Fig.~\ref{f:korte:neutral}.  This indicates that the weak convection
is driven from the diffuse interface within the film.

%
%
Finally, we consider the case of asymmetrical bias.
Fig.~\ref{f:diff-conv-asym} above gives results for the energetically
favorable $n=1/2$ branch at $a^+=0.2$ and $a^-=0$ in comparison to
the purely diffusive case.
The overall behavior resembles strongly the one described for
symmetrical bias above in Section~\ref{sec-hydro-sbs} for the
influence of convection on the $n=0$ branch: up to $h\approx4.5$ the
profile is only slightly more unstable with convection than
without. In contrast, for $h>4.5$ the film is much more unstable with
convection. Without convective motion the film stabilizes at $h_c
\approx 5.8$. Allowing for convection and therefore for the evolution
of a surface deflection the film remains laterally unstable above
$h_c$.  The growth rate and wavenumber of the most unstable mode still
decrease roughly exponentially with increasing film thickness.

The rationale behind this finding is that convective motion does not
only add a second transport process to the dynamics but it as well
allows the film to realize a different class of solutions, namely,
films with laterally modulated concentration \textit{and} thickness
profiles. This qualitative change is as well reflected in a change of
the character of the perturbation velocity profile
(cf.~Fig.~\ref{f:diff-conv-asym}(c)). At $h=3.5$ it shows
non-monotonous behavior indicating driving from the diffuse interface
inside the film whereas at larger thicknesses (shown for $h=5$)
driving comes from the free surface.  This corresponds to the
transition from bulk Korteweg driving with perturbation fields that
resemble Fig.~\ref{f:korte:neutral} to surface Korteweg (or Marangoni)
driving with perturbation fields that resemble Fig.~\ref{f:korte:sym}.
The relative strength of the surface deflection $h_1$ (not shown)
shows similar behaviour as for the $n=0$ branch in the
symmetric case (see solid line in Fig.~\ref{f:defle-sym}(b)).

We conclude this section with a remark on the character of the
dispersion relation. For larger film thicknesses we find wavenumber
ranges corresponding to complex modes. Although they seem to be of
minor importance (as we find the most dangerous mode to be always
real) they are interesting from a theoretical point of view. They seem
to be closely related to the possibility of surface deflection,
however, their physical interpretation remains elusive. They appear in
parameter ranges where several existing real modes re-connect in a
different way. However, we carefully checked that they are no
numerical artifact: they do not depend on details of the used
discretization and are reproducible over several orders of magnitude
of numerical tolerances in the employed continuation procedure. Open
questions related to this point will be the subject of further
investigations.
%
\subsection{Influence of Re, Re/Ps and S}
\mylab{sec:pars}
%

\begin{figure}[t!]
\hspace*{-0.5cm} \includegraphics[width=0.76\hsize]{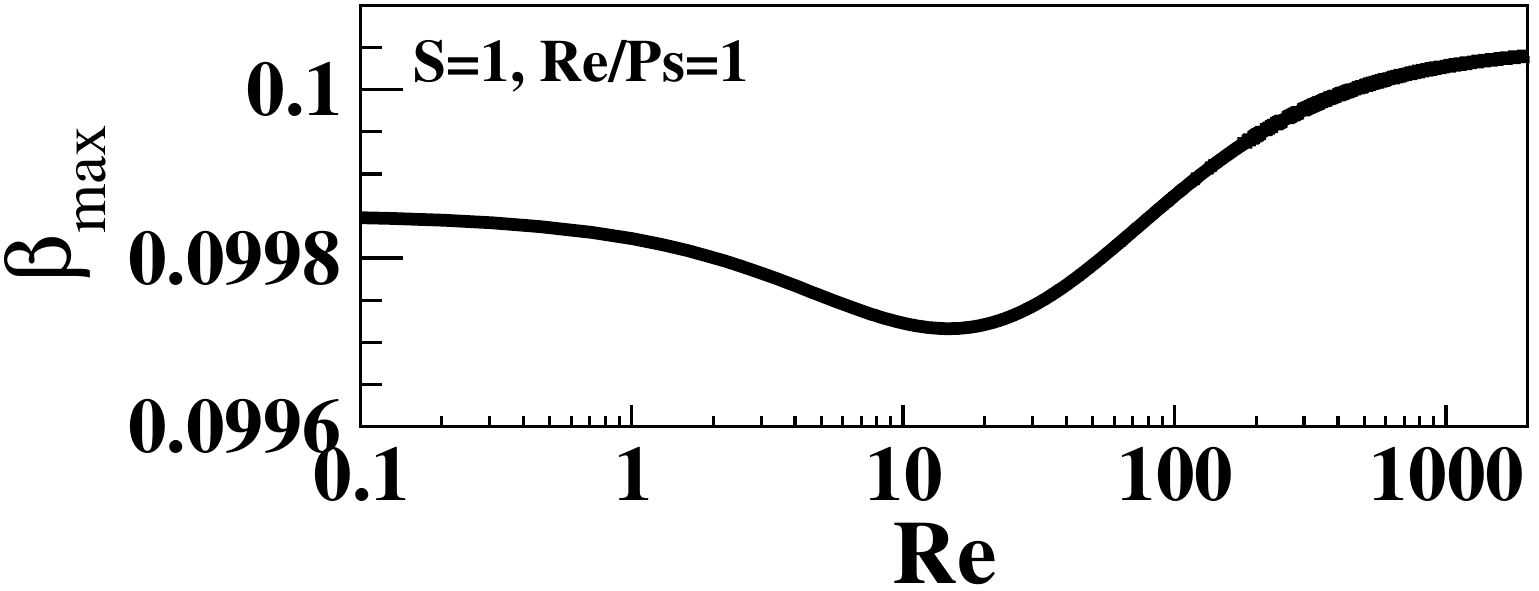}
\includegraphics[width=0.74\hsize]{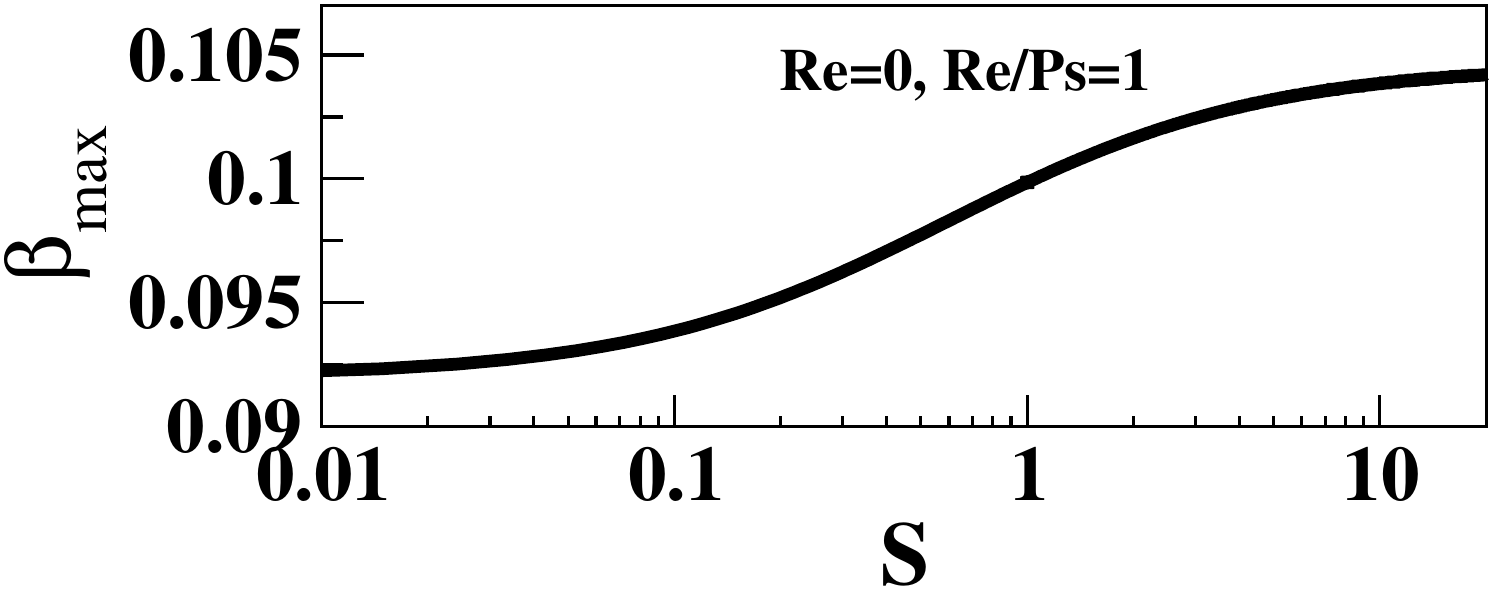}
\hspace*{0.5cm} \includegraphics[width=0.7\hsize]{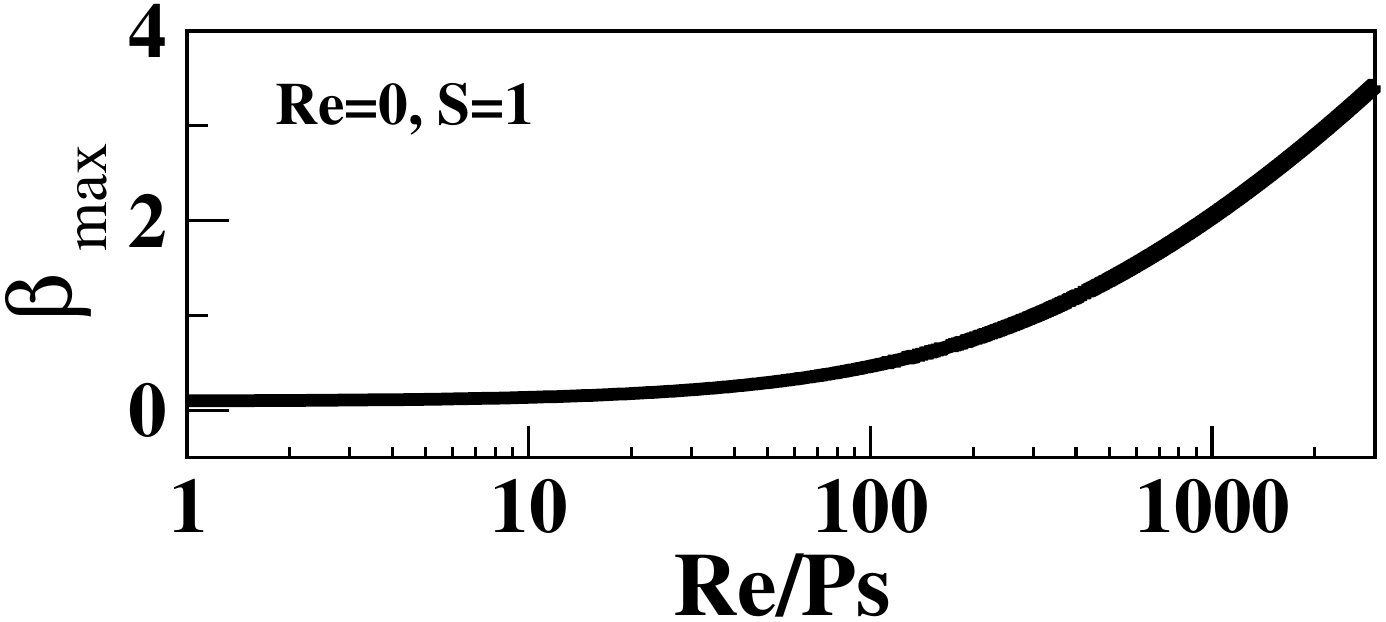}
\caption{Dependence of the maximal growth rate as a function of (a) the Reynolds number Re, (b)
  the surface number S, and (c) the ratio Re/Ps. The curves are
  obtained for a stratified film of the branch $n=1/2$ with $h=3.5$
  and neutral surfaces $a^\pm=b^\pm=0$.
  \mylab{f:ReS} }
\end{figure}

One can expect that for different types of binary mixtures such as
n-pentane/n-decane mixtures, polystyrene/cyclohexane mixtures, or
ionic binary mixtures such as triethyl n-hexyl borate in diphenyl
ether the parameters that we have kept fixed up to here, e.g., the
Reynolds number Re, the ratio Re/Ps, and the dimensionless surface
tension S will strongly differ. A full scale study of the influence of
those parameters is beyond the scope of the present work. However, we
would like to indicate the tendencies that are to be expected by
looking at one particular example.  We show in Fig.~\ref{f:ReS}(a) to
(c) the influence of changing Re, S and Re/Ps, respectively, on the
growth rate of the most dangerous lateral instability mode for the
case of neutral surfaces and a film thickness of $h=3.5$.  Wavenumbers
are only discussed in the text.

When changing the Reynold number (Fig.~\ref{f:ReS}(a); keeping the
other parameters including Re/Ps fixed), the growth rate at first (for
Re$ < 14.5$) slightly decreases before increasing until it saturates
at about Re$=1000$.  The corresponding wavenumber increases
monotonically with Re and also saturates for Re $\simeq 1000$.  Note,
however, that the differences between the minimal and maximal value of
the growth rate and the wavenumber in the studied Re range are only
about $0.3\%$ and $0.5\%$, respectively. Therefore, in the present
scaling and the studied case the variation of the Reynolds number
alone has practically no influence on the stability of a mixture with
unbiased surfaces. Remember, that the same velocity enters the
definition of Re and the time scale that enters $\beta$. Fixing Re/Ps
furthermore implies that the ratio of diffusive and convective
transport remains constant.

When increasing the surface tension number (Fig.~\ref{f:ReS}(b))
the growth rate and the wavenumber both increase monotonically. The
growth rate saturates at about S$\simeq 5$ and the maximum value of
the growth rate and wavenumber are close to the ones found for high
Re.  Although the increase of S  has a slightly larger impact than
the increase of Re, the overall variation is still quite small: growth rate
$\approx 10\%$ and wavenumber $\approx 3\%$).

The ratio Re/Ps has a much larger impact on the stability of the
mixture (c.f. Fig.~\ref{f:ReS}(c)): The increase of the growth rate is
orders of magnitude larger than when increasing Re or S. The growth
rate increases monotonously with increasing Re/Ps, whereas the
associate wavenumber decreases. As discussed above Re/Ps corresponds
to the ratio of the typical velocity of the viscose flow driven by
Korteweg stresses and the typical velocity of diffusive processes.
Therefore, an increase of Re/Ps shifts the transition towards a larger
influence of convective transport to smaller film thicknesses and
results in the behavior presented in Fig.~\ref{f:ReS}(c).
%

\section{Conclusions}
\mylab{sec:conclusions}
%
We have studied the linear stability with respect to lateral
perturbations of homogeneous and layered films of polymer mixtures
that have a free deformable surface and are bound on the other side by
a rigid solid substrate. The paper represents the second part of a
series of works that develops and applies a version of model-H
suitable for the study of confined systems involving free surfaces.

In the first part (Ref.~\onlinecite{TMF07}) we had derived a
generalized model-H coupling transport equations for momentum, density
of one component of the mixture and entropy. The model was then
simplified for isothermal systems. Furthermore, we had modeled an
energetic bias towards one of the components at the surfaces, had
discussed how to include the evolving free surface and had finally
determined stratified film base states for various types of
energetic bias at the surfaces.

The present second part has focused on the determination of the linear
stability properties of the stratified film states studied in
Ref.~\onlinecite{TMF07}. To this aim, we have linearized the governing
equations with respect to small perturbations in the velocity and
concentration fields, and the surface profile. We have emphasized that
hydrodynamic flow has to be accounted for even in the case of
extremely slow creeping flow as otherwise surface deflections can not
evolve. The resulting linear system of equations and boundary
conditions plays a similar role as the coupled Orr-Sommerfeld-type
equations for velocity and temperature perturbations describing the
linear surface-tension driven instability of a horizontal liquid layer
with a deformable free-surface of a heated layer of simple liquid
\cite{Taka81}.

We have performed a numerical analysis of the stability with respect
to lateral modes of homogeneous and stratified base states (quiescent
films) for several different cases of energetic bias at the surfaces,
corresponding to linear and quadratic solutal Marangoni effects. In
passing we have elucidated the close relation between surface Korteweg
driving (forcing by Korteweg stresses in the bulk momentum equation
for the diffuse layer adjacent to the free surface) and Marangoni
driving. The numerical analysis has been performed to a high precision
relying on continuation techniques. The latter have, in particular,
allowed to follow the most dangerous linear mode over a wide range of
a number of important parameters.

For neutral surfaces two types of steady film solutions exist:
homogeneous and stratified ones. For homogeneous films, perturbations
of the composition and velocity fields are linearly decoupled. In
consequence, the stability of the films is ruled by the linear modes
of a Cahn-Hilliard equation in a confined slab-type geometry. We have
found that the film is laterally unstable for all thicknesses in
agreement with literature.\cite{FMD97,FMD98}
For layered films, the films become exponentially less unstable with
increasing thickness in the cases without and with inclusion of
convective transport. In the latter case the decrease is slightly
slower, i.e., hydrodynamics further destabilizes the films. We have
found that for neutral surfaces and critical mixtures the convective
motion is driven by a bulk Korteweg effect, i.e., a forcing through
the Korteweg stresses in the diffuse interface region between the two
components of the mixture.

Homogeneous films are still a base state for quadratically
energetically biased surfaces.  We have investigated the case
as a benchmark for comparison with literature
\cite{FMD97,FMD98,Kenz01}. As expected we have found that both -
symmetrical and antisymmetrical energetic bias -- stabilize the
lateral instability below respective threshold thicknesses. This is,
however, not the case for an asymmetrical bias; a question that needs
further investigation.

We have found that the stability behavior of layered films under different
types (symmetrical, antisymmetrical and asymmetrical) of energetic bias
is rather rich.  For purely diffusive transport, depending on the
type of branch and type of bias an increase in film thickness either
(i) exponentially decreases the lateral instability or (ii) it
entirely stabilizes the film when reaching a threshold
thickness. Including convective transport leads in many cases to a
small or strong destabilization as compared to the purely diffusive
case without changing the behavior qualitatively. Most remarkably,
however, in some cases the inclusion of convective transport and the
related widening of available configuration space for the film (it may
then change its profile) changes the stability qualitatively. For
instance, an asymmetrical energy bias results in a stabilization at a
threshold thickness when no convection is allowed. With convection the
film remains unstable well above this threshold, although it still
becomes exponentially more stable with increasing film thickness.

Finally, we have presented results regarding the dependence of the
instability on several other parameters, namely, the Reynolds number, the
Surface tension number and the ratio of typical velocities of
convective transport driven by Korteweg stresses and by diffusive
transport. We have identified the latter parameter as the most
influential one. 

Note that we have entirely focused on the case of a critical mixture,
i.e., the case of zero mean concentration.  A full scale analysis for
off-critical systems, for instance, of the dependence of stability on
mean concentration has been outside the scope of the present paper and
will be pursued elsewhere.

In general, our results have shown that the possibility of a change in
the height profile of the fill alters the stability behavior of a film
of a binary mixture not only quantitatively but as well
qualitatively. Therefore such a possibility has always to be taken
into account.

Although, the performed stability analysis yields results regarding
stability thresholds, and time- and length-scales of the most dangerous
modes, it does not allow to predict the nonlinear short- and long-time
evolution of an unstable film of a binary mixture. Results obtained on
related two-layer films using long-wave sharp interface
models\cite{PBMT04,PBMT05,PBMT06} show that the nonlinear behavior
may be quite unexpected. An example are the morphological transitions
that may occur in the process of coarsening\cite{PBMT05,PBMT06}.  To
fully account for the behavior the free surface model-H should be
used to study the time evolution of films of binary mixtures in the
nonlinear regime. An alternative approach could focus solely on the
final stable structures, i.e., steady state films that are
characterized by a steady non-flat surface profiles and internal
non-homogeneous concentration profiles.

Finally, we would like to point out that the given analysis
only presents a first step towards a fully operative thin film model
that is able to describe in a quantitatively correct way the coupled
dewetting and decomposition observed in many experiments using films
of polymer blends of thicknesses well below 100nm.\cite{GeKr03}. A
basic ingredient of thin film physics -- the effective molecular
interactions between film and substrate -- has not yet been taken into
account. The approach followed in thin film models of simple liquids
is to include those interactions as an additional pressure term that
depends on film thickness and models wettability.\cite{deGe85,KaTh07}.

Nevertheless, the present theory allows to formulate a hypothesis that
may explain why rather thick films (100-200nm) of polymer blends can
rupture rather fast:\cite{noteUlli} The presently investigated
decomposition-caused convection-mediated lateral instability leads to
a growing lateral modulation of the film surface. This allows to
'bypass' the rather slow linear phase of the dewetting instability as
the film is taken directly into the non-linear regime of it.

\acknowledgments

This research was supported by a Marie Curie European Reintegration
Grant within the 7$^{th}$ European Community Framework Programme, by 
the European Union, Deutsche
Forschungsgemeinschaft, and Universidad Polit\'ecnica de Madrid under
grants PERG04-GA-2008-234384, MRTN-CT-2004-005728, SFB 486 B13 and AL09-P(I+D)
respectively. UT thanks MPIPKS Dresden and Peter H\"anggi
(Universit\"at Augsburg) for support at various stages of the project.


\begin{thebibliography}{10}
\bibitem{note1}
 The parameters are $a=(N_a\,\rho \,k_B T/M_w)\,(-\chi/2+N')$ and $b=(N_a\,\rho
  \,k_B T/M_w)\,(N'/3)$, with $M_w$ molecular weight, $N_a$ Avogadro number,
  $\rho$ density, $k_B$ Boltzman constant, $N'$ number of monomers. The
  parameter $\chi$ is determined in Flory-Huggins theory to be
  $\chi(c,T)=\left(P_0+P_1/T \right) \left(1+P_2(1+C)(1+C)/4\right)$, where
  $P_0$, $P_1$ and $P_2$ are phenomenological constants. It is determined
  experimentally by Morariu and Steiner\cite{noteUlli}.
\end{thebibliography}
\end{document}